\documentclass[preprint,pteplogo]{ptephy_v2}

\preprintnumber{UTHEP-803, UTCCS-P-166, HUPD-2504, KEK-TH-2716} 
\usepackage{hyperref}

\usepackage{mathrsfs}

\usepackage{multirow}
\usepackage{graphicx}
\usepackage{slashed} 
\usepackage{tablefootnote} 

\begin{document}

\title{Investigating the axial structure of the nucleon based on large-volume lattice QCD at the physical point}


\author[a]{Ryutaro Tsuji}
\author[b]{Yasumichi Aoki}
\author[c]{Ken-Ichi Ishikawa}
\author[d]{\\ Yoshinobu Kuramashi}
\author[e]{Shoichi Sasaki}
\author[f]{Kohei Sato}
\author[d,g]{Eigo Shintani}
\author[h]{Hiromasa Watanabe}
\author[i,d]{Takeshi Yamazaki}
\affil[ ]{\normalsize{\bf \sffamily (PACS Collaboration)}}

\affil[a]{High Energy Accelerator Research Organization (KEK),
  305-0801, Tsukuba, Japan \email{rtsuji@post.kek.jp}}

\affil[b]{RIKEN Center for Computational Science,
  650-0047, Kobe, Japan}

\affil[c]{Graduate School of Advanced Science and Engineering, Hiroshima University,
  739-8526, Higashi-Hiroshima, Japan}

\affil[d]{Center for Computational Sciences, University of Tsukuba,
  305-8577, Tsukuba, Japan}

\affil[e]{Department of Physics, Tohoku University,
  980-8578,Sendai, Japan}

\affil[f]{Seikei University, Office of the President, 180-8633,
Tokyo, Japan}

\affil[g]{Graduate School of Engineering, University of Tokyo, 113-0033, Tokyo, Japan}

\affil[h]{Department of Physics, and Research and Education Center for Natural
Sciences, Keio University, 223-8521, Kanagawa, Japan}

\affil[i]{Institute of pure and Applied Sciences, University of Tsukuba,\\
  305-8571, Tsukuba, Japan}


\begin{abstract}%
We present a short summary for the calculations of the nucleon \textit{isovector} form factors,
which are relevant to improving the accuracy of the current neutrino oscillation experiments.
The calculations are carried out with two of three sets of the $2+1$ flavor lattice QCD configurations generated at the physical point in large spatial volumes by the PACS Collaboration. 
The two gauge configurations are generated with the six stout-smeared $O(a)$ improved Wilson quark action and Iwasaki gauge action
at the lattice spacing of $0.09$ fm and $0.06$ fm.
We summarize the results for three form factors as well as the nucleon axial-vector ($g_A$), induced pseudoscalar ($g_P^*$) and pion-nucleon ($g_{\pi NN}$) couplings.
Although our couplings agree with the experimental data,
a firm conclusion should be drawn only after a continuum limit extrapolation is taken.
We investigate the partially conserved axial-vector current (PCAC) relation in the context of the nucleon correlation functions. 
The low-energy relations arising from the PCAC relation can be used to verify whether the lattice QCD data correctly reproduce the physics in the continuum within the statistical accuracy.
It is demonstrated that our \textit{new analysis} reduces the systematic uncertainty for the induced pseudoscalar and pseudoscalar form factors to a greater extent than the \textit{traditional analysis},
and the results offer a theoretical insight into the pion-pole dominance model.
Finally,
we examine the applicable $q^2$ region for the low-energy relations. 
\end{abstract}

\subjectindex{Nucleon structure, Lattice QCD}

\maketitle

\section{Introduction}
\label{sec:introduction}

The low-energy physics of quantum chromodynamics (QCD) can be rigorously described in discretized Euclidean space-time and numerically solved by lattice QCD with High Performance Computing (HPC).
In recent years,
lattice QCD simulations have achieved a very precise determination of observables that can be compared to experimental measurements.
It is noteworthy that it has achieved an unprecedented level of accuracy in calculating not only the QCD parameters and properties of the QCD vacuum, such as the running coupling ($\alpha_s$), the quark masses, and the chiral condensate, but also the properties of hadrons, including
hadron mass spectroscopy, the matrix elements that arise in the semileptonic decay
(\textit{e.g.} $K\to \pi l\nu$, $D_{(s)}\to X_{(s)} l\nu$, and $B_{(s)}\to X_{(s)}l\nu$),
and its consequences for the Cabibbo-Kobayashi-Maskawa (CKM) matrix elements~\cite{FlavourLatticeAveragingGroupFLAG:2024oxs, Barone:2023tbl, Kellermann:2025pzt, DeSantis:2025qbb, DeSantis:2025yfm}.
Furthermore, it is worth mentioning that
there are ongoing developments in the field of algorithms.
These developments involve the utilization of artificial intelligence (AI) and quantum computing (QC) to investigate strongly coupled systems~\cite{Bauer:2022hpo, Choi:2024ryu, 
Lawrence:2025rnk, Tomiya:2025quf}.

Lattice QCD is the sole \textit{ab initio} approach for studying the nucleon structure in terms of the quark-gluon dynamics.
This is because many important features concerning the nucleon structure are primarily determined by the nature of non-perturbative QCD, which cannot be expressed in perturbation theory.
In fact, recent lattice QCD simulations with realistic physical parameters (\textit{e.g.}, the number of dynamical quarks and their quark masses) on finer lattices with larger spatial extents
have reproduced the nucleon axial-vector coupling ($g_A$) with a few percent of the statistical precision~\cite{FlavourLatticeAveragingGroupFLAG:2024oxs}.
We are now ready to begin a high-precision investigation of the nucleon structure based on large-volume lattice QCD at the physical point.

In the context of QCD, the nucleon structure is of particular interest,
because the structure itself is a nontrivial consequence of quark-gluon dynamics.
The proton radius puzzle,
which has become well known as the discrepancy in experimental measurements of electric root-mean-square (RMS) radius of the proton,
is one of the subjects to be solved.
The proton radius is of great interest due to the fact that it can be determined by various and different methods,
including spectroscopy using both muonic and ordinary hydrogen atoms, as well as elastic scatterings of both electrons and muons off protons.
This feature is of considerable significance in the validation of Quantum Electrodynamics (QED) and QCD with regard to hadron and nuclear physics.

For example,
the 21 cm line in hydrogen (1S hyperfine splitting) has been employed in the testing of QED.
However, while hydrogen maser frequencies have been experimentally measured to 12 digits~\cite{Essen:1971},
the QED test is limited to only 6 digits due to the proton structure effects~\cite{Eides:2007exa}.
This indeed indicates that
the structure of the nucleon has not yet been fully clarified from the perspective of the QCD theory due to its non-perturbative features.

Studies of nucleon structure are closely related to the
primary aim of the future Electron-Ion Collider (EIC) facility at Brookhaven National Laboratory,
which is to understand the quark-gluon structure of matter –– proton and neutron~\cite{NAP25171, AbdulKhalek:2021gbh}.
In particular,
there are three immediate and profound questions about the nucleon (\textit{i.e.}, the proton or the neutron):
\begin{itemize}
    \item \textit{How does the mass of the nucleon arise?}
    \item \textit{How does the spin of the nucleon arise?}
    \item \textit{What are the emergent properties of dense systems of gluons?}
\end{itemize}
The answer to these questions requires a better understanding of the nucleons as quantum many-body systems of strongly interacting quarks and gluons.
In other words,
knowledge gained from the EIC provides insight into the fundamental questions by offering a deeper understanding of how the color force holds protons and neutrons together in nuclei.

Among the questions to be addressed by the EIC,
the origin of the nucleon mass and spin is of particular interest.
The total nucleon mass is experimentally measured as $M_N^\mathrm{exp} = 0.939\ [\mathrm{GeV}]$.
However the portion carried by the masses of the valence quarks is only about 1\%, which is directly related to the origin of the Higgs condensation.
The rate of quark and gluon contributions to nucleon mass (reffered to as the mass decomposition) can only be determined by solving QCD nonperturbatively or through experimental data.

As well as the mass decomposition,
there is a long-standing problem known as the \textit{spin crisis}, which is related to the origin of the spin.
In the quark model,
the spin of the nucleon is simply the sum of the spins of the three valence quarks. 
However, the deep inelastic scattering (DIS) European Muon Collaboration (EMC) experiment of polarized muons on polarized protons found that
the quark spin ($\Delta\Sigma$) contributes very little to the proton spin (see Refs.~\cite{Hughes:1983kf, EuropeanMuon:1987isl, EuropeanMuon:1989yki}).
Although the recent global analysis of experiments gives the $\Delta\Sigma$ at about 30\% level~\cite{deFlorian:2009vb, Nocera:2014gqa, Ethier:2017zbq}, the rest of the proton spin remains unknown.

To reveal the origin of the mass and the spin,
gravitational form factors (GFFs)~\cite{Ji:2021mtz, Burkert:2023wzr} play an essential role.
The GFFs are defined as the matrix elements of the energy-momentum tensor (EMT).
These provide direct access to the mass, spin~\cite{Ji:1996ek, Yang:2014xsa, Hatta:2018sqd, Tanaka:2018nae,  Metz:2020vxd, Liu:2021gco, Tanaka:2022wzy} and
angular momentum~\cite{Lorce:2017wkb, Granados:2019zjw, Schweitzer:2019kkd} distributions,
as well as the pressure and shear distributions~\cite{Polyakov:2002wz, Polyakov:2002yz, Pasquini:2014vua, Lorce:2015lna, Lorce:2018egm,  Kumano:2017lhr, Burkert:2018bqq}, inside hadrons.
According to Refs.~\cite{Ji:1996ek, Polyakov:2018zvc, Tanaka:2018nae},
the matrix element of the symmetric QCD energy-momentum tensor $T^{\mu\nu}_{q,g}$ ($q$:quark, $g$:gluon),
using the nucleon states $| N(p)\rangle$ and $| N(p^\prime)\rangle$ with
the four-momenta $p$ and $p^\prime$,
is given by
\begin{align}
    \label{eq:ggf}
    \langle N(p^\prime) |  T^{\mu\nu}_{q,g} | N(p)\rangle
    =
    \bar{u}_N(p^\prime)
    &
    \left[ 
    A_{q,g}(t)\gamma^{( \mu}\bar{P}^{\nu )} 
    + B_{q,g}(t) \frac{\bar{P}^{(\mu}i\sigma^{\nu)\alpha}\Delta_\alpha}{2M_N}
    \right.\nonumber \\
    & \quad \left.
    + D_{q,g} \frac{\Delta^\mu\Delta^\nu-\eta^{\mu\nu}t}{4M_N}
    + \bar{C}_{q,g}(t)M_N\eta^{\mu\nu}
    \right]
    u(p)_N
\end{align}
with the GFFs $A_{q,g}(t), B_{q,g}(t), D_{q,g}(t)$ and $\bar{C}_{q,g}(t)$,
where $M_N$ and $u(p)_N$ are the nucleon mass and spinor,
$\eta_{\mu\nu}$ is the metric tensor, $\eta_{\mu\nu}=\mathrm{diag}(1,-1,-1,-1)$,
and other kinematical variables are given by $\bar{P}=\frac{p^\mu+p^{\prime\mu}}{2}$, $\Delta^\mu=p^{\prime\mu}-p^\mu$, and $t=\Delta^2$.
The values of $A_{q,g}(t)$ and $B_{q,g}(t)$ at $t=0$ lead to
the following relations:
\begin{align}
    \label{eq:parton_momentum_fraction}
    \langle x \rangle_{q,g} & = A_{q,g}(0), \\
    \label{eq:parton_total_spin}
    J_{q,g} & = \frac{1}{2}\left[ A_{q,g}(0) + B_{q,g}(0) \right],
\end{align}
where $\langle x \rangle_{q,g}$ and $J_{q,g}$ represent the momentum fraction and the total spin carried by the quark or gluons inside a nucleon, respectively.
First, as for the mass decomposition,
the rest mass of the nucleon is decomposed into 
\begin{align}
    \label{eq:nucleon_rest_mass}
    M_N
    & =
    \left.
    \frac{1}{2M_N}\langle p(p) | T^{00} | p(p) \rangle 
    \right|_{\boldsymbol{p}=\boldsymbol{0}}
    =
    M_p
    \left(
    A_q(0) + A_g(0) + \bar{C}_q(0) + \bar{C}_g(0)
    \right)
\end{align}
in terms of the values of GFFs at $t=0$.
The proton mass decomposition can be investigated by combining the values of the GFFs and some non-perturbative inputs, such as the quantity $\langle N(p) | m_q\bar{q}q | N(p) \rangle$ for each quark $q$ up to the charm~\cite{Hatta:2018sqd, Tanaka:2022wzy}.
The total spin of the nucleon $J$ can be investigated through Eq.~(\ref{eq:parton_total_spin}) as $J=J_q+J_g$.
In particular,
the quark contribution $J_q$ is also decomposed into 
the contributions from the intrinsic quark spin $\Delta\Sigma$ and the quark orbital angular momentum $L_q$
as $J_q=\frac{1}{2}\Delta\Sigma+L_q$~\cite{Jaffe:1989jz, Ji:1996ek}.
$\Delta\Sigma$ can be obtained from
the nucleon matrix element of the flavor-singlet axial-vector current $A_\alpha^0=\sum_f\bar{q_f}\gamma_\alpha\gamma_5 q_f$,
which leads to $\langle N(p,s) | A^0_\alpha | N(p,s) \rangle = 2M_Ns_\alpha\Delta\Sigma$
with
the nucleon covariant spin vector $s_\alpha$ normalized as $s_\alpha^2=-1$.

Lattice QCD has the potential to contribute to the EIC physics by calculating the aforementioned non-perturbative inputs from first principles.
It is important to note that there are some pioneering studies on the mass and spin decompositions~\cite{Fukugita:1994fh, Dong:1995rx, Alexandrou:2017oeh, Liang:2018pis, Lin:2018obj, Yang:2018nqn, Liu:2021gco, Liu:2021lke, FlavourLatticeAveragingGroupFLAG:2024oxs}.
However,
it is still necessary to investigate the possible systematic uncertainties associated with the nucleon structure studies using lattice QCD.
One of the challenges for the lattice QCD computation of the GFFs is constructing 
the renormalized EMT on the lattice.
This is because the loss of translational invariance is inevitable due to the discretization of the space-time,
whereas the EMT is proportional to the derivative of the Noether current with respect to the scale transformations.
Although
the gradient Yang-Mills flow approach has been proposed as a means of constructing the renormalized EMT operator from the flowed fields~\cite{Suzuki:2013gza} (the so-called ``small flow time expansion") to overcome this problem,
detailed studies are necessary to investigate the systematic uncertainties.
Another challenge for flavor-singlet quantities requires a calculation involving the ``disconnected contribution'',
where the current is inserted into the quark loop to yield the vacuum polarization contribution (see Fig.~\ref{fig:quarkline_diagrams} in Sec.~\ref{sec:lattice_method_to_determine_nucleon_matrix_elements}).
Computing the disconnected contributions is costly and subject to large statistical fluctuations. 
Therefore, a new algorithm has been proposed to calculate them more accurately~\cite{Giusti:2019kff, Gerardin:2023naa}.
In general,
the systematic uncertainties from excited-state contamination, the finite-size effect, and discretization effect must also be examined.
Thus, the high-precision calculations for the GFFs and the related non-purturbative inputs remain challenging and are still in progress.

Under these circumstances, as part of the PACS Collaboration,
we have computed the nucleon form factors associated with the electromagnetic and weak interactions.
To this end, high-precision calculations with statistical accuracy of a few percent have become available~\cite{Ishikawa:2018jee, Ishikawa:2018rew, Shintani:2018ozy, Ishikawa:2021eut, Tsuji:2022ric, Tsuji:2023llh, Tsuji:2024scy, Aoki:2025taf},
while the pilot study of the momentum and helicity fractions of the nucleon, which are related to the EIC, is currently underway, with about 10\% uncertainty~\cite{Tsuji:2021bdp}.

We have been performing large-volume lattice QCD calculations of the nucleon form factors with physical light and strange quark masses (referred to hereafter as the physical point),
using gauge configurations generated by the PACS Collaboration.
In a series of studies, 
certain essential features are retained, particularly in those employing the PACS10 gauge configurations:
1) we performe \textit{fully dynamical lattice QCD simulation,}
where both valence and sea quarks are treated by the same action,
2) the physical spatial size is maintained at larger than $10\ \mathrm{fm}$, where the finite-size effect can be sufficiently ignored and furthermore the nonzero minimum value of the momentum transfer reaches about $q^2\sim 0.01\ [\mathrm{GeV}^2]$,
3) the quark masses are carefully tuned to the physical point, indicating that our simulations are free from
an extrapolation from an unphysical simulation point to the physical point (referred to as the chiral extrapolation).
These specific features allow us to overcome systematic uncertainties due to chiral and infinite-volume extrapolations and access the low $q^2$ region, $q^2\lesssim 0.1\ [\mathrm{GeV}^2]$.

This paper is organized as follows. In Sec.~\ref{sec:axial_structure_of_the_nucleon},
we first present a brief summary of the neutrino oscillation experiments,
where the high-precision determination of the nucleon axial structure plays an important role.
The axial matrix elements of the weak current and the axial Ward--Takahashi identity 
and its related low-energy relations are also introduced.
Definitions and notations for the nucleon form factors are also summarized in this section.
In Sec.~\ref{sec:lattice_method_to_determine_nucleon_matrix_elements},
we provide a brief introduction to
the lattice QCD methodology for calculating nucleon two- and three-point correlation functions and determining the nucleon matrix elements, including our \textit{new analysis} to evaluate the nucleon form factors.
Furthermore, the PCAC quark mass is introduced to verify the Partially Conserved Axial Current (PCAC) relation using the nucleon.
Section~\ref{sec:simulation_details} is devoted to the details of our Monte Carlo simulations, including a brief description of the gauge configurations generated by the PACS Collaboration.
In Sec.~\ref{sec:numerical_results_i},
we present a highlight of our basic results reported in our previous studies~\cite{Shintani:2018ozy, Tsuji:2023llh, Tsuji:2024scy, Sasaki:2025qro}.
In Sec.~\ref{sec:numerical_results_ii} presents the results of the PCAC quark mass calculated with two sets of the PACS10 ensembles. We also discuss the implications of our finding from a direct comparison of two types of the PCAC quark mass.
Subsequently, Section~\ref{sec:numerical_results_iii} is devoted to
examining the results of low-energy relations, such as the generalized Goldberger-Treiman relation and the pion-pole dominance model. In Sec.~\ref{sec:discussion}, the validity region of $q^2$ in the context of the low-energy relations will be addressed.
Finally, we close with a summary and outlook in Sec.~\ref{sec:summary_and_outlook}.

In this paper, hereafter, the matrix elements are given in the Euclidean metric convention. $\gamma_5$ is defined by $\gamma_5 \equiv \gamma_1\gamma_2\gamma_3\gamma_4=-\gamma_5^M$, which has the opposite sign relative to that in the Minkowski convention ($\vec{\gamma}^M=i\vec{\gamma}$ and $\gamma_0^M=\gamma_4$) adopted in the particle data group.
The sign of all the form factors is chosen to be positive. The Euclidean four-momentum squared $q^2$ corresponds to the spacelike momentum squared as $q^2_M=-q^2<0$ in Minkowski space.

\section{Axial structure of the nucleon}
\label{sec:axial_structure_of_the_nucleon}

\subsection{Nucleon structure and neutrino oscillation experiments}
\label{ssec:nucleon_structure_and_neutrino_oscillation_experiments}
In this paper,
before proceeding to the study of
the distribution of mass, spin, and pressure inside the nucleon,
we will focus especially on the axial structure of the nucleon, such as the axial ($F_A$), induced pseudoscalar ($F_P$), pseudoscalar ($G_P$) form factors, and the associated couplings determined by these form factors at a specific $q^2$ value.
The $q^2$ dependence of the $F_A$ and $F_P$ form factors can be used as an important input for neutrino oscillation experiments,
which is required to improve the accuracy of neutrino energy reconstruction in charged-current quasi-elastic scattering off nuclear targets.
This is because these form factors describe the weak process associated with the neutrino-nucleus scattering~\cite{Kronfeld:2019nfb, Meyer:2022mix, Ruso:2022qes, Ashkenazi:2025ust}.

The associated couplings
have relevant contributions to a wide range of physics such as 
the search for new physics and phenomenological implications.
The axial-vector coupling ($g_A$), 
which can be determined from $g_A=F_A(q^2=0)$, 
is associated with the neutron lifetime puzzle~\cite{Czarnecki:2018okw}. 
Since the discrepancy between the results of beam experiments and storage experiments remains unsolved, 
it is still an open question that deserves further investigation.
The induced pseudoscalar coupling $g_P^*$ and pion-nucleon coupling $g_{\pi NN}$ are defined through 
$ g_P^* = m_\mu F_P(q^2=0.88m_\mu^2)$ and $g_{\pi NN} \equiv \lim_{q^2 \to -M_\pi^2}\frac{M_\pi^2 + q^2}{2F_\pi}F_{P}(q^2)$
with muon mass $m_\mu$, 
pion mass $M_\pi$, and pion decay constant $F_\pi$.
Although
the $g_P^*$ and $g_{\pi NN}$ are obtained by the experiments of 
the muon capture experiments~\cite{MuCap:2015boo}
and the pion-nucleon scattering~\cite{Babenko:2016idp, Limkaisang:2001yz}
with some help of the theoretical analysis using 
the low-energy chiral effective models,
the values are much less known than $g_A$.

A very precise knowledge of the $F_A$ and $F_P$ form factors
plays crucial roles in the modern and future long-baseline neutrino oscillation experiments 
such as T2K/HK, DUNE and NOvA~\cite{Nunokawa:2007qh, T2K:2011ypd, T2K:2011qtm, Hyper-KamiokandeProto-:2015xww, DUNE:2015lol, DUNE:2020jqi, DUNE:2020ypp, Meyer:2022muy, T2K:2024wfn, Ashkenazi:2025ust} (see Table~\ref{tab:neutrino_experiments} and Ref.~\cite{Ashkenazi:2025ust} for a summary including other categories).
The precise measurements for the neutrino oscillation are one of the driving forces
to investigate 
absolute masses of the neutrino, the ordering of the mass states, and the charge-parity violating phase.
The neutrino oscillation probability can be obtained from the ratio of the neutrino flux measured at the near and far detectors.
However, the experiments measure only the interaction rate~\cite{CLAS:2021neh}.
Thus,
the neutrino flux $\Phi$ should be reconstructed by the Monte-Carlo event generator as
\begin{align}
    N_f(E_{\mathrm{rec}},L)
    \propto
    \sum_i\int
    \Phi_f(E_\nu,L)
    \sigma_i(E_\nu)
    g_{\sigma_i}(E_\nu,E_{\mathrm{rec}})
    dE_\nu,
\end{align}
where $N_f(E_{\mathrm{rec}},L)$ denotes the interaction rate for lepton flavor $f$ with the reconstructed energy $E_{\mathrm{rec}}$ and beam line $L$.
A subscript $i$ represents the processes of the neutrino interaction, such as quasi-elastic scattering and deep inelastic scattering, and then $\sigma_i(E_\nu)$ indicates the neutrino-nucleon interaction cross section with the neutrino incident energy $E_\nu$.
In addition, 
$g_{\sigma_i}(E_\nu,E_{\mathrm{rec}})$ is the smearing matrix of $E_\nu$ and $E_{\mathrm{rec}}$,
which is subject to variation
due to experimental and nuclear interaction effects.

\begin{table}[tb]
\centering
\footnotesize
\begin{tabular}{l|l|p{6.5cm}}
\hline
\hline
\textbf{Category} & \textbf{Experiment} & \textbf{Main Goals \& Features} \\
\hline
\multirow{3}{*}{Long-Baseline} 
  & T2K/HK~\cite{T2K:2011ypd, T2K:2011qtm, Hyper-KamiokandeProto-:2015xww, T2K:2023smv, T2K:2024wfn} & Precision neutrino oscillation, CP violation \\
  & DUNE~\cite{DUNE:2015lol, DUNE:2020jqi, DUNE:2020ypp} & CP violation, mass hierarchy, supernova neutrinos, proton decay \\
  & NOvA~\cite{NOvA:2019cyt} & $\nu_\mu \rightarrow \nu_e$ appearance, mass ordering, CP violation \\
\hline
\multirow{4}{*}{Short-Baseline} 
  & MicroBooNE~\cite{MicroBooNE:2015bmn, MicroBooNE:2019nio, MicroBooNE:2023tzj} & Investigate possible systematics associated with the published MiniBooNE data\\
  & SBND~\cite{McConkey:2017dsv, Machado:2019oxb} / ICARUS~\cite{Farnese:2019xgw} & Sterile neutrino search, short-baseline oscillations \\
  & ProtoDUNE~\cite{Brizzolari:2024xma} & Liquid Ar time projection chamber (TPC) testbed for DUNE, beam studies \\
  & NuSTORM~\cite{nuSTORM:2012jbd} & Muon storage ring for precision flux, cross-sections, BSM search \\
  & MINERvA~\cite{MINERvA:2013kdn, MINERvA:2023avz} & Neutrino-nucleus interaction measurements, cross-section measurements \\
\hline
\multirow{3}{*}{Telescopes / Forward} 
  & IceCube~\cite{IceCube:2013low, IceCube:2019dqi} & High-energy astrophysical neutrinos \\
  & KM3NeT~\cite{KM3Net:2016zxf, KM3NeT:2018wnd, KM3NeT:2021ozk} & Mass ordaring (ORCA), and detection of high-energy neutrinos from astrophysical sources (ARCA) \\
  & FASER$\nu$~\cite{FASER:2023zcr, FASER:2024hoe} & Detect high-energy forward neutrinos from LHC across all three flavors: electron, muon, and $\tau$ \\
\hline
Future Project 
  & ESSnuSB~\cite{ESSnuSB:2013dql, ESSnuSB:2021azq} & High-power superbeam for precision CP violation study \\
\hline
\hline
\end{tabular}
\caption{A brief summary of modern and future accelerator- and telescope-based neutrino experiments.
}
\label{tab:neutrino_experiments}
\end{table}

This reconstruction is accomplished by three steps in experimental and theoretical procedures:
1) extracting information of the $F_A$ form factor from neutrino-nuclei scattering experiments, 2) calculating theoretical cross sections for certain target nuclei using the nuclear many body theory~\cite{Kronfeld:2019nfb}
and 3) evaluating the neutrino flux by the Monte-Carlo event generator using the cross sections as input.

However, 
the reconstruction inevitably includes a major source of systematic uncertainty in the extraction of the $F_A$ form factor and also in the calculation of the final cross section.
To evaluate the neutrino-nucleon quasi-elastic cross sections,
the decomposition of the unpolarized differential cross section is often done using the structure-dependent functions of $A$, $B$ and $C$~\cite{LlewellynSmith:1971uhs, Tomalak:2023pdi} as
\begin{align}
\label{eq:dcs}
    \frac{d \sigma}{d q^2}\left(E_\nu, q^2\right)=\frac{G_{F}^2\left|V_{u d}\right|^2}{2 \pi} \frac{M_N^2}{E_\nu^2}\left[\left(\tau+r_{\ell}^2\right) A\left(\nu, q^2\right)-\frac{\nu}{M_N^2} B\left(\nu, q^2\right)+\frac{\nu^2}{M_N^4} \frac{C\left(\nu, q^2\right)}{1+\tau}\right],
\end{align}
where three kinematic variables are given by $\nu=E_\nu/M_N - \tau - r_\ell^2$, $\tau=q^2/4M_N^2$ and $r_\ell=m_l/2M_N$ with a mass of lepton $m_l$.
In addition,
$V_{ud}$ and $G_F$ represent the CKM matrix element and the Fermi coupling constant, respectively.
The structure dependent functions $A$, $B$ and $C$ are expressed in terms of four types of nucleon form factors ($G_E^v, G_M^v, F_A, F_P$) that are responsible for
describing the structure of the nucleon as
\begin{align}
A & =\tau\left(G_M^v\right)^2-\left(G_E^v\right)^2+(1+\tau) F_A^2
 -r_{\ell}^2\left[\left(G_M^v\right)^2+F_A^2+4 F_P\left(F_A-\tau F_P\right)\right], 
 \label{eq:dcs_A}\\
B & =4 \eta \tau G_M^v F_A, \\
C & =\tau\left(G_M^v\right)^2+\left(G_E^v\right)^2+(1+\tau) F_A^2,
\end{align}
where $\eta=+1(-1)$ denotes the (anti)neutrino scattering.
The total quasi-elastic cross sections are obtained by integrating Eq.~(\ref{eq:dcs}) over the kinematically allowed region of the momentum transfer: $q^2_{-}\le q^2 \le q^2_{+}$ with
\begin{align}
q_{\pm}^2 & =\frac{2 M_N E_\nu^2}{M_N+2 E_\nu}-4 M_N^2 \frac{M_N+E_\nu}{M_N+2 E_\nu} r_{\ell}^2
\pm \frac{4 M_N^2 E_\nu}{M_N+2 E_\nu} \sqrt{\left(\frac{E_\nu}{2 M_N}-r_{\ell}^2\right)^2-r_{\ell}^2}.
\end{align}

Current neutrino-deuteron and neutrino-nucleon scattering experiments have an uncertainty of 23-34\% in determining the axial radius~\cite{Kuzmin:2007kr, Meyer:2016oeg, MINERvA:2023avz, Tomalak:2023pdi}:
\begin{align}
    \label{eq:ra}
    \langle r_A^2 \rangle
    =
    -\frac{6}{F_A(0)}
    \left. \frac{dF_A}{dq^2} \right|_{q^2=0},
    \end{align}
which is a useful quantity to compare
the accuracy of measurements in the low-$q^2$ region.
In particular, accurate information of the $F_A$ form factor is essential to improve the analysis of the T2K experiment in Japan~\cite{T2K:2011qtm, T2K:2011qtm, Hyper-KamiokandeProto-:2015xww, T2K:2023smv, T2K:2024wfn}.
The T2K experiment mainly performs a search for $\mu$ neutrino to $e$ neutrino oscillation~\footnote{
Not only the neutrino oscillation studies using the atomospheric neutrino,
T2K also performs a search for $\mu$ neutrino to $\tau$ neutrino oscillation,
and measures the difference between the square of the mass of the $\mu$ neutrino and that of the $\tau$ neutrino.
}. 
The neutrino beam is produced by a pion decay which is provided in collisions between a proton beam and a graphite target at Tokai.
The energy range of the T2K neutrino beam is centered at 600 MeV, since $\mu$ neutrinos of this energy tends to oscillate while traveling between Tokai and Kamioka (295 km).
The dominant process of the neutrino-nucleus scattering in this energy range is quasi-elastic scattering,
therefore it is crucial to understand the systematics of the neutrino-nucleus scattering in order to improve accuracy.

One of the current obstacles 
in the systematic uncertainty of the T2K experiment concerns
the determination of the $F_A$ form factor~\cite{T2K:2023smv}.
In fact,
the $F_A$ form factor carries the largest uncertainties on the final cross section among the four form factors.
In the analysis of experiments, the $F_A$ form factor is often parameterized using the dipole ansatz as
\begin{align}
    \label{eq:fa_dipole}
    F_A(q^2) = \frac{F_A(0)}{(1+\frac{q^2}{M_A^2})^2},
\end{align}
where $F_A(0)=g_A$ is the nucleon axial-vector coupling, and $M_A$ denotes the axial mass,
which is also widely used to discuss the $q^2$ dependence of the $F_A$ form factor.
Under this dipole ansatz,
the axial radius is given as
\begin{align}
    \label{eq:ra_dipole}
    \langle r_A^2 \rangle
    =
    \frac{12}{M_A^2}.
\end{align}
Experimental data of $F_A(q^2)$ 
were extracted from the neutrino-deuteron scattering data provided in the 1980s by the bubble chamber experiments~\cite{Baker:1981su, Miller:1982qi} and a dipole mass of $M_A=1.026(21)$ [GeV] 
was determined in Ref.~\cite{Bernard:2001rs}.

The conventional value of $M_A\sim1.0$ [GeV]
has been widely used in the neutrino oscillation experiments.
However, since the late 1990s, 
experiments using heavier nucleus have found some discrepancies at $1.0-1.3$ [GeV] for the axial mass $M_A$~\cite{ MiniBooNE:2010bsu, Golan:2012wx, Kakorin:2021axo}.
Especially,
the recent study performed by T2K also reports $M_A\sim1.2$ [GeV]~\cite{T2K:2023smv, T2K:2024wfn}.
These relatively new experimental values are indeed larger than the conventional value,
but this discrepancy could be 
due to a possible modification of the $F_A$ form factor associated with nuclear effects
and no firm conclusion has yet been reached.
In Ref.~\cite{T2K:2023smv},
the authors also reported that this discrepancy causes large systematic uncertainties, especially for the T2K beam energies and therefore more accurate and detailed studies of the $F_A$ form factor are needed.

In order to improve the accuracy of the long-baseline neutrino oscillation experiments,
the $F_A$ form factor determined from first principles calculation using lattice QCD can significantly reduce systematic uncertainties. 
The fact that neutrinos rarely interact with the target makes it difficult to accurately measure the axial radius in experiments, 
whereas the first-principles calculation has no such problem.
Notably, 
one of collaborations in the lattice QCD community, the PNDME Collaboration,
recently has reported a comparison of experimental and theoretical determinations of (anti)neutrino-nucleon total quasi-elastic cross sections~\cite{Tomalak:2023pdi}.
The authors concluded that both lattice QCD results and antineutrino-hydrogen measurements provide increasingly precise data leading to increased confidence in the region of $0.2\ [\mathrm{GeV}^2] \le q^2 \le 1\ [\mathrm{GeV}^2]$.
On the other hand,
the authors also pointed out that 
lattice QCD can provide robust data of the $F_A$ form factor with good statistical precision in the region of $q^2\lesssim 0.2\ [\mathrm{GeV}^2]$,
where large experimental uncertainties remain.

Regarding the uncertainties for the long-baseline neutrino oscillation experiments using (anti-)$\mu$ neutrino,
the uncertainty of the $F_A$ form factor has a dominant contribution to the total cross section
This is because
the $F_P$ form factor appears only in the third term of Eq.~(\ref{eq:dcs_A}),
which is proportional to the lepton mass squared and then becomes a subleading contribution to the total cross section.
The $F_P$ form factor has not been considered important, even though there is a large uncertainty due to a lack of experimental information.
However,
the uncertainty in the $F_P$ form factor can no longer be ignored,
if we consider the $\tau$ neutrino.
Indeed,
the uncertainty due to the $\tau$ neutrino cross section has a largest systematic uncertainty for the neutrino oscillation studies using atmospheric neutrino.
In order to reduce the uncertainty during the analysis of experiments,
the $q^2$ dependence of the $F_P$ form factor is often parametrized using the pion-pole dominance (PPD) model as $F^{\mathrm{PPD}}_{P}(q^2) \propto F_A(q^2)/({q^2+M_\pi^2})$,
however, the model is justified only 
in the chiral limit ($M_\pi \rightarrow 0$) and has no justification that is applicable to the data at the physical point.
This means that a first-principles calculation is required to provide the theoretical basis for applying the PPD model to the analysis of experiments.

\subsection{Axial matrix element of the weak current}
\label{ssec:axial_matrix_elements_of_the_weak_currents}

Let us consider
the nucleon matrix elements of the axial-vector current which is included in the weak current that describes the weak process, such as
the standard $\beta$-decay, muon capture, neutrino scattering.
Although the primary concern is the axial structure of the nucleon, the pseudoscalar current is considered here as well as the axial vector current, since we will later consider the axial Ward-Takahashi identity between the two currents.

The nucleon matrix elements of a renormalized quark bilinear \textit{isovector} currents
$J^O_\alpha=\bar{u}\Gamma_\alpha d$
with $\Gamma_\alpha=\gamma_\alpha\gamma_5, \gamma_5$ for the axial-vector ($A_\alpha)$ and pseudoscalar ($P$) currents
are decomposed into the following relativistically covariant forms in terms of three form factors ($F_A(q^2), F_P(q^2)$ and $G_P(q^2)$):
\begin{align}
    \label{eq:nme_fap}
    \langle p(\boldsymbol{p}^{\prime})| A_\alpha(x) | n(\boldsymbol{p}) \rangle 
    & =
    \bar{u}_p(\boldsymbol{p}^{\prime})
    \left(
    \gamma_\alpha\gamma_5 F_A(q^2) + iq_\alpha\gamma_5 F_P(q^2)
    \right)
    u_n(\boldsymbol{p}) e^{iq\cdot x}, \\
    \label{eq:nme_gp}
    \langle p(\boldsymbol{p}^{\prime})| P(x) | n(\boldsymbol{p}) \rangle 
    & =
    \bar{u}_p(\boldsymbol{p}^{\prime})
    \left(
    \gamma_5 G_P(q^2) 
    \right)
    u_n(\boldsymbol{p}) e^{iq\cdot x},
\end{align}
where $|p(\boldsymbol{p})\rangle$ and $|n(\boldsymbol{p})\rangle$ are the proton ($p$) and neutron ($n$) ground states with the three-dimensional momentum $\boldsymbol{p}$.
In the above equations,
the four-dimensional momentum transfer $q$ between the proton and neutron is given by $q=P-P^\prime$ with
$P^{\prime}=(E_p(\boldsymbol{p}^{\prime}), \boldsymbol{p}^{\prime})$ and $P=(E_n(\boldsymbol{p}), \boldsymbol{p})$.
It is important to note that this matrix element is 
flavor changing process,
where there is no disconnected contribution.

Theoretically,
in the axial part of the weak processes at low energies,
the physics of chiral symmetry and its spontaneous breaking induced by the strong interaction play important roles for providing low-energy relations associated with the axial structure of the nucleon.
The spontaneous chiral symmetry breaking
ensures the presence of pseudo Nambu-Goldstone particles such as the pion.
This concept was suggested by
the PCAC hypothesis~\cite{Nambu:1960xd, Gell-Mann:1960mvl},
in which the divergence of the axial-vector current is proportional to the pion field.
Applying this idea to the nucleon matrix element with the axial-vector current as shown in Eq.~(\ref{eq:nme_fap}),
a specific relation, known as the Goldberger--Trieman (GT) relation~\cite{Goldberger:1958tr, Gasser:1987rb}, is derived 
between the axial-vector coupling $g_A$ and the residue of the pion-pole structure of $F_P(q^2)$ together with the nucleon mass $M_N$.
Although the GT relation is exact only in the chiral limit,
this relation reproduces the experimental data with a level of accuracy of less than 3\%.

In a context of the low-energy physics of the axial structure of the nucleon,
the axial Ward--Takahashi (AWT) identity,
which is phenomenologically referred to as the PCAC relation,
plays essential roles.
The AWT identity with a \textit{renormalized} degenerate up and down quark mass $m=m_u=m_d$ is represented as
\begin{align}
\label{eq:awti}
\partial_\alpha A_\alpha(x)=2mP(x).
\end{align}

By sandwiching the AWT identity between the nucleon ground states,
the generalized Goldberger-Trieman (GGT) relation is provided~\cite{Weisberger:1966ip, Bernard:1994wn, Sasaki:2007gw}. 
The nucleon matrix element of the divergence of the axial-vector current is represented in the
following form:
\begin{align}
    \label{eq:nme_diva}
    \langle p(\boldsymbol{p}^{\prime})| \partial_{\alpha}A_\alpha(x) | n(\boldsymbol{p}) \rangle 
    & =
    \bar{u}_p(\boldsymbol{p}^{\prime})
    \left(
    i(\slashed p^\prime - \slashed p)
    F_A(q^2) - q^2 F_P(q^2)
    \right)\gamma_5
    u_n(\boldsymbol{p}) e^{iq\cdot x} \cr
    & = 
    \left[2M_NF_A(q^2) - q^2 F_P(q^2)\right]
    \bar{u}_p(\boldsymbol{p}^{\prime})\gamma_5
    u_n(\boldsymbol{p}) e^{iq\cdot x},
\end{align}
where the second line is obtained using
the Dirac equations, $\bar{u}_p(\boldsymbol{p}^{\prime})(i\slashed p^\prime+M_N)=0$ for the proton and $(i\slashed p+M_N)u_n(\boldsymbol{p})=0$ for the neutron 
with the nucleon mass $M_N=M_p=M_n$ under the isospin limit.
Therefore, the $q^2$ dependence of these three form factors, $F_A(q^2)$, $F_P(q^2)$ and $G_P(q^2)$,
should be constrained by the GGT relation:
\begin{align}
\label{eq:ggt}
    2M_NF_A(q^2) - q^2 F_P(q^2)
    =
    2mG_P(q^2).
\end{align}
In addition,
as a consequence of the AWT identity and GGT relation,
the pion-pole dominance (PPD) model of $F_P(q^2)$,
in which the $F_P$ form factor approximately behaves like
\begin{align}
\label{eq:ppd}
    F^{\mathrm{PPD}}_{P}(q^2)
=
\frac{2M_NF_A(q^2)}{q^2+M_\pi^2}
\end{align}
with the mass of pion ($M_\pi$) and nucleon ($M_N$),
is predicted.
It is important to note that this model becomes exact only in the chiral limit
where the pion is massless~\cite{Nambu:1960xd} (see Appendix B of Ref.~\cite{Sasaki:2007gw}).

According to the GGT relation of Eq.~(\ref{eq:ggt}) and the PPD model for $F_P(q^2)$ in Eq.~(\ref{eq:ppd}),
one may expect that the $G_P$ form factor also has the pion-pole structure:
\begin{align}
\label{eq:ppd_gp}
    2mG_P^{\mathrm{PPD}}(q^2)
    =
    2M_NF_A(q^2) \frac{M_\pi^2}{q^2+M_\pi^2},
\end{align}
which should be dominant in the low $q^2$ region~\cite{Nambu:1960xd, Gasser:1983yg,Bernard:1994wn}.
Consequently,
the ratio of $G^{\mathrm{PPD}}_P(q^2)$ and $F^{\mathrm{PPD}}_P(q^2)$ will not depend on $q^2$, and provides the low-energy constant $B_0$ as 
\begin{align}
\label{eq:gmor}
\frac{{G}_P^{\mathrm{PPD}}(q^2)}{F_P^{\mathrm{PPD}}(q^2)}
=
B_0
\end{align}
with the help of the Gell-Mann--Oakes--Renner (GMOR) relation for the pion mass: $M^2_\pi=2B_0m$~\cite{Gell-Mann:1968hlm}.
Therefore, if the PPD model is valid
in the low $q^2$ region, the ratio of $G_P(q^2)$ and $F_P(q^2)$ may not show an appreciable $q^2$ dependence.
The violation of this relation (\ref{eq:gmor}) away from
small $q^2$ is investigated 
in Ref.~\cite{Bernard:2001rs}, 
where the analysis of the Goldberger--Trieman discrepancy using the chiral perturbation theoty (ChPT) shows a linear dependence on $q^2$ in the ratio of $G_P(q^2)$ and $F_P(q^2)$ at relatively large $q^2$.

\section{Lattice method to determine nucleon matrix elements}
\label{sec:lattice_method_to_determine_nucleon_matrix_elements}

\subsection{Measurements with lattice QCD}
\label{ssec:measurements_with_lattice_qcd}

In general,
the partition function of lattice QCD is given as
\begin{align}
    Z & =
    \int \mathcal{D}[\overline{\psi}] \mathcal{D}[\psi] \mathcal{D}[U]
    \mathrm{exp}\left[
    -a^4 \sum_{f}^{N_f} \sum_{n, m} \overline{\psi}^{(f)}(n) D_{n, m}(U,m^{(f)}) \psi^{(f)}(m)-S_{\mathrm{G}}^{\mathrm{lat.}}(U)
    \right]\\
    & =
    \label{eq:partition_function_lattice}
    \int \mathcal{D}[U]
    \left[
    \prod_f^{N_f}
    \mathrm{det}\left(D(U,m^{(f})\right)
    \right]
    \mathrm{exp}\left[-S_{\mathrm{G}}^{\mathrm{lat.}}(U)\right],
\end{align}
where $\psi^{(f)}$ and $m^{(f)}$ denote the quark fields and their mass for $f$ flavor quark.
The gauge action ($S_{\mathrm{G}}^{\mathrm{lat.}}(U)$) and the Dirac operator ($D_{n, m}(U,m^{(f)})$) on the lattice depend on link variables
$U$.
The link variables correspond to gauge fields on the lattice.
The gauge part of the action remains 
in the form of
the Boltzmann factor, 
whereas the fermion part is finally described by the determinant of the Dirac operators by performing the Grassmann integral
as shown in the second line of the above equation.
However, 
such fermionic contributions can be rewritten
in the form of the Boltzmann factor with a non-local bosonic action by using the pseudofermion approach~\cite{Finkenrath:2013soa} as
\begin{align}
    \mathrm{det}\left(D(U,m^{(f)})\right)
    =
    \int \mathcal{D}[\eta]
    \mathrm{exp}
    \left[
    -\eta^{\dagger}
    D(U,m^{(f)})^{-1}
    \eta
    \right].
\end{align}
Hence,
all integral parts of Eq.~(\ref{eq:partition_function_lattice})  with respect to link variables $U$ can be treated as the Boltzmann factor of QCD.
This implies that the expectation value of a certain observable $\mathcal{O}$ can be statistically evaluated as
\begin{align}
\label{eq:lattce_approach}
\langle \mathcal{O} \rangle
=\int \mathcal{D}[U] \mathcal{O} P[U]
\quad \mathrm{with} \quad
P[U]  =
\frac{1}{Z}
\prod_f^{N_f}
\mathrm{det}\left(D(U,m^{(f)})\right)
\mathrm{exp}\left[-S_{\mathrm{G}}^{\mathrm{lat.}}(U)\right],
\end{align}
where $P[U]$ represents the probability distribution.

It is important to note that  the path integral given in Eq.~(\ref{eq:lattce_approach}) is formulated on the discretized space-time (lattice).
This indicates that QCD in the continuum can be reproduced 
after taking the continuum limit where the lattice spacing $a$ goes to zero, while the physical volume is kept constant for changes in the gauge coupling $\beta$ with the quark masses appropriately adjusted at the physical point.
This procedure is known as ``a line of constant physics".

We can numerically evaluate $\langle \mathcal{O} \rangle$ by the Monte-Carlo integration approach with the importance sampling~\footnote{
Importance sampling approach generates points only in important regions for the integration as below.
\begin{enumerate}
    \item Suppose the integration
    $
    \begin{aligned}
        I = \int_V d^dx f(x) = \int_V d^dx \frac{f(x)}{g(x)}g(x).
    \end{aligned}
    $
    \item Generate points $\left\{x_1,\cdots,x_n\right\}$ in $V$ with probability $g(x)$.
    \item Compute 
    $
    \begin{aligned}
        h_i=\frac{f(x_i)}{g(x_i)}
    \end{aligned}
    $
    for $i=1,\cdots,n$.
    \item Monte-Carlo theory gives
    $
    \begin{aligned}
        I = \int_V d^dx h(x)g(x) = \frac{1}{n}\sum_i^nh_i.
    \end{aligned}
    $
\end{enumerate}
} as
\begin{align}
\label{eq:numerical_int}
    \langle \mathcal{O} \rangle
    =
    \frac{1}{N_{\mathrm{conf}}}\sum_{i}\mathcal{O}[U_i] \pm 
    O\left( \frac{1}{\sqrt{N_{\mathrm{conf}}}} \right),
\end{align}
where $U_i$ is a set of the link variables (denoted as ``gauge configuration"),
and $N_{\mathrm{conf}}$ represents the number of gauge configurations.
A sequence of gauge configurations $\{U_i\}$ is generated using the Markov-Chain-Monte-Carlo (MCMC) method.
The most common MCMC procedure follows three steps.
First,
a trial gauge configuration $U'$ is generated with a conditional weight $Q[U]$ of the transition probability $T[U\rightarrow U']$ from the configuration $U$. 
Next, when the new $U'$ is accepted with a probability (denoted as ``acceptance ratio") of
\begin{align}
    P_{\mathrm{acc.}}[U,U']
    =
    \mathrm{min}
    \left[
    1,\frac{P[U']Q[U]}{P[U]Q[U']}
    \right],
\end{align}
we obtain a new ensemble $U'$.
Finally, only ensembles that are in thermodynamic equilibrium with probability $P[U]$ are obtained, since the MCMC procedure satisfies the fixed-point or stability condition
\begin{align}
    \int \mathcal{D}[U] T[U\rightarrow U'] P[U] = P[U'] \quad \mathrm{for\ all} \ U',
\end{align}
otherwise 
the numerical integration to calculate the result of $\langle \mathcal{O}\rangle$ given
in Eq.~(\ref{eq:numerical_int}) dose not work properly.
In this approach,
the path integral of QCD can be evaluated without any perturbative approach
and thus lattice QCD is regarded as
{\it a nonperturbative calculation}. 
In practice, however
$N_\mathrm{conf}$ is finite and then $\langle \mathcal{O} \rangle$ suffers from the statistical uncertainties (noises) at $O(1/\sqrt{N_\mathrm{conf}})$.
This indicates that the observables evaluated in lattice QCD simulations have the statistical errors.

\subsection{Nucleon correlation functions}
\label{ssec:nucleon_correlation_functions}

\begin{figure*}[tb]
\centering
\includegraphics[width=0.8\textwidth]{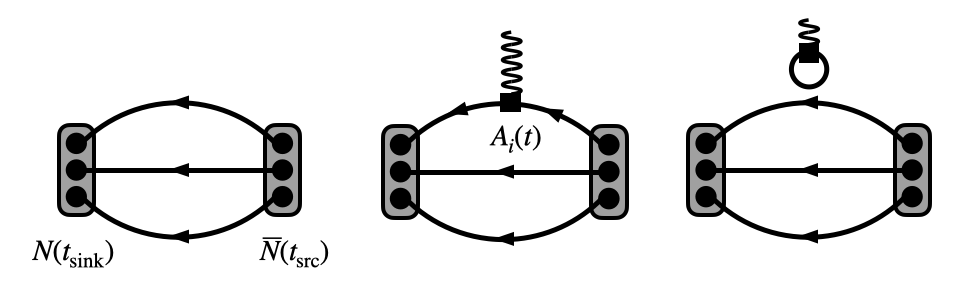}
\caption{
Quark line diagrams for the nucleon two- (left) and three-point (middle and right) correlation functions,
where the quark line is shown as solid line.
Shaded area with three filled circles represent the nucleon source (or sink) operator with three valence quark fields,
while the current insertion point is represented as filled square.
}
\label{fig:quarkline_diagrams}
\end{figure*}

The most common approach to studying the structure of hadrons is to calculate the correlation functions of the target hadron.
In detail,
one needs appropriate two- and three-point correlation functions to calculate the nucleon matrix elements.
The left panel of Fig.~\ref{fig:quarkline_diagrams} schematically shows a quark line diagram for the nucleon two-point correlation function.
For the nucleon three-point correlation function, there are two types of the quark line diagrams:
quark-line connected
diagram (middle panel) and quark-line disconnected diagram (left panel), as shown in Fig.~\ref{fig:quarkline_diagrams}.
All correlation functions can be computed by connecting 
the nucleon source and sink operators,
and also the current operator with
the quark propagator.

%
%

In this study,
the nucleon two-point function is constructed with 
smeared~\footnote{
The exponentially smeared quark operator $q_S(t,\boldsymbol{x})=\sum_{\boldsymbol{y}}A\mathrm{e}^{-B|\boldsymbol{x}-\boldsymbol{y}|}q(t,\boldsymbol{y})$
with the Coulomb gauge fixing is used for the construction of the nucleon interpolating operator as well as a local quark operator $q(t,\boldsymbol{x})$.
For detail, see Ref.~\cite{Tsuji:2023llh} and references therein. 
}
(\textit{S}) or local (\textit{L}) types of the nucleon interpolating operators located at 
the source-time position $(t_{\rm src})$ and the sink-time position $(t_{\rm sink})$ as 
\begin{align}
    \label{eq:two_pt_func}
    C_{XS}(t_{\mathrm{sink}}-t_{\mathrm{src}}; \boldsymbol{p})
    =
    \frac{1}{4}\mathrm{Tr}
    \left\{
        \mathcal{P_{+}}\langle N_X(t_{\mathrm{sink}}; \boldsymbol{p})
        \overline{N}_S(t_{\mathrm{src}}; -\boldsymbol{p})  \rangle
    \right\}\ \mathrm{with}\ X=\{S,L\},
\end{align}
where $\mathcal{P}_+=(1+\gamma_4)/2$ is the projection operator to eliminate
contributions from the opposite-parity 
excited state of the nucleon~\cite{Sasaki:2001nf,Sasaki:2005ug}.
The local nucleon operator with a three-dimensional momentum $\boldsymbol{p}$ is given for the proton state by
\begin{align}
    \label{eq:interpolating_op}
N_L(t,\boldsymbol{p})&
    =\sum_{\boldsymbol{x}}\mathrm{e}^{-i\boldsymbol{p}\cdot\boldsymbol{x}}\varepsilon_{abc}
    \left[
        u^{T}_{a}(t,\boldsymbol{x})C\gamma_5d_b(t,\boldsymbol{x})
    \right]
    u_c(t,\boldsymbol{x})
\end{align}
with the charge conjugation matrix, $C=\gamma_4\gamma_2$. The superscript $T$ denotes a transposition, while the indices $a$, $b$, $c$ and
$u$, $d$ label the color and the flavor, respectively.
Using the Euclidean time evolution of the nucleon operator ($N(t) = \mathrm{e}^{Ht}N(0)\mathrm{e}^{-Ht}$) with the Hamiltonian $H$,
one can read off the ground-state energy of the nucleon from an asymptotic behavior of the two-point correlation function at large $t$, since
the higher-energy states decay faster.

Here, let us define a useful quantity called the effective energy (mass) using the two-point correlation
function defined in Eq.~(\ref{eq:two_pt_func}) as below
\begin{align}
    \label{eq:effective_mass}
    E^{\mathrm{eff}}_N(t,\boldsymbol{p})
    =
    -
    \frac{1}{a}
    \mathrm{ln}
    \left( 
    \frac{C_{XS}(t+a; \boldsymbol{p})}{C_{XS}(t; \boldsymbol{p})}
    \right)
\end{align}
with lattice spacing $a$.
When the ground-state contribution dominates the two-point correlation function, the effective energy exhibits a plateau behavior as a function of $t$, indicating the ground-state energy.
The energy splittings $\Delta E_N(\boldsymbol{p})\equiv E_N(\boldsymbol{p})-E_N(\boldsymbol{0})$ can be evaluated
in the similar way, if the ratio of the nonzero and zero momentum two-point correlation functions 
\begin{align}
    \label{eq:energy_splitting}
    R_X^\mathrm{2pt.}(t;\boldsymbol{p})
    =
    \frac{C_{XS}(t;\boldsymbol{p})}{C_{XS}(t;\boldsymbol{0})},
\end{align}
is used in Eq.~(\ref{eq:effective_mass}) instead of $C_{XS}(t;\boldsymbol{p})$. 
In this study, we determine the ground-state energies $E_N(\boldsymbol{p})$ for each $\boldsymbol{p}$ using the ground-state mass $M_N=E_N(\boldsymbol{0})$ and
$\Delta E_N(\boldsymbol{p})$, which are
evaluated with the smear-local ($X=L$) combination for Eq.~(\ref{eq:effective_mass}) and Eq.~(\ref{eq:energy_splitting}).

On the other hand, 
when the momentum $\boldsymbol{p}$ and $\boldsymbol{p}^\prime$ are given to the initial and final states, respectively,
the nucleon three-point function with fixed source
and sink separation ($t_{\mathrm{sep}}\equiv t_{\mathrm{sink}}-t_{\mathrm{src}}$) is given by
\begin{align}
    \label{eq:three_pt_func}
    C^{5z}_{\mathcal{O}_\alpha}
    (t; \boldsymbol{p}^{\prime}, \boldsymbol{p})
    =
    \frac{1}{4} \mathrm{Tr}
    \left\{
        \mathcal{P}_{5z}\langle N_S(t_{\mathrm{sink}}; \boldsymbol{p}^{\prime})        \widetilde{J}^{\mathcal{O}}_{\alpha}
        (t; \boldsymbol{q})
        \overline{N}_S(t_{\mathrm{src}}; -\boldsymbol{p})  \rangle
    \right\}
\end{align}
with the spatially smeared source and
sink operators of the nucleon.  
The local current $\widetilde{J}^{\mathcal{O}}_{\alpha}$~\footnote{
Hereafter, the currents and the form factors with and without tilde indicates bare and renormalized ones.}
located at the time slice $t$ carries the momentum transfer $\boldsymbol{q}=\boldsymbol{p}-\boldsymbol{p}^{\prime}$. 
The projection operator $\mathcal{P}_{5z} = \mathcal{P}_{+}\gamma_5\gamma_3$ appearing in Eq.~(\ref{eq:three_pt_func}) means the $z$ direction is chosen as the polarization direction.

As illustrated in Fig.~\ref{fig:quarkline_diagrams},
there are two types of the quark line diagrams for the nucleon three-point correlation function.
Recall, however, that in the $2+1$ flavor lattice QCD, where
the $SU(2)$ isospin symmetry ($m_u=m_d$) is exact,
the nucleon three-point correlation function with the \textit{isovector} current receives no contributions from the disconnected diagrams of all quark flavors thanks to their mutual cancellation.

In the standard plateau method, the following ratio is constructed
by appropriate combinations of two-point function of Eq.~(\ref{eq:two_pt_func}) and three-point function of Eq.~(\ref{eq:three_pt_func})
with the source-sink separation ($t_{\mathrm{sep}}\equiv t_{\mathrm{sink}}-t_{\mathrm{src}}$) in order to extract the form factors~\cite{Gockeler:2003ay, Hagler:2003jd}:
\begin{align}
\mathcal{R}^{5z}_{O_{\alpha}}
\left(t; \boldsymbol{p}^{\prime}, \boldsymbol{p}\right) = \frac{
C_{O_{\alpha}}^{5z}
\left(t; \boldsymbol{p}^{\prime}, \boldsymbol{p}\right)}{C_{2}\left(t,t_{\mathrm{sep}}; \boldsymbol{p}^{\prime}, \boldsymbol{p}\right)},
\end{align}
where $C_{2}\left(t,t_{\mathrm{sep}}; \boldsymbol{p}^{\prime}, \boldsymbol{p}\right) \propto e^{-E_N(\boldsymbol{p})(t-t_{\mathrm{src}})}e^{-E_N(\boldsymbol{p}^\prime)(t_{\mathrm{sink}}-t)}$ is given by the following combination:
\begin{align}
&C_2(t, t_{{\mathrm{sep}}}; \boldsymbol{p}^\prime, \boldsymbol{p})\cr
&\equiv
\frac{C_{N}(t_{\mathrm{sink}}-t_{\mathrm{src}}; \boldsymbol{p}^{\prime})}{
\sqrt{\frac{C_{LS}(t_{\mathrm{sink}}-t; \boldsymbol{p}) C_{SS}(t-t_{\mathrm{src}}; \boldsymbol{p}^{\prime}) C_{LS}(t_{\mathrm{sink}}-t_{\mathrm{src}}; \boldsymbol{p}^{\prime})}{C_{LS}(t_{\mathrm{sink}}-t; \boldsymbol{p}^{\prime}) C_{SS}(t-t_{\mathrm{src}}; \boldsymbol{p}) C_{LS}(t_{\mathrm{sink}}-t_{\mathrm{src}}; \boldsymbol{p})}}
},
\end{align}
which is representative of $t$ dependence of the nucleon three-point correlation function due to the contribution of the nucleon ground state. All $t$ dependence due to the contribution of the nucleon
ground state can be eliminated in the ratio $\mathcal{R}^{5z}_{O_{\alpha}}\left(t; \boldsymbol{p}^{\prime}, \boldsymbol{p}\right)$, being independent of the choice of $t_{{\mathrm{sep}}}$. 

For simplicity, we consider only the rest
frame of the final state with $\boldsymbol{p}^\prime=\boldsymbol{0}$, which leads to the condition of $\boldsymbol{q}=\boldsymbol{p}-\boldsymbol{p}^\prime=\boldsymbol{p}$. Therefore, the squared four-momentum transfer is given by
$q^2=2M_N(E_N(\boldsymbol{q})-M_N)$ where $M_N$
and $E_N(\boldsymbol{q})$ represent the nucleon mass and energy with the momentum $\boldsymbol{q}$. 
In this kinematics, $\mathcal{R}^{5z}_{O_{\alpha}}\left(t; \boldsymbol{p}^{\prime}, \boldsymbol{p}\right)$ and
$C^{5z}_{O_{\alpha}}\left(t; \boldsymbol{p}^{\prime}, \boldsymbol{p}\right)$
are rewritten by a simple notation $\mathcal{R}^{5z}_{O_{\alpha}}\left(t; \boldsymbol{q}\right)$ and
$C^{5z}_{O_{\alpha}}\left(t; \boldsymbol{q}\right)$.
Hereafter the nucleon energy $E_N(\boldsymbol{q})$ is simply abbreviated by the shorthand notation $E_N$. 

The ratio $\mathcal{R}^{5z}_{O_{\alpha}}\left(t; \boldsymbol{q}\right)$ gives the following asymptotic values including the respective bare form factors in the asymptotic region~\cite{Sasaki:2007gw}:
\begin{align}
    \label{eq:fa_def}
    \mathcal{R}^{5z}_{A_i}(t;\boldsymbol{q})
    & =
    \sqrt{\frac{E_N+M_N}{2E_N}}
    \left[
        \widetilde{F}_A(q^2)\delta_{i3}-\frac{q_iq_3}{E_N+M_N}\widetilde{F}_P(q^2)
    \right]+\cdots, \\
    \label{eq:fa4_def}
    \mathcal{R}^{5z}_{A_4}(t;\boldsymbol{q})
    &=
    \frac{iq_3}{\sqrt{2E_N(E_N+M_N)}}
    \left[
        \widetilde{F}_A(q^2)-(E_N-M_N)\widetilde{F}_P(q^2)
    \right]+\cdots,\\
    \label{eq:gp_def}
    \mathcal{R}_{P}^{5z}(t; \boldsymbol{q}) 
    & =
    \frac{iq_3}{\sqrt{2 E_{N}(E_N+M_N)}} \widetilde{G}_{P}(q^2)+\cdots,
\end{align}
where the ellipsis denotes the excited-state contributions
that are supposed to be ignored 
in the asymptotic region ($t_{\mathrm{sep}}/a\gg (t-t_{\mathrm{src}})/a \gg 1$).
Three target form factors can be read off from an asymptotic plateau of the ratio $\mathcal{R}_{\cal{O}_\alpha}^{5z}(t; \boldsymbol{q})$, being independent of the choice of the source-sink separation $t_{\mathrm{sep}}$.
The respective value of the form factor can be extracted by the correlated or uncorrelated constant fit in an appropriate fit range where the good plateau behavior is observed with respect to $t$.
This approach, hereafter referred to as the \textit{traditional analysis} in this paper, is the most common method for extracting the desired form factor.

Recall here that in the special case of $\boldsymbol{q}=\boldsymbol{0}$, only the ratio $\mathcal{R}^{5z}_{A_3}(t; \boldsymbol{0})$ is available for kinetic reasons.
Eq.~(\ref{eq:fa_def}) with $i=3$ and $\boldsymbol{q}=\boldsymbol{0}$
leads to
\begin{align}
    \label{eq:demo_ga}
    \mathcal{R}^{5z}_{A_3}(t;\boldsymbol{0})
    & = \widetilde{F}_A(q^2=0)+\cdots
    = \widetilde{g}_A+\cdots,
\end{align}
which provides the precise determination of the bare axial-vector coupling $\widetilde{g}_A$ as the most accurate quantity for the nucleon axial structure.

Indeed, for the $F_A$ form factor, the \textit{traditional analysis} using Eq.~(\ref{eq:fa_def}) in the standard plateau method has been quite successful in reproducing known experimental values 
after proper renormalization, including the renormalized axial-vector coupling $g_A$.
On the other hand, the $F_P$ and $G_P$ form factors obtained by the \textit{traditional analysis} with Eqs.~(\ref{eq:fa_def}) and (\ref{eq:gp_def})
do not reproduce the experimental values well due to a strong $\pi N$ excited-state contamination~\cite{Ishikawa:2018rew, Shintani:2018ozy, Tsuji:2023llh, Gupta:2024krt}.
In Refs.~\cite{Sasaki:2025qro, Aoki:2025taf}, we 
have proposed a simple
subtraction method for removing the leading $\pi N$-state contribution 
which is considered to be the most significant contributions of the excited states in the $F_P$ and $G_P$ form factors.

The leading $\pi N$ subtraction method~\cite{Sasaki:2025qro, Aoki:2025taf}, 
which is hereafter referred to as the \textit{new analysis},  is described as follows.
We assume that the residual
$t$ dependence of the correlator ratios $\mathcal{R}_{\cal{O}_{\alpha}}(t;\boldsymbol{q})$ can be
expressed by
the functions $\Delta_{\pm}(t,t_{\mathrm{sep}};\boldsymbol{q})$, 
which encode the leading $\pi N$ contributions with a given
$t_{\mathrm{sep}}$ into the \textit{traditional analysis} as
\begin{align}
    \label{eq:fa_newdef}
    \overline{\mathcal{R}}^{5z}_{A_i}(t;\boldsymbol{q})
    & =
    \mathcal{R}^{5z}_{A_i}(t;\boldsymbol{q})
    -
    \delta_{i3}\mathcal{R}^{5z}_{A_i}(t;\boldsymbol{q}_0)\cr
    & =
    -q_iq_3K^{-1}
    \left[
        \widetilde{F}_P(q^2)
        - \Delta_+(t,t_\mathrm{sep};\boldsymbol{q})
    \right], \\
    \label{eq:fa4_newdef}
    {\mathcal{R}}^{5z}_{A_4}(t;\boldsymbol{q})
    &=
    iq_3K^{-1}
    \left[
        \widetilde{F}_A(q^2)-(E_N-M_N)\widetilde{F}_P(q^2)
        + E_\pi \Delta_-(t,t_\mathrm{sep};\boldsymbol{q})
    \right],\\
    \label{eq:gp_newdef}
    {\mathcal{R}}_{P}^{5z}(t; \boldsymbol{q}) 
    & =
    iq_3K^{-1}
    \left[
    \widetilde{G}_{P}(q^2) - Z_A B_0\Delta_{+}(t,t_\mathrm{sep};\boldsymbol{q})
    \right]
\end{align}
with $K=\sqrt{2 E_{N}(E_N+M_N)}$ and $\boldsymbol{q}_0=(q_1, q_2, 0)$ satisfying $|\boldsymbol{q}_0|=|\boldsymbol{q}|$.
The functions for the $\pi N$ contribution, $\Delta_{\pm}(t,t_\mathrm{sep};\boldsymbol{q})$
can be expressed by the following form with $t$-independent coefficients $B(\boldsymbol{q})$ and $C(\boldsymbol{q})$
\begin{align}
\Delta_{\pm}(t,t_\mathrm{sep};\boldsymbol{q})=B(\boldsymbol{q})e^{-(E_\pi+M_N-E_N)t}\pm C(\boldsymbol{q})e^{-(E_\pi-M_N+E_N)(t_\mathrm{sep}-t)},
\end{align}
which has the following property with respect to the time derivative~\footnote{
The time derivative $\partial_4$ is defined by $\partial_4 f(t)=\frac{1}{2a}\left(f(t+a)-f(t-a)\right)$ so that Eq.~(\ref{eq:time_derivative_Delta}) is valid up to ${\cal O}(a^2)$.}:
\begin{align}
\label{eq:time_derivative_Delta}
\partial_4 \Delta_{\pm}(t,t_\mathrm{sep};\boldsymbol{q})=
-E_\pi \Delta_{\mp}(t,t_\mathrm{sep};\boldsymbol{q})
+(E_N-M_M)\Delta_{\pm}(t,t_\mathrm{sep};\boldsymbol{q}),
\end{align}

Taking advantage of the above property, the $\pi N$ contribution
$\Delta_{\pm}(t,t_\mathrm{sep};\boldsymbol{q})$
can be subtracted from the correlator ratios ${\mathcal{R}}^{5z}_{A_\alpha}(t;\boldsymbol{q})$
by using two types of the time-derivative of the correlator ratio, $\partial_4\overline{\mathcal{R}}^{5z}_{A_i}(t;\boldsymbol{q})$ and 
$\partial_4{\mathcal{R}}^{5z}_{A_4}(t;\boldsymbol{q})$. 
In the \textit{new analysis}, the $F_P$ and $G_P$ form factors are extracted as follows,
\begin{align}
\label{eq:new_FP}
\widetilde{F}_P(q^2)
&=-K\frac{{\overline{\mathcal{R}}}^{5z}_{A_i}(t, {\boldsymbol q})}{q_iq_3}
+\frac{K}{(\Delta E_N)^2-(E_\pi)^2}
\left[
\Delta E_N\frac{\partial_4 {\overline{\mathcal{R}}}^{5z}_{A_i}(t, {\boldsymbol q})}{q_iq_3}
+\frac{\partial_4 {{\mathcal{R}}}^{5z}_{A_4}(t, {\boldsymbol q})}{iq_3}
\right],
\\
\label{eq:new_GP}
\widetilde{G}_P(q^2)
&=K\frac{{\mathcal{R}}^{5z}_{P}(t, {\boldsymbol q})}{iq_3}
+\frac{Z_A B_0 K}{(\Delta E_N)^2-(E_\pi)^2}
\left[
\Delta E_N\frac{
\partial_4 {\overline{\mathcal{R}}}^{5z}_{A_i}(t, {\boldsymbol q})
}{q_iq_3}
+\frac{\partial_4 {{\mathcal{R}}}^{5z}_{A_4}(t, {\boldsymbol q})}{iq_3}
\right]
\end{align}
with the energy splitting $\Delta E_N = E_N - M_N$.
It should be noted that
the first terms appearing in Eq.~(\ref{eq:new_FP})
and Eq.~(\ref{eq:new_GP}) correspond to the $F_P$ and $G_P$ form factors obtained by the \textit{traditional analysis}, respectively.

Here, it is worth mentioning that
the functional form of the $\pi N$ contribution in ${\mathcal{R}}_{P}^{5z}(t; \boldsymbol{q})$ is the same as in $\overline{\mathcal{R}}_{A_i}^{5z}(t; \boldsymbol{q})$ except for the coefficient $Z_AB_0$. This was proved in Ref.~\cite{Sasaki:2025qro} by using the axial Ward-Takahashi identity in terms of the nucleon three-point correlation function,
which is summarized in Sec.~\ref{ssec:the_pcac_relation_with_nucleon_correlation_functions}.

Finally,
recall that the quark local currents on the lattice receive finite renormalizations relative to their continuum counterparts.
The form factors are renormalized as
\begin{align}
    F_A(q^2) & = Z_A \widetilde{F}_A(q^2), \\
    F_P(q^2) & = Z_A \widetilde{F}_P(q^2), \\
    G_P(q^2) & = Z_P \widetilde{G}_P(q^2),
\end{align}
where the renormalization factors 
$Z_{\cal{O}}({\cal{O}}=A,P)$ are defined through
the renormalization of the quark local currents, $J^{\cal{O}}_\alpha=Z_{\cal{O}}\widetilde{J}^{\cal{O}}_\alpha$.

\subsection{Parameterization of the dependence on the four-momentum squared $q^2$}
\label{ssec:parameterization_of_the_dependence_on_the_four_momentum_squared_q^2}

To parameterize the $q^2$ dependence of the form factors,
the $z$-expansion method~\cite{Hill:2010yb} is used in this study. 
The $z$-expansion method is known as a model-independent analysis and has been widely used in the analysis of the form factors in both experiments and lattice calculations.
In this method,
the given form factor $G(q^2)$ is fitted by the following functional form
\begin{align}
    \label{eq:z-expansion}
    G(q^2)
    &=
    \sum_{k=0}^{k_{\mathrm{max}}}c_kz(q^2)^k=
    {c_0+c_1z(q^2)+c_2z(q^2)^2+c_3z(q^2)^3}+\dots,
\end{align}
where a new variable $z$ is defined by a 
conformal mapping from $q^2$ as
\begin{align}
    \label{eq:conformal-map}
    z(q^2)
    =\frac{\sqrt{t_{\rm cut}+q^2}-\sqrt{t_{\rm cut}-t_0}}
    {\sqrt{t_{\rm cut}+q^2}+\sqrt{t_{\rm cut}-t_0}}
\end{align}
with $t_{\rm cut}=9M_\pi^2$ for the $F_A$ and $F_P$ form factors,
while $t_{\rm cut}=4M_\pi^2$ for the $G_E$ and $G_M$ form factors.
Since respective values of $t_{\rm cut}$ are associated with the three-pion continuum
or the two-pion continuum, the value of $M_\pi$ is set to be the simulated pion mass. 
A parameter $t_0$ can be taken arbitrarily within the range of $t_{\rm cut}>t_0$.
For simplicity,
$t_0=0$ is chosen in this study~\footnote{
The optimal choice of $t_0$ is given by
$t_0^{\rm opt}
=
t_\mathrm{cut}
\left(
1-\sqrt{1+q^2_{\mathrm{max}}/t_{\mathrm{cut}}}
\right)$ for minimizing the maximum size of $|z|^2$ when
the value of $q^2$ ranges from 0 to $q_{\rm max}^2$~\cite{Meyer:2016oeg}. 
However, the maximum of the momentum transfer $q^2_\mathrm{max}\approx 0.1\ [\mathrm{GeV}^2]$ used in this study is so small that the fit result is insensitive to the choice of either $t_0=0$ or $t_0=t_0^{\rm opt}$.
}.
The transformation of Eq.~(\ref{eq:conformal-map}) maps the analytic domain of $q^2$ inside a unit circle $|z|<1$ in the $z$ plane so that Eq.~(\ref{eq:z-expansion}) is supposed to be
a convergent Taylor series in terms of $z$. 

The RMS radius for the $G_E^v$, $G_M^v$ and $F_A$ form factors is given by 
$\sqrt{\langle r_l^2\rangle }=\sqrt{-6(c_1/c_0)(4t_\mathrm{cut}})$.
In particular,
to evaluate the values of $g_P^*$ and $g_{\pi NN}$,
which are defined through the $F_P$ form factor as
\begin{align}
   \label{eq:couplings}
   g_P^* = m_\mu F_P(q^2=0.88m_\mu^2),\ \mathrm{and}\quad
   g_{\pi NN} \equiv \lim_{q^2 \to -M_\pi^2}\frac{M_\pi^2 + q^2}{2F_\pi}F_{P}(q^2)
\end{align}
with the muon mass $m_\mu$, 
the $z$-expansion method is employed in the form of $(q^2+M_\pi^2)F_P(q^2)$.
The factor $(q^2+M_\pi^2)$ is involved in our analysis to factor out the pole singularities associated with the pion in the timelike region.
The details for our analysis with the $z$-expansion method, see Ref.~\cite{Tsuji:2023llh}.

\subsection{The PCAC relation in terms of the nucleon three-point correlation functions}
\label{ssec:the_pcac_relation_with_nucleon_correlation_functions}

The axial Ward-Takahashi (AWT) identity is often used to determine a quark mass (regarded as PCAC quark mass) in lattice QCD~\footnote{The AWT identity on the lattice may be represented by
$\partial_\alpha A_\alpha(x) = 2 m P(x)$ with the renormalized currents  $A_\alpha=Z_A\widetilde{A}_\alpha$ and $P=Z_P\widetilde{P}$. Thus the quark mass $m$ appearing in the 
 AWT identity represents the bare value unless $P$ is renormalized.}.
For this purpose, the pion two-point correlation function is often used because of its correspondence with the PCAC relation.
The resulting quark mass is denoted as $m_{\mathrm{PCAC}}^{\mathrm{pion}}$
in this paper. 

Here, according to Refs.~\cite{Bali:2018qus, Tsuji:2023llh}, we introduce the following ratio
that incorporates the nucleon three-point correlation functions
 for an alternative candidate of a PCAC quark mass (denoted as $m_{\mathrm{PCAC}}^{\mathrm{nucl}})$ as 
\begin{align}
\label{eq:m_awti_pcac}
m_{\mathrm{PCAC}}^{\mathrm{nucl}}
=
\frac{
Z_A\partial_{\alpha} C^{5z}_{A_{\alpha}}(t;\boldsymbol{q})
}{
2 C^{5z}_{P}(t;\boldsymbol{q})
}
=
\frac{1}{2}
\frac{
\langle N_S(t_{\mathrm{sink}}) \partial_{\alpha} A_{\alpha}(t)
\overline{N}_S(t_{\mathrm{src}}) \rangle
}{
\langle N_S(t_{\mathrm{sink}}) \widetilde{P}(t) \overline{N}_S(t_{\mathrm{src}}) \rangle
},
\end{align}
which can be evaluated at the level of the nucleon correlation function without
isolating the ground-state contribution. 
This is in contrast to the usage of the GGT relation~\cite{Sasaki:2007gw}, 
which requires the spectral decomposition.
As long as the value of $m_{\mathrm{PCAC}}^{\mathrm{nucl}}$ exhibits $q^2$ independent behavior as a function of $q^2$,
this ratio can be regarded as an alternative \textit{bare} quark mass,
which should be identical to the previous one, $m_{\rm PCAC}^{\mathrm{pion}}$.

It is worth mentioning here that a comparison of the two types of {\textit{bare}} quark masses 
provides
a framework for investigating whether the nucleon three-point correlation functions in lattice QCD correctly reproduce the physics in the continuum within the statistical accuracy.
If $m_\mathrm{PCAC}^\mathrm{nucl}$ coincides with $m_\mathrm{PCAC}^\mathrm{pion}$
at the finite lattice spacing
without an $O(a)$-improvement of the axial-vector current, $\widetilde{A}^{\mathrm{imp}}_\alpha = \widetilde{A}_\alpha+ac_A\partial_\alpha \widetilde{P}$, 
the lattice discretization artifacts are well under control.
This is simply because the second term of $\partial_\alpha \widetilde{P}$ 
appearing in $\widetilde{A}^{\mathrm{imp}}_\alpha$ affects $m_\mathrm{PCAC}^\mathrm{nucl}$,
while it is not the case for $m_\mathrm{PCAC}^\mathrm{pion}$ determined by the \textit{zero-momentum projected} two-point functions of the pion.

There are a few technical remarks. 
In Eq.~(\ref{eq:m_awti_pcac}),
the derivatives of the nucleon three-point function 
with respect to the coordinate are evaluated by
\begin{align}
\label{eq:del_three-pt_time}
\partial_4 C^{5z}_{A_4}(t;\boldsymbol{q})
& =
\frac{1}{2a}
\left\{
C^{5z}_{A_4}(t+a;\boldsymbol{q})
-C^{5z}_{A_4}(t-a;\boldsymbol{q})
\right\}
\end{align}
for the time component and 
\begin{align}
\label{eq:del_three-pt_space}
\partial_k C^{5z}_{A_k}(t;\boldsymbol{q})
& =
iq_k C^{5z}_{A_k}(t;\boldsymbol{q})
\end{align}
for the spatial components ($k=1,2,3$).
Here, we adopt the naive discrete momentum $q_k=\frac{2\pi}{aL}n_k$ ($n_k=0, 1, 2, \cdots, (L-1)$).
This is simply because we would like to
treat the momentum in a manner equivalent to the analysis for the nucleon 
form factors which are extracted from the common three-point functions.

\section{Simulations details}
\label{sec:simulation_details}

\subsection{PACS gauge configurations}
In this paper,
we reanalyze the data previously reported in Refs.~\cite{Shintani:2018ozy, Ishikawa:2021eut, Tsuji:2023llh}.
We mainly use two sets of the PACS10 gauge ensemble generated by the PACS Collaboration
with the six stout-smeared ${O}(a)$ improved Wilson-clover quark action 
and Iwasaki gauge action~\cite{Iwasaki:1983iya} 
with lattice volume larger than $(10\ \mathrm{fm})^4$
at $\beta=1.82$ and $2.00$ corresponding to the lattice spacings of $0.09$ fm (coarse) and $0.06$ fm (fine), 
respectively~\cite{Shintani:2018ozy, Tsuji:2022ric, Tsuji:2023llh}.
In addition,
we also use other gauge ensemble generated by the PACS Collaboration.
The gauge configurations are generated with the same parameters as the coarse PACS10 ensemble, except for the lattice size,
as this ensemble is generated on about $(5.9\ \mathrm{fm})^4$ volume.
This smaller volume ensemble can not only be used to check for systematic uncertainty due to the finite-size effect, but also provide statistically accurate data due to the lower computational cost.

The simulated pion mass ($M_\pi$) in each ensemble is tuned close to the physical point, so that the resulting pion decay constant ($F_\pi$)~\footnote{
We use a traditional convention as $F_\pi=f_\pi/\sqrt{2}\sim 93$ MeV, while $f_\pi$ is quoted in Ref.~\cite{Ishikawa:2022ulx}.} is in good agreement with the experimental value.
A brief summary of each gauge ensemble is given in Table~\ref{tab:simulation_details}.
The stout-smearing parameter is set to $\rho=0.1$~\cite{Morningstar:2003gk},
and the improved coefficient of $c_\mathrm{SW}$
is nonperturbatively determined using the Schr\"odinger functional (SF) scheme~\cite{Taniguchi:2012gew}. 
The hopping parameters of $(\kappa_{ud},\kappa_{s})$ are 
carefully chosen to be almost at the physical point. 
The simulation parameters are summarized in Table~\ref{tab:simulation_parameters}.
The scale $a^{-1}$ [GeV] is determined from the $\Xi$ baryon mass input $M_\Xi = 1.3148$ [GeV] ~\cite{Shintani:2019wai}.

It is worth mentioning here that
lattice QCD simulations with a spatial size of more than 10 fm using the PACS10 gauge ensemble offer a distinctive  opportunity to explore the nucleon structure.
For instance,
the finite-size effects on the nucleon mass and the nucleon matrix elements have been examined in our previous work, {\it e.g.}, Refs.~\cite{PACS:2019ofv, Ishikawa:2021eut}.
It was determined that these effects are negligible with our large-volume lattice QCD calculation.
Especially, the large spatial volume of more than
$(10~\mathrm{fm})^3$ allows us to investigate the form factors in the small momentum transfer region, since the spatial momentum is quantized as $\boldsymbol{q}=\frac{2\pi}{aL}\boldsymbol{n}$ with the spatial extent $L$ and an integer three-dimensional vector $\boldsymbol{n}$ on the lattice under the periodic boundary condition.
For a spatial size of about 10 fm,
the lowest nonzero momentum transfer reaches the value of $q^2\sim0.01\ \mathrm{[GeV^2]}$, which is smaller than $M_\pi^2$ even at almost physical pion mass.
In this study, 
for nonzero spatial momentum,
we have selected seven lowest values of $\boldsymbol{q}\ne \boldsymbol{0}$ as listed in Table~\ref{tab:qsq_list}.

%
%
\begin{table*}[tb]
\caption{
Summary for the PACS $2+1$ flavor lattice QCD ensembles:
two sets of the PACS10 gauge ensemble and an additional
ensemble with a smaller volume.
See Refs.~\cite{{Shintani:2018ozy},{Ishikawa:2021eut},{Tsuji:2022ric},{Tsuji:2023llh},{Ishikawa:2022ulx}} for further details.
\label{tab:simulation_details}}
\centering
\begin{tabular}{cccccccc}
\hline \hline
$\beta$ & Label & $L^3\times T$&$La$ [fm] &$M_\pi$ [GeV] & $aF_\pi$ \\
          \hline
 1.82 & PACS10/L128& $128^3\times 128$ &10.8& 0.135 & 0.040244(62) \\
      & PACS5/L64  &   $64^3\times 64$ & 5.9& 0.138 & 0.040099(101) \\
 2.00 & PACS10/L160& $160^3\times 160$ &10.1& 0.138 & 0.030387(22)  \\
\hline \hline
\end{tabular}
\end{table*}
%

%
%
\begin{table*}[tb]
\caption{
Simulation parameters for the PACS $2+1$ flavor lattice QCD ensembles.
See Refs.~\cite{{Shintani:2018ozy},Ishikawa:2021eut,{Tsuji:2022ric},{Tsuji:2023llh}} for further details.
\label{tab:simulation_parameters}}
\centering
\begin{tabular}{lccccccc}
\hline \hline
$\beta$ & $a^{-1}$ [GeV] & $a$ [fm] &$\kappa_{ud}$ & $\kappa_{s}$ &$c_{\mathrm{SW}}$ & $\rho$\\
          \hline
 1.82 &2.3162(44)& 0.08520(16) &0.126177  & 0.124902  & 1.11 & 0.1\\
 2.00 &3.1108(70)& 0.06333(14) &0.125814  & 0.124925  & 1.02 & 0.1\\
\hline \hline
\end{tabular}
\end{table*}

\subsection{All-mode average}

As discussed in Sec.~\ref{sec:lattice_method_to_determine_nucleon_matrix_elements},
in lattice QCD, 
the appropriate correlation functions of the target hadron
are numerically evaluated by the Monte-Carlo integration approach with an importance sampling technique.
Consequently, it is inevitable that the lattice QCD results will contain statistical uncertainties.
Furthermore,
in the computation of the nucleon correlation functions,
the signal-to-noise ratios of the correlation functions decay exponentially in Euclidean time with a rate predicted by Refs.~\cite{Parisi:1983ae, Lepage:1989hd}.
This means that a statistical improvement is necessary for the nucleon structure studies by lattice QCD.
However,
the numerical costs of generating a large number of gauge configurations are so expensive that
it is difficult to naively reduce the statistical noise,
especially in a system where a lattice spacing is small,
while a volume is large with physical quark masses for the lighter quarks.
In fact, it is known that the computational cost roughly increases as the power laws
$\propto a^{-d_0}m^{-d_1}L^{d_2}$ with a quark mass $m$ and coefficients $d_0,d_1,d_2>1$~\cite{Finkenrath:2023sjg}.

In the calculation of nucleon two-point and three-point functions, the all-mode-averaging (AMA) technique~\cite{Blum:2012uh, Shintani:2014vja, vonHippel:2016wid, Bali:2009hu} (denoted as ``AMA" or the Truncated Solver + Bias correction method known as ``TSM") is used to significantly reduce the computational cost of multiple measurements and to achieve a much higher statistical accuracy.
As a consequence of AMA,
we compute the combination of the observables with high-precision $\mathcal{O}^{(\mathrm{org})}$ 
and low-precision $\mathcal{O}^{(\mathrm{approx})}$ at each gauge configuration as
\begin{align}
\label{eq:ama}
    \mathcal{O}^{(\mathrm{AMA})} 
    =
    \frac{1}{N_{\mathrm{org}}}\sum^{N_{\mathrm{org}}}_{f\in G}
    \left(\mathcal{O}^{(\mathrm{org})f} - \mathcal{O}^{(\mathrm{approx})f} \right)
    +
    \frac{1}{N_{G}}\sum^{N_{G}}_{g\in G}\mathcal{O}^{(\mathrm{approx})g},
\end{align}
where 
$O^{\mathrm{(AMA)}}$ represents the statistically improved observable, and
the superscripts $f, g$ denote the transformation under the lattice symmetry $G$.
In our calculations, it is translational symmetry, {\it e.g.}, changing the position of the source operator
as in Refs.~\cite{Ishikawa:2018rew,  Ishikawa:2018jee, PACS:2019ofv, PACS:2019hxd}.
$N_{\mathrm{org}}$ and $N_{G}$ are the numbers for $\mathcal{O}^{(\mathrm{org})}$ and $\mathcal{O}^{(\mathrm{approx})}$, respectively.
In order to reduce the computational cost in the PACS10 configurations,
the quark propagators are obtained with
the deflated Schwarz Alternating Procedure  (SAP)~\cite{Luscher:2003qa} and 
Generalized Conjugate Residual (GCR) \cite{Luscher:2007se}
for the measurements as shown in our previous works~\cite{Shintani:2018ozy,Ishikawa:2021eut, Tsuji:2023llh}.
The statistical errors are estimated by the single elimination jackknife method with the respect to the number of configurations.
A brief summary of the measurement numbers is given in Table~\ref{tab:measurements}.

\subsection{Smearing procedure}
The smearing procedure for nucleon interpolation operators plays an important role in accurately extracting the matrix elements of the nucleon ground state from the nucleon three-point functions.
In this study, the nucleon interpolating operators defined
in Eq.~(\ref{eq:interpolating_op}) are constructed by using the exponentially smeared quark operator with two spearing parameters $(A, B)=(1.2,0.16)$ for PACS10/L128 and PACS5/L64, and $(A,B)=(1.2,0.11)$ for PACS10/L160.
It is important to note that, as we will show later in Fig.~\ref{fig:ga_tdep_p-n_1XXc_exp}, the smearing parameters are carefully tuned so that the contribution of the nucleon ground state is dominant in the nucleon three-point correlation functions and the excited state contributions are maximally suppressed.
To construct the nucleon three-point correlation functions,  the sequential source method is employed with a fixed source-sink separation $t_\mathrm{sep}$~\cite{Martinelli:1988rr,Sasaki:2003jh}.
The effect of the excited-state contamination is studied by systematically varying $t_\mathrm{sep}$ within the range of $t_{\rm sep}/a=\{12,14,16\}$ for PACS10/L128 and $t_{\rm sep}/a=\{13,16,19\}$ for PACS10/L160.

\subsection{Non-perturbative renormalization}

As pointed out in Sec.~\ref{ssec:axial_matrix_elements_of_the_weak_currents},
the quark local currents on the lattice receives finite renormalizations
relative to their continuum counterparts. Since the renormalization
constants for vector and axial-vector currents are scale independent,
both are precisely determined with the Schr\"odinger functional (SF) scheme at vanishing quark mass. See Appendix E of Ref.~\cite{Tsuji:2023llh} for details. 
As for the renormalization constant of the axial-vector current, $Z_A$,
the resultant values are $Z_A=0.9650(68)(95)$ for $\beta=1.82$ and $Z_{A}=0.9783(21)(81)$ for $\beta=2.00$,
where the first error represents statistical one and the second error represents the systematic one that is evaluated from the difference between two volumes.
However, the second errors are simply ignored in the later analysis, since we choose the larger volume to set the physical scale.

%
%
\begin{table*}[tb]
\caption{
Details of the measurements: the spatial extent ($L$), 
temporal extent ($T$),
time separation ($t_{\mathrm{sep}}$),
the stopping condition of quark propagator for the conjugate gradient method in the high- and low-precision calculations ($\epsilon_{\mathrm{high}}$ and $\epsilon_{\mathrm{low}}$),
the number of measurements for 
the high- and low-precision calculations ($N_{\mathrm{org}}$ and $N_{G}$), the number of configurations ($N_{\mathrm{conf}}$) and the total number of the measurements ($N_{\mathrm{meas}}=N_{G}\times N_{\mathrm{conf}}$), respectively. 
As for the PACS10 configurations,
the low-precision calculation uses a fixed number of iterations for the stopping condition:
$5$ GCR iterations using $8^4$ SAP domain size with $50$ deflation fields for PACS10/L128,
and
$6$ GCR iterations using $10^4$ SAP domain size with $40$ deflation fields for PACS10/L160.
\label{tab:measurements}}
\centering
\begin{tabular}{cccccccccc}
\hline \hline
Label &  $t_{\mathrm{sep}}$ & $\epsilon_\mathrm{high}$ & $\epsilon_\mathrm{low}$& $N_{\mathrm{org}}$
    & $N_{G}$ & $N_{\mathrm{conf}}$ & $N_{\mathrm{meas}}$ & Fit range\\
\hline
    PACS10/L128& 12& $10^{-8}$& ---
    & 1& 256& 20& 5,120 & [4:8]\\
     & 14& $10^{-8}$& --- & 2& 320& 20& 6,400 & [5:9]\\
                   & 16& $10^{-8}$& --- & 4& 512& 20& 10,240 & [6:10]\\
          \hline
PACS10/L160 & 13& $10^{-8}$& ---& 1& 64& 76& 4,864 & [4:8]\\
    & 16& $10^{-8}$& --- & 3& 192& 76& 14,592& [6:10]\\
    & 19& $10^{-8}$& --- & 4& 768& 76& 58,368& [7:11]\\
    \hline
PACS5/L64 & 12 & $10^{-8}$& $0.005$& 4& 256& 100& 25,600 & [4:8]\\
   & 14 & $10^{-8}$& $0.005$& 4& 1,024& 100& 102,400 & [5:9]\\
   & 16 & $10^{-8}$& $0.002$& 4& 2,048& 100& 204,800 & [6:10]\\
\hline \hline
\end{tabular}
\end{table*}
%

%
%
\begin{table*}[tb]
    \caption{Choices for the nonzero spatial momenta: $\boldsymbol{q}=\frac{2\pi}{La}\times \boldsymbol{n}$. The bottom row shows the degeneracy of $\boldsymbol{n}$ due to the permutation symmetry between $\pm x,\pm y,\pm z$ directions.
    \label{tab:qsq_list}}
\centering
\begin{tabular}{ccccccccc}
\hline \hline
    Label& Q0& Q1& Q2& Q3& Q4& Q5& Q6& Q7 \\
          \hline
    $\boldsymbol{n}$& (0,0,0)& (1,0,0)& (1,1,0)& (1,1,1)& (2,0,0)& (2,1,0)& (2,1,1)& (2,2,0)\\
    $|\boldsymbol{n}|^2$ & 0& 1& 2& 3& 4& 5& 6& 8 \\
    Degeneracy& 1& 6& 12& 8& 6& 24& 24& 12 \\
\hline \hline
\end{tabular}
\end{table*}

\section{Numerical results I: Basic results}
\label{sec:numerical_results_i}

We summarize our basic findings,
which include the results reported in our previous studies~\cite{Shintani:2018ozy, Tsuji:2023llh, Tsuji:2024scy, Aoki:2025taf}.
In Sec.~\ref{ssec:nucleon_dispersion_relation},
the nucleon dispersion relation is presented.
A typical analysis is demonstrated for evaluating the nucleon matrix elements from the nucleon three-point correlation functions using the axis-vector coupling $g_A$ in Sec.~\ref{ssec:axial_vector_coupling}.
Secs.~\ref{ssec:the_fa_fp_and_gp_form_factors} and \ref{ssec:the_ge_and_gm_form_factors} are intended to present
the final results concerning five types of the \textit{isovector} nucleon form factors:
electric ($G_E^v$), magnetic ($G_M^v$), axial-vector ($F_A$), induced pseudoscalar ($F_P$) and pseudoscalar ($G_P$) form factors,
without providing exhaustive discussion of the analysis. 
Instead, for a detailed analysis of each form factor, see Refs.~\cite{{Tsuji:2023llh},{Aoki:2025taf}}.

\subsection{Nucleon dispersion relation}
\label{ssec:nucleon_dispersion_relation}

\begin{figure*}
\centering
\includegraphics[width=0.8\textwidth,bb=0 0 792 692,clip]{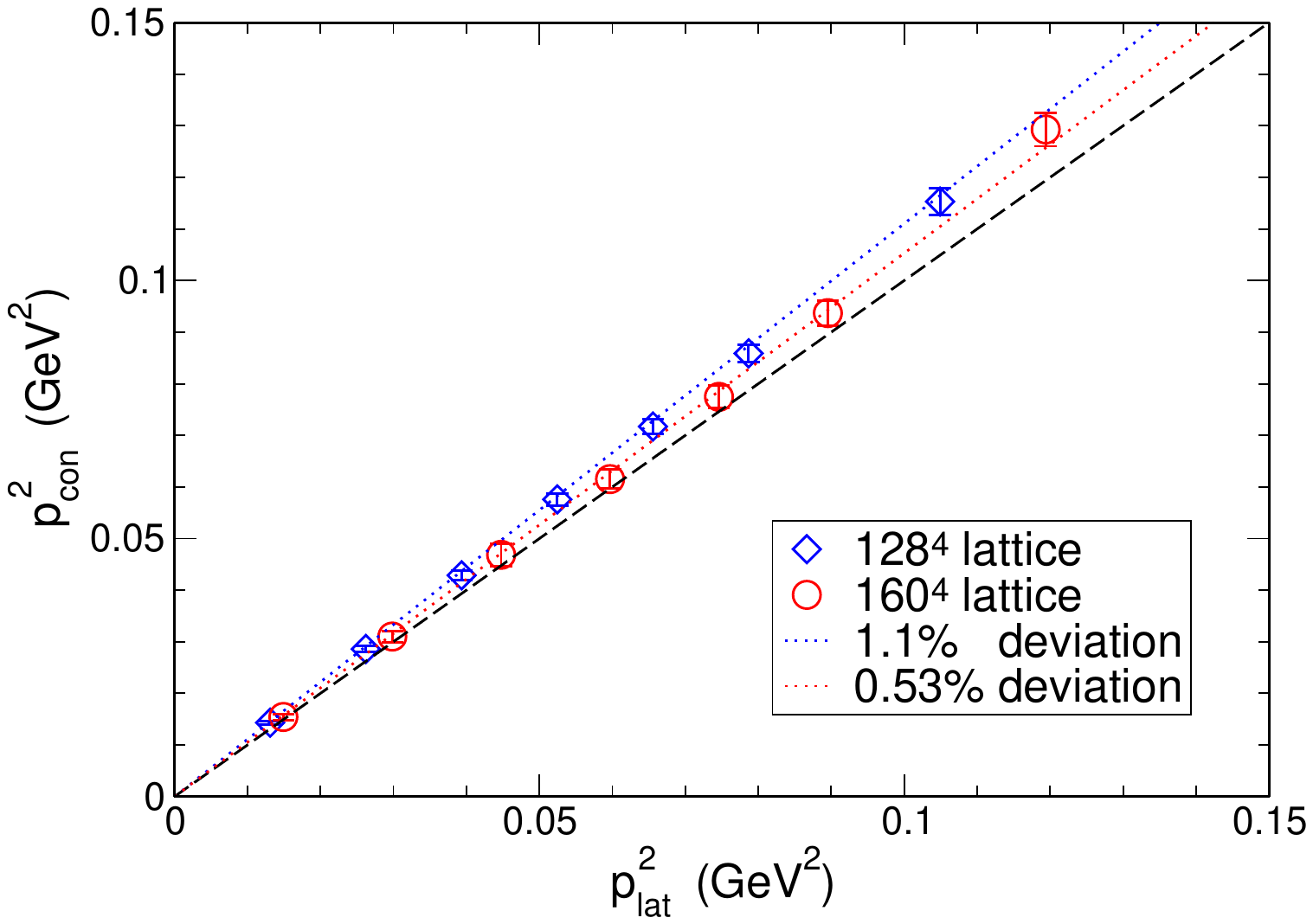}
\caption{
Check of the dispersion relation for the nucleon,
in which the values of $p_\mathrm{con}^2$ are evaluated with the values of $\Delta E_N$.
A dashed black line represents the relativistic continuum dispersion relation, 
while red and blue are given by linear fits of each data set.
}
\label{fig:dispersion_relation_nucleon}
\end{figure*}

In order to discuss the 
lattice discretization artifacts on 
the nucleon energies $E_N(\boldsymbol{p})$ with finite momentum $\boldsymbol{p}$,
let us consider the dispersion relation of the nucleon.
In Fig.~\ref{fig:dispersion_relation_nucleon},
the horizontal axis represents $p_\mathrm{lat}^2 = \left(\frac{2\pi}{aL}\right)^2\times |\boldsymbol{n}|^2$,
while the vertical axis shows $p_\mathrm{con}^2=\Delta E_N(\Delta E_N + 2M_N)$ with the energy splitting ($\Delta E_N$), which can 
be precisely evaluated by using Eq.~(\ref{eq:energy_splitting}).
The results obtained from PACS10/L128 and PACS10/L160 are shown as blue diamonds and red circles, respectively. 
The discrepancy between these data points and the relativistic continuum dispersion relation represented by black dashed line 
becomes visible at large $p^2$.
A linear fit applied to each data set results in a deviation
of 1.1\% (0.53\%) for the coarse (fine) lattice.

The size of the deviations is roughly consistent with the $O(a^2)$-corrections on the speed of light,
which is expected from the nonperturbatively $O(a)$ improved Wilson fermions used in the lattice QCD calculation.
This indicates that the observed discretization effect is sufficiently small to allow the continuum dispersion relation employed in evaluating $E_N=\sqrt{M_N^2+(2\pi\boldsymbol{n}/(La))^2}$ and $q^2=2M_N\left(\sqrt{M_N^2+(2\pi\boldsymbol{n}/(La))^2}-M_N\right)$ in Eqs.~(\ref{eq:fa_def})-(\ref{eq:gp_def}) and (\ref{eq:fa_newdef})-(\ref{eq:gp_newdef}) in the analysis of the nucleon form factors.

%
%

\subsection{Axial-vector coupling $g_A$}
\label{ssec:axial_vector_coupling}

\begin{figure*}
\centering
\includegraphics[width=0.8\textwidth,bb=0 0 792 692,clip]{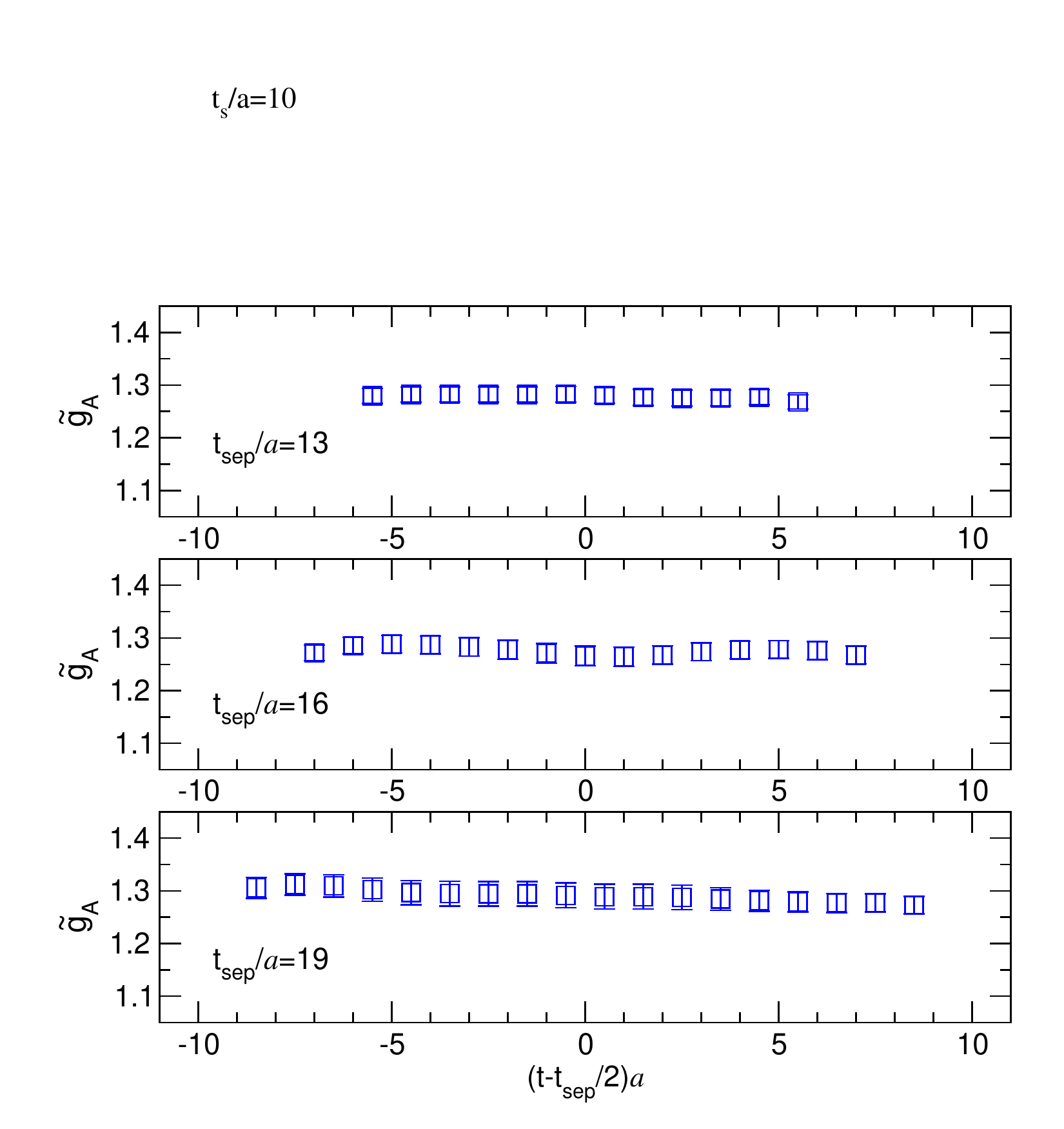}
\caption{
A typical example of the standard plateau method: 
the \textit{bare} axial-vector coupling calculated from PACS10/L160 with three different 
values of the source-sink
separation $t_{\mathrm{sep}}$ as a function of the current operator insertion point $t$. 
No visible dependence on both $t$ and $t_{\mathrm{sep}}$ indicates that the ground state contribution is certainly dominant.}
\label{fig:ga_tdep_p-n_1XXc_exp}
\end{figure*}

First, let us take a closer look at the axial-vector coupling $g_A$ as a typical example of evaluating the nucleon matrix elements from the nucleon three-point functions using the standard ratio method.
As described in Sec.~\ref{ssec:measurements_with_lattice_qcd},
$g_A=F_A(q^2=0)$ can be obtained by using the ratio $\mathcal{R}^{5z}_{A_i}(t;\boldsymbol{q}=\boldsymbol{0})$ of Eq.~(\ref{eq:fa_def}).
For instance, as shown in Fig.~\ref{fig:ga_tdep_p-n_1XXc_exp}, 
the ratio $\mathcal{R}^{5z}_{A_i}(t;\boldsymbol{q}=\boldsymbol{0})$ calculated from the $160^4$ lattice ensemble for all choices of $t_\mathrm{sep}$ are plotted as a function of the current operator insertion point $t$.
Since the asymptotic plateau that is independent of $t$
corresponds to the ground-state contribution in the nucleon three-point correlation function, 
the value of $\widetilde{g}_A$ can be extracted by a constant fit with respect to $t$ in an appropriate fit range.
In the \textit{traditional analysis} with the standard plateau method, one should calculate the ratio with several choices of $t_\mathrm{sep}$, and then make sure
whether the evaluated value does not change with a variation 
of $t_\mathrm{sep}$ within a certain precision. 

As shown in Fig.~\ref{fig:ga_tdep_p-n_1XXc_exp}, it is clearly seen that no visible dependence on both $t$ and $t_{\mathrm{sep}}$ indicates that the ground state contribution is certainly dominant within our statistical precision.
Recall that in our study
the smearing parameters of the nucleon interpolating operator
were highly optimized to eliminate as much as possible the contribution of excited states in the nucleon two-point correlation
function.

\begin{figure*}
\centering
\includegraphics[width=0.8\textwidth,bb=0 0 792 692,clip]{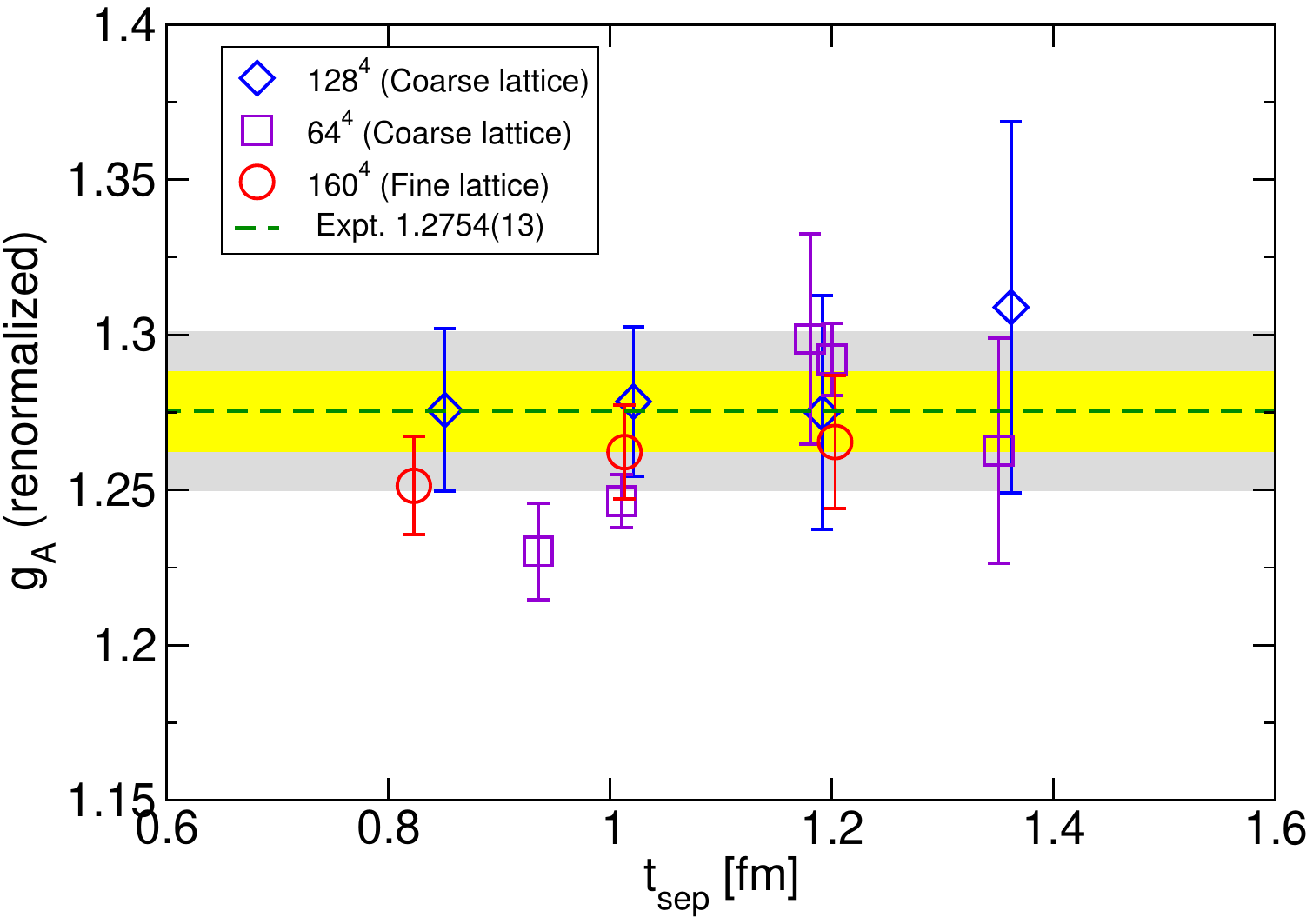}
\caption{
Source-sink separation ($t_\mathrm{sep}$) dependence of the \textit{renormalized} values of the nucleon axial-vector coupling ($g_A$).
The horizontal axis represents $t_\mathrm{sep}$ in physical unit.
The red circles represent the results obtained from PACS10/L160~\cite{Tsuji:2023llh},
while the blue diamonds and violet squares are obtained from 
PACS10/L128 and PACS5/L64~\cite{Tsuji:2022ric}, respectively.
The horizontal dashed line shows the experimental value~\cite{ParticleDataGroup:2022pth}, while yellow and gray bands display 1\% and 2\% deviations from the experimental
value.
}
\label{fig:ga_tsdep_p-n_1XXc_exp}
\end{figure*}

Figure~\ref{fig:ga_tsdep_p-n_1XXc_exp} compiles all results of
the \textit{renormalized} axial-vector coupling $g_A=Z_A\widetilde{g}_A$ obtained from both the $128^4$ and
$64^4$ lattice ensembles at the coarse lattice spacing
and the $160^4$ lattice ensemble at the fine lattice spacing
as a function of $t_{\mathrm{sep}}$.
For a comparison, the experimental value~\cite{ParticleDataGroup:2022pth} is represented as
a dashed line with yellow and gray bands, which display 1\% and
2\% deviations from the experimental value.
We then confirmed that for the axial-vector coupling $g_A$, 
\begin{enumerate}
\item systematic uncertainty stemming from the excited-state 
contamination is well under control for $t_{{\mathrm{sep}}}\gtrsim1.0\ [\mathrm{fm}]$ on the larger volumes
(PACS10/L128 and PACS10/L160).
\item systematic uncertainty due to the finite-size effect is
negligible at the level of the statistical precision of less than
2\% in two large spatial extents of about 10 and 5 fm (PACS10/L128 and PACS5/L64). 
\item systematic uncertainty due to the lattice discretization
effect is smaller than the statistical error of $g_A$. 
\end{enumerate}
Our first-principles calculations of $g_A$ achieve control of all major systematic uncertainties from chiral extrapolation, finite-size effect, excited-state contamination, and discretization effect at the current statistical precision of less than 2\%, and reproduce the experimental value with an accuracy of a few percent.

\subsection{Form factors: $F_A$, $F_P$ and $G_P$}
\label{ssec:the_fa_fp_and_gp_form_factors}

\begin{figure*}
\centering
\includegraphics[width=0.8\textwidth,bb=0 0 792 692,clip]{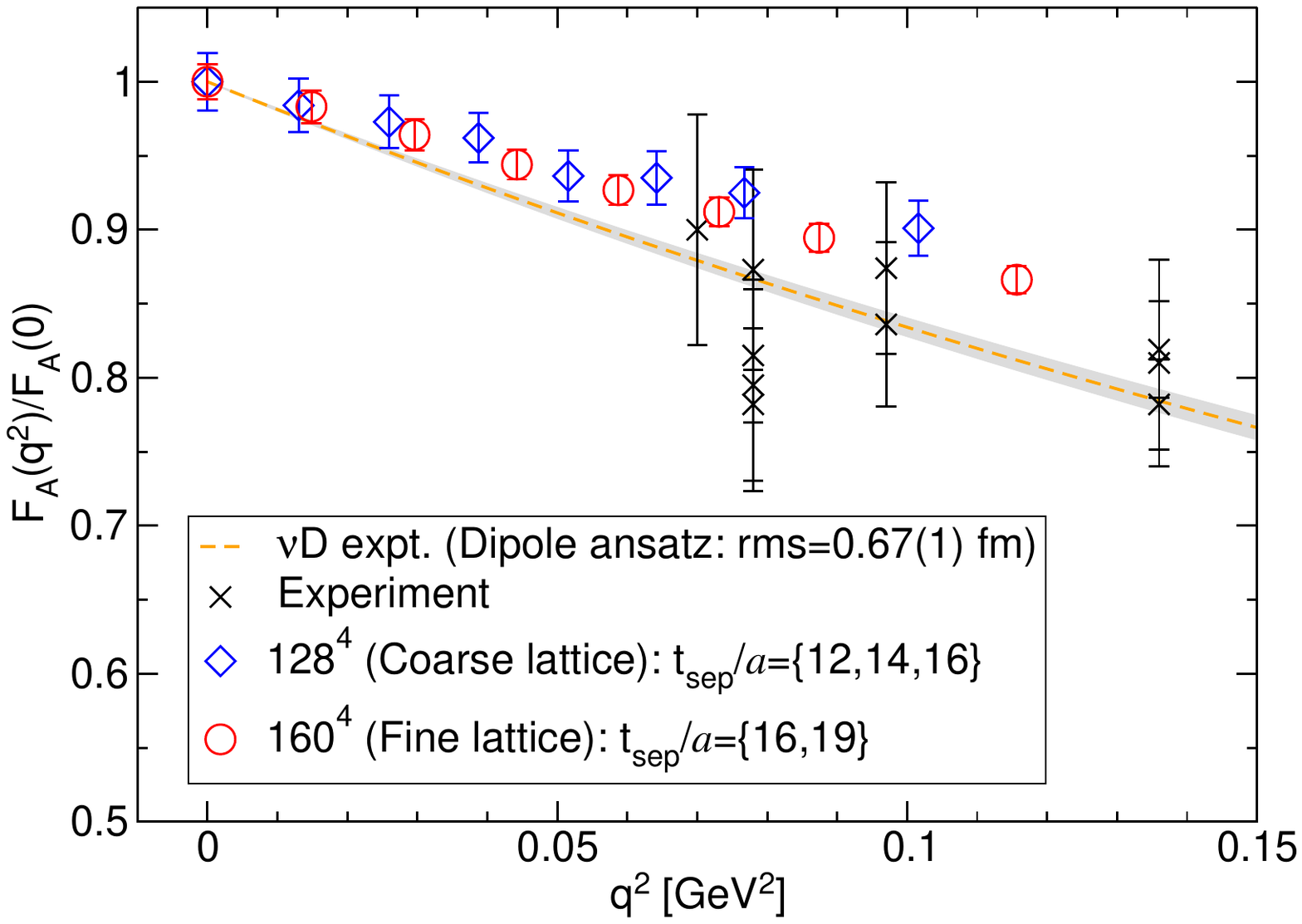}
\caption{
Results of the $F_A$ form factor normalized by $F_A(q^2=0)$ as a functions of four-momentum squared $q^2$ for each combined data of $t_\mathrm{sep}/a=\{13,16,19\}$ for PACS10/L160 and $t_\mathrm{sep}/a=\{14,16\}$ for PACS10/L128.
The orange dashed line with gray band represents a mean and an uncertainty of $q^2$ dependence with the dipole ans\"atz,
while the cross symbols represent the actual experimental values~\cite{Baker:1981su, Miller:1982qi}.
}
\label{fig:fa_wslope_qsqr}
\end{figure*}

Figure~\ref{fig:fa_wslope_qsqr} shows our lattice data of the $F_A$ form factor normalized at $q^2=0$,
which is obtained by the \textit{traditional analysis}
similar to the case of $g_A$ as described in Sec.~\ref{ssec:axial_vector_coupling}.
The blue diamonds and red circles are plotted for each combined data of $t_\mathrm{sep}/a=\{13,16,19\}$ for PACS10/L160 and $t_\mathrm{sep}/a=\{14,16\}$ for PACS10/L128,
while the band shows a dipole form factor~\cite{Bernard:2001rs} using the old bubble chamber data,
which are indicated by black crosses~\cite{Baker:1981su, Miller:1982qi}.
Compared to the experimental results measured in
the range of $q^2\gtrsim 0.07\ [\mathrm{GeV}^2]$,
our lattice data provide higher resolution in the lower $q^2$ region. 
Our lattice data obtained in the range of $0.01\ [\mathrm{GeV}^2]\lesssim q^2\lesssim 0.12\ [\mathrm{GeV}^2]$ are located slightly above the dipole form factor regardless of the lattice spacing, resulting in heavier dipole masses.
However, no firm conclusions can be drawn at present.
This is due to the fact that, in contrast to axial-vector coupling, there is a systematic uncertainty of approximately 10\% related to the lattice discretization effects in the case of the dipole mass (axial radius). Therefore this uncertainty makes it difficult to make quantitative comparisons. Obviously, additional lattice simulations at much finer lattice spacings are required for a comprehensive study of discretization uncertainties.
Our new simulations towards the continuum limit are in progress on a $256^4$ lattice with finer lattice spacing 
($a\approx 0.04$ fm)~\cite{Tsuji:2024scy}.

\begin{figure*}
\centering
\includegraphics[width=0.45\textwidth,bb=0 0 792 692,clip]{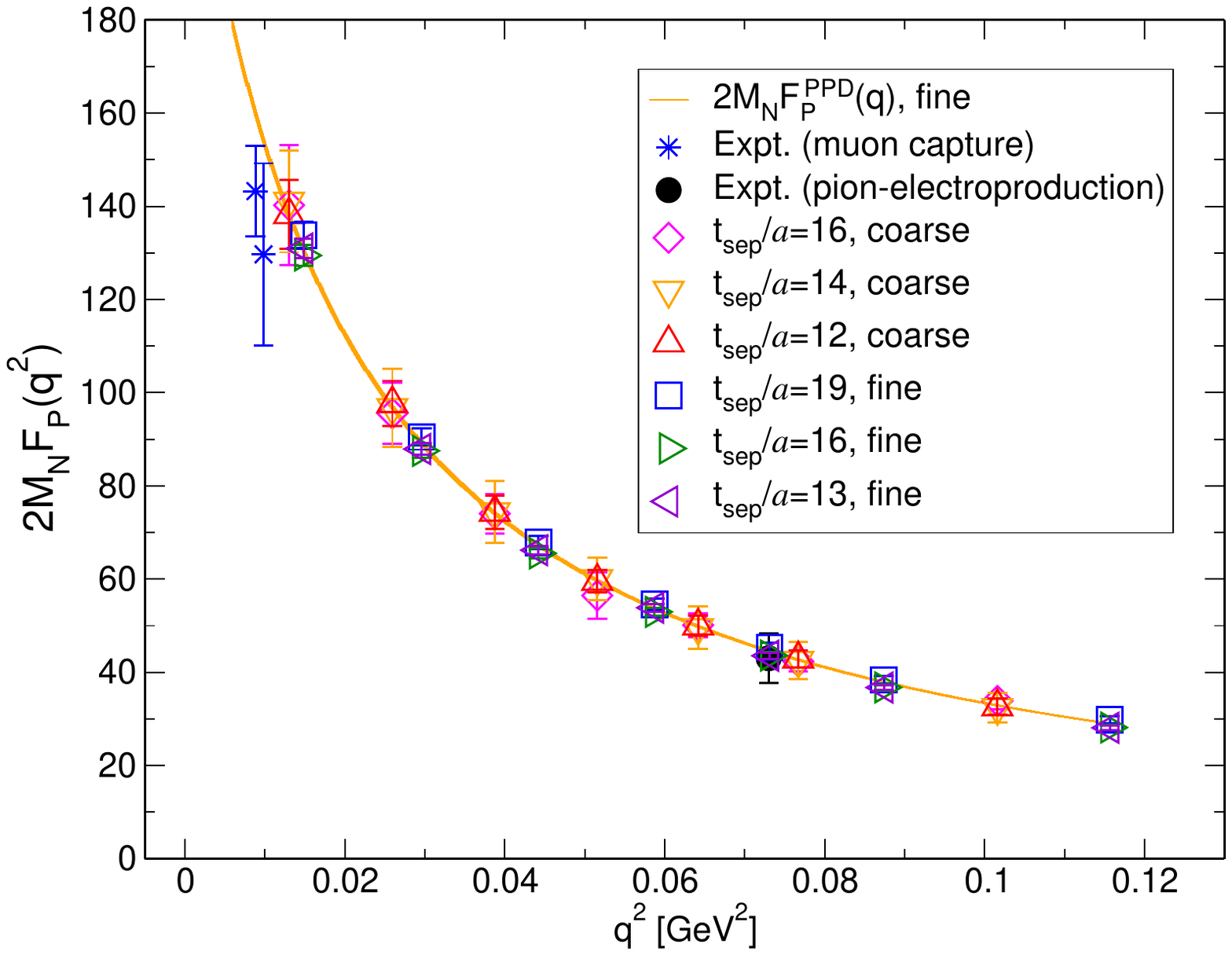}
\includegraphics[width=0.45\textwidth,bb=0 0 792 692,clip]{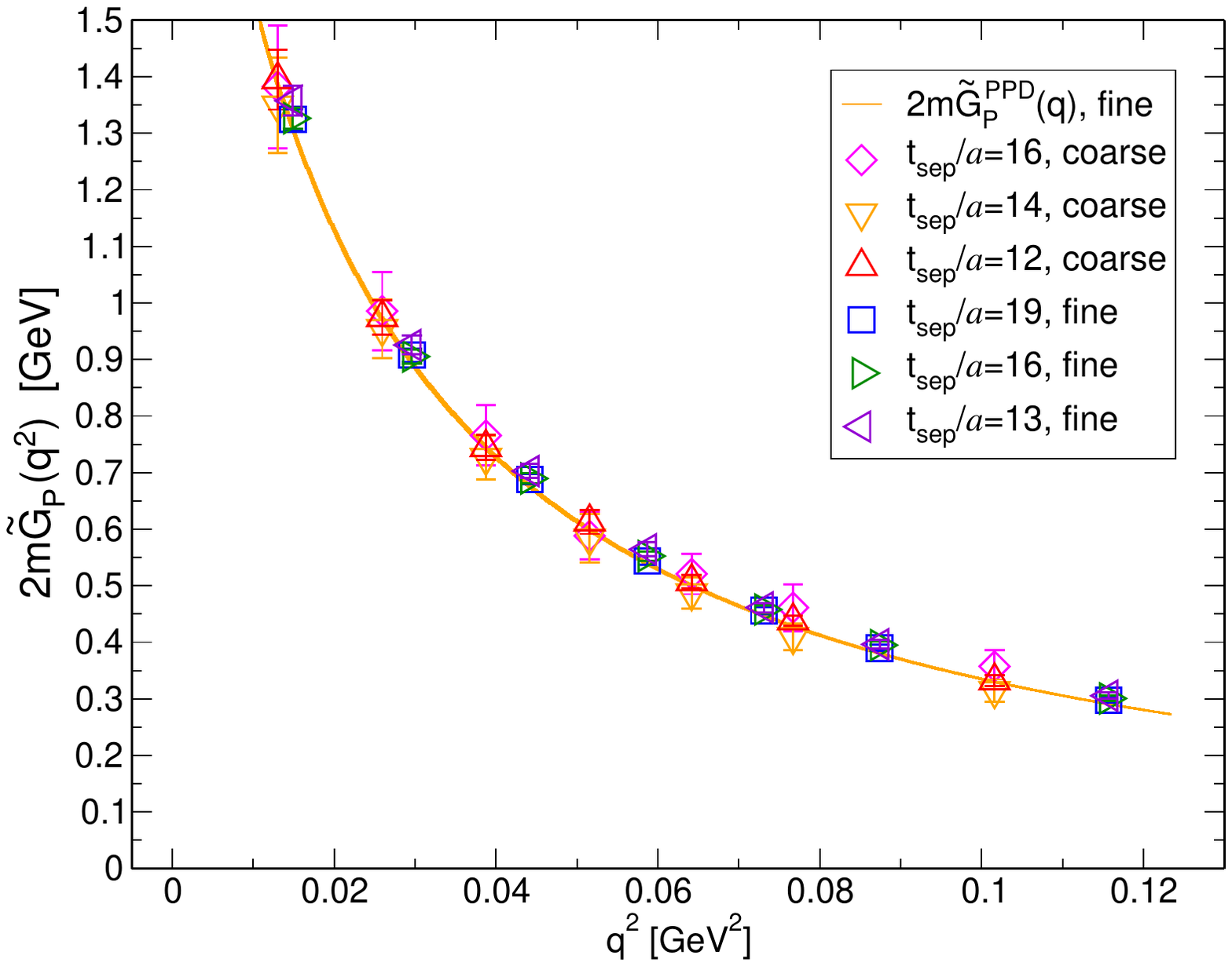}
\caption{
Results of $2M_NF_P(q^2)$ (left) and $2m_\mathrm{PCAC}\widetilde{G}_P(q^2)$ (right) as a functions of four-momentum squared $q^2$ for each data set of $t_\mathrm{sep}/a=\{13,16,19\}$ for PACS10/L160 and $t_\mathrm{sep}/a=\{12,14,16\}$ for PACS10/L128.
The solid curves in each panel is given by the PPD model estimates giving in Eqs.~(\ref{eq:ppd}) and (\ref{eq:ppd_gp}), 
while the filled circle and star symbols in the left panel represent the experimental values~\cite{Gorringe:2002xx, Choi:1993vt}.
}
\label{fig:fp_2mgp_qsqr}
\end{figure*}

Next,
in Fig.~\ref{fig:fp_2mgp_qsqr},
the $q^2$ dependence is shown for each data of the $F_P$ and $G_P$ form factors obtained by our \textit{new analysis}, where the leading $\pi N$ contribution
induced by the pion pole structure of $F_P(q^2)$ and $G_P(q^2)$ is eliminated~\cite{Sasaki:2025qro, Aoki:2025taf}.
The open symbols represent our lattice data with sets of $t_\mathrm{sep}/a=\{13,16,19\}$ for PACS10/L160 and $t_\mathrm{sep}/a=\{12,14,16\}$ for PACS10/L128,
while each solid curve represents the PPD model estimates giving in Eqs.~(\ref{eq:ppd}) and (\ref{eq:ppd_gp}).
Two experimental results of the muon capture~\cite{Gorringe:2002xx} and the pion-electroproduction~\cite{Choi:1993vt} are plotted as blue
filled diamonds and black filled circles.
As can be easily seen, the results of $F_P(q^2)$ are consistent with the experimental values.
Moreover, both the results of $F_P(q^2)$ and $G_P(q^2)$
are in good agreement with the predictions of the PPD model evaluated with the lattice data of $F_A(q^2)$ together with
the simulated values of $M_N$, $M_\pi$ and $F_\pi$.
This observation suggests that the PPD model,
which is only valid in the chiral limit,
continues to serve as a valuable functional form for describing the $q^2$ dependence of the $F_P$ and $G_P$ form factors at the physical point, at least up to $q^2\sim 0.1\ [\mathrm{GeV}^2]$.

It should be here emphasized that the excited-state contamination has been a long-standing obstacle to the accurate determination of $F_P(q^2)$ and $G_P(q^2)$.
It is noteworthy that several approaches have been proposed
to eliminate the effects,
such as the utilization of the temporal $A_4$ current~\cite{Jang:2019vkm} and
proper projection determined by the variational analysis with the explicit $\pi N$-operator~\cite{Barca:2022uhi, Alexandrou:2024tin}.
In our calculations, where the smearing parameters for the nucleon ground state have been carefully tuned,
we have succeeded in completely removing the leading $\pi N$ contribution from the standard ratios for the $F_P$ and $G_P$ form factors
in our \textit{new analysis}~\cite{Sasaki:2025qro,Aoki:2025taf}.

\begin{figure*}
\centering
\includegraphics[width=0.45\textwidth,bb=0 0 792 692,clip]{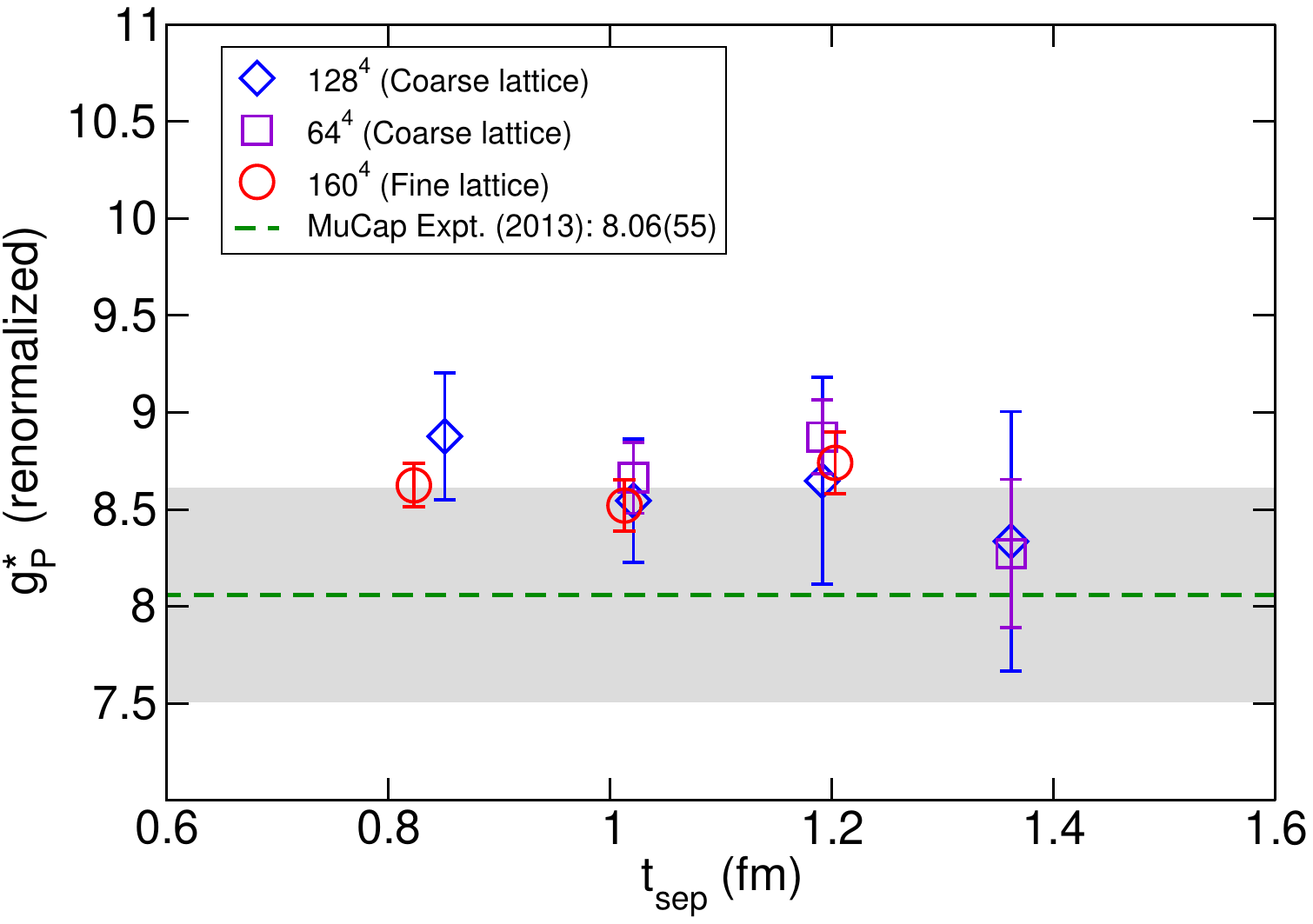}
\includegraphics[width=0.45\textwidth,bb=0 0 792 692,clip]{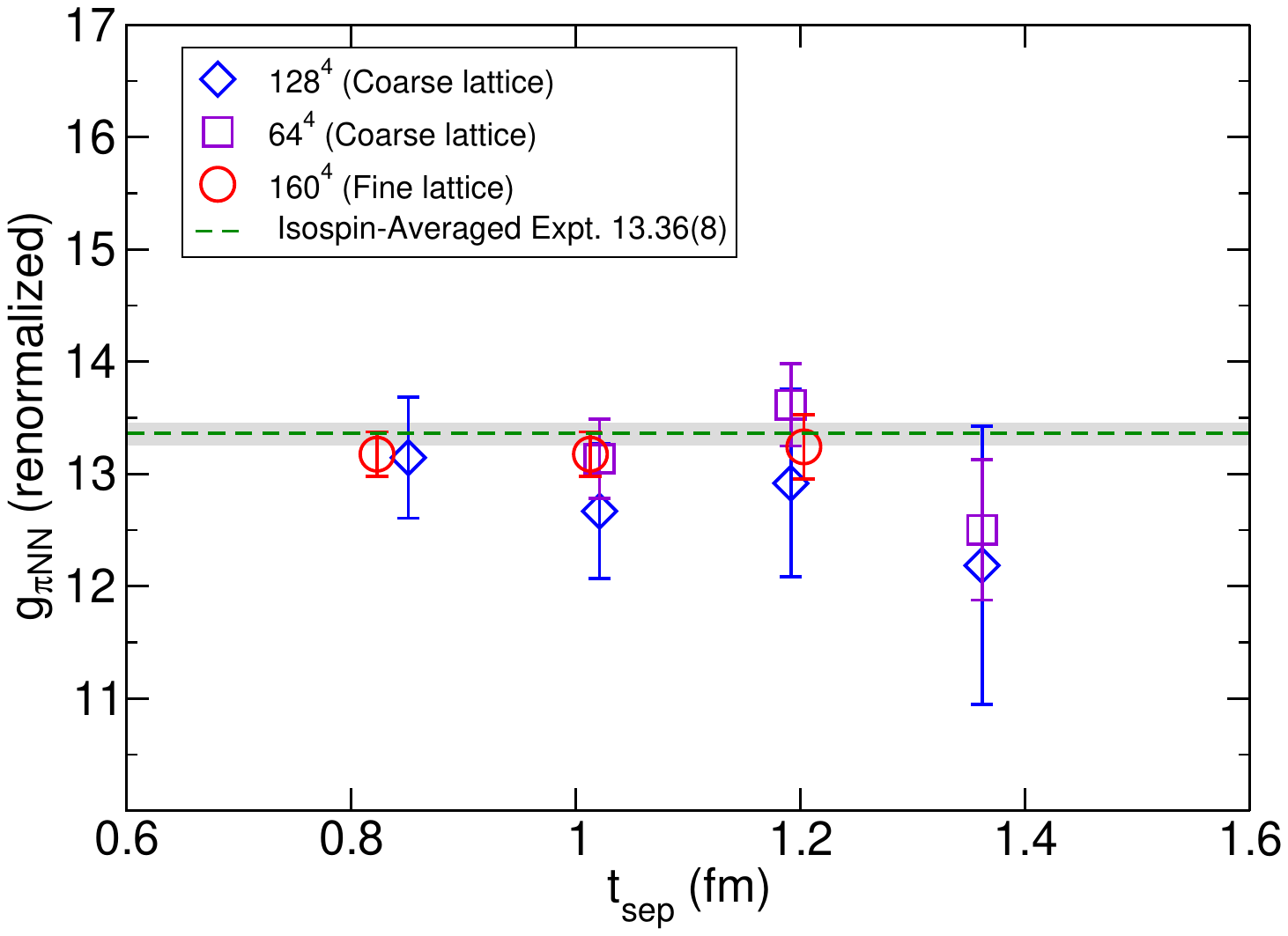}
\caption{
Source-sink separation ($t_\mathrm{sep}$) dependence of the \textit{renormalized} values of $g_P^*$ (left) and $g_{\pi NN}$ (right).
The horizontal axis represents $t_\mathrm{sep}$ in physical unit.
The red circles represent the results obtained from PACS10/L160, while the blue diamonds are obtained from PACS10/L128 and PACS5/L64~\cite{Tsuji:2024scy}, respectively.
The horizontal dashed line and gray band shows the experimental value and its uncertainty~\cite{MuCap:2012lei, Babenko:2016idp, Limkaisang:2001yz}.
These figures are reprinted from Ref.~\cite{,Aoki:2025taf}.
}
\label{fig:gp_gpNN}
\end{figure*}

The induced pseudoscalar coupling $g_P^*$ and pion-nucleon coupling $g_{\pi NN}$ are defined through the $F_P$ form factor at some finite momentum transfer $q^2$ as Eq.~(\ref{eq:couplings}).
In this study,
we use the z-expansion method to parameterize the $q^2$ dependence of $(q^2+M_\pi^2)F_P(q^2)$ to evaluate the two couplings.
Figure~\ref{fig:gp_gpNN} shows the $t_\mathrm{sep}$ dependence of the evaluated values of the induced pseudoscalar coupling $g_P^*$ (left) and pion-nucleon coupling $g_{\pi NN}$ (right) obtained with sets of $t_\mathrm{sep}/a=\{13,16,19\}$ for PACS10/L160 and $t_\mathrm{sep}/a=\{12,14,16\}$ for both of PACS10/L128 and PACS5/L64.
In each panel, the horizontal dashed line together with gray band indicates the corresponding experimental result and
its uncertainty.
The experimental value of $g_P^*$ is given by the MuCap experiment~\cite{MuCap:2007tkq,MuCap:2012lei}, while
the experimental value of $g_{\pi NN}$ is quoted as the isospin average value of the experimental results~\cite{Babenko:2016idp, Limkaisang:2001yz} as $g_{\pi NN}^2=\frac{1}{3}\left(g_{\pi^0 NN}^2+2g_{\pi^\pm NN}^2\right)$. 

As can be easily seen in Fig.~\ref{fig:gp_gpNN}, 
the results of $g_P^*$ and $g_{\pi NN}$ obtained from all the coarse ($128^4$ and $64^4$) and fine ($160^4$) lattice ensembles 
show no significant dependence on $t_\mathrm{sep}$ and agree
well with each other, suggesting that the three main
systematic uncertainties due to excited-state contamination,
finite-size effect and discretization effect are sufficiently small compared to the present statistical accuracies, as is the axial-vector coupling $g_A$.

Furthermore, the results are consistent with the experimental value (denoted by the horizontal dashed line and bands).
It is important to note that the lattice determination of 
$g_P^*$ and $g_{\pi NN}$ using the {\it new analysis} 
results in a smaller error than the experimental one for $g_P^*$ and a comparable error for $g_{\pi NN}$. 
More rigorous comparison to experimental values requires taking the continuum limit of the target quantities.

\begin{figure*}
\centering
\includegraphics[width=0.45\textwidth,bb=0 0 528 612,clip]{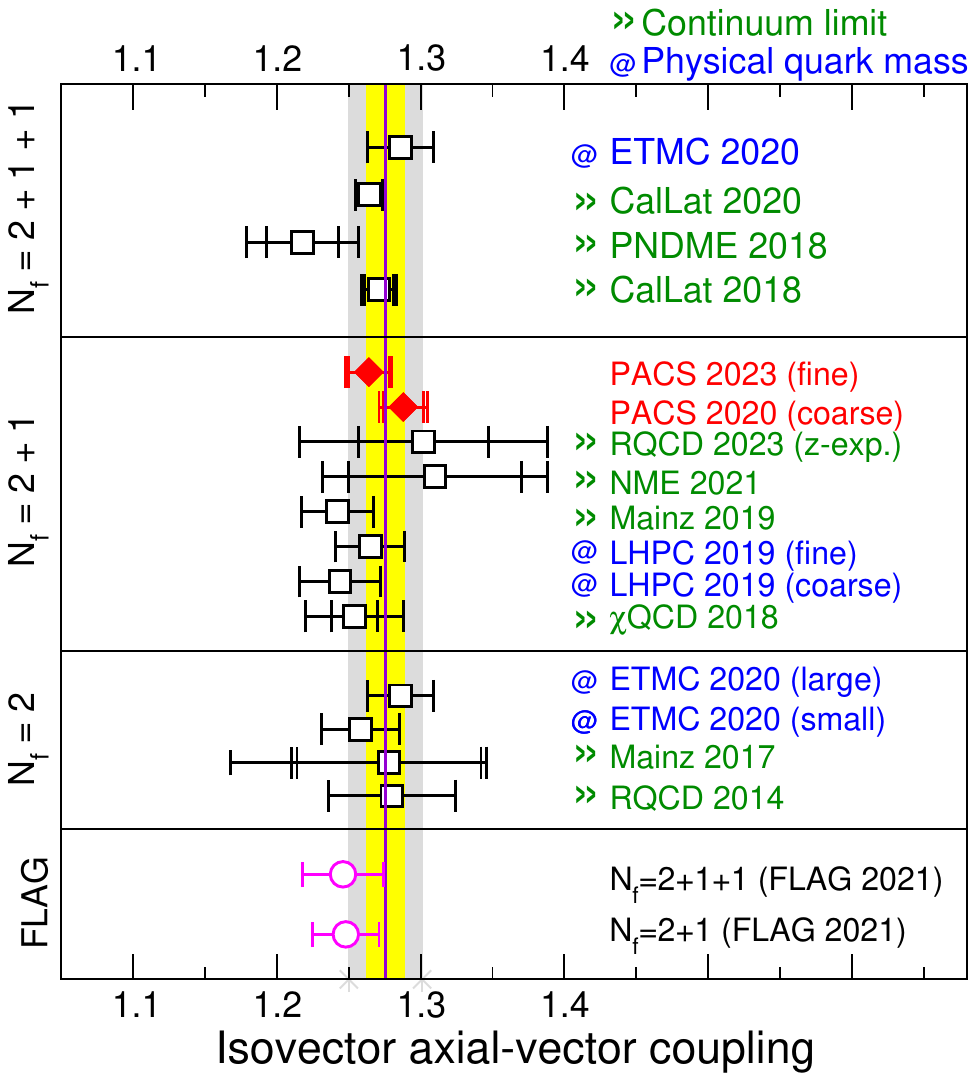}
\includegraphics[width=0.45\textwidth,bb=0 0 528 612,clip]{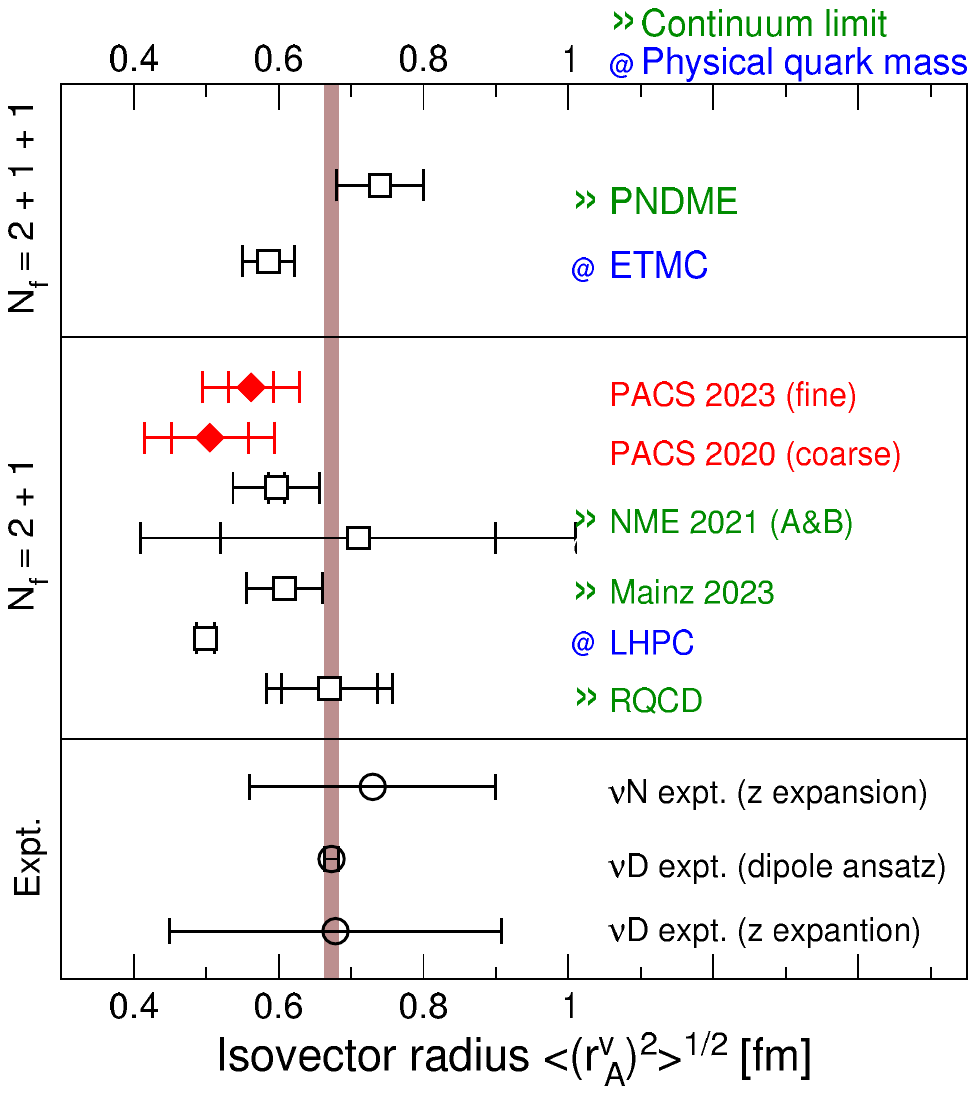}
\includegraphics[width=0.45\textwidth,bb=0 0 528 612,clip]{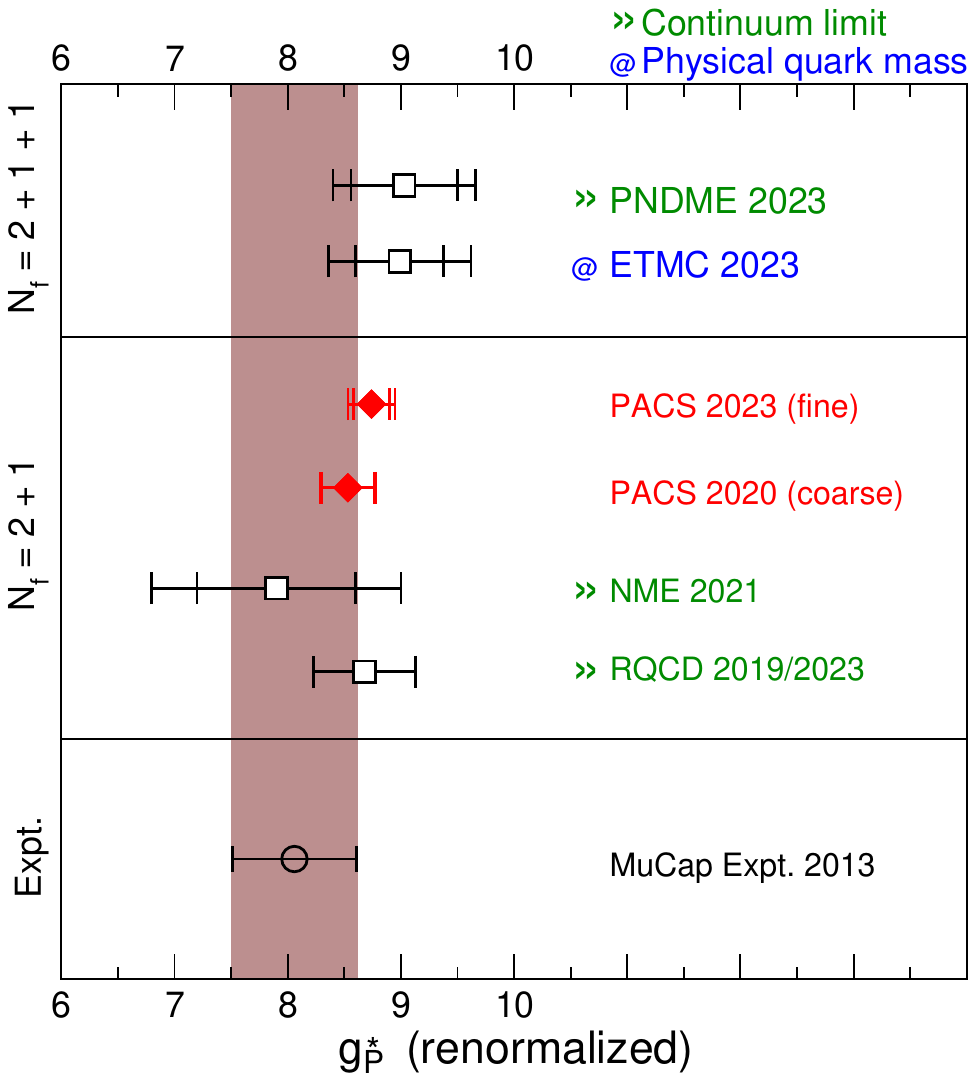}
\includegraphics[width=0.45\textwidth,bb=0 0 528 612,clip]{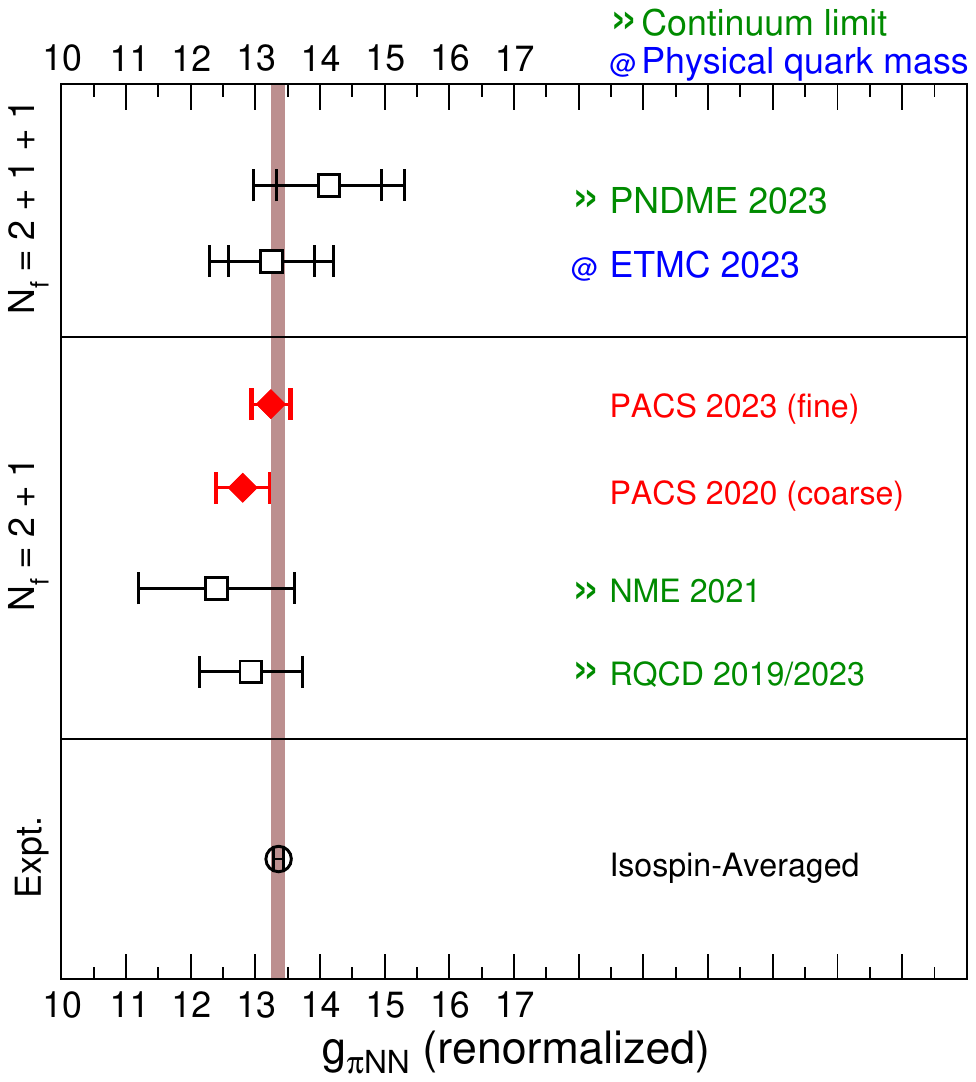}
\caption{Summary plot for our current status with the experimental values and other lattice QCD results: $g_A$ (top left), $\sqrt{\langle r_A^2\rangle}$ (top right), $g_P^{\ast}$ (bottom left) and $g_{\pi NN}$ (bottom right).
The inner error bars represent the statistical errors,
while the outer error bars are evaluated by both the statistical and systematic errors added in quadrature.
The experimental values denoted as a brown band in each panel.
The values of $g_A$ and $g_P^*$ are taken from Ref.~\cite{ParticleDataGroup:2022pth} and Ref.~\cite{MuCap:2012lei}
while the value for $g_{\pi NN}$ shown here is given by the isospin average
as described in the text.
}
\label{fig:comparison}
\end{figure*}

Finally,
Figure~\ref{fig:comparison} shows summary for our results of $g_A$ (top left), $\sqrt{\langle r_A^2\rangle }$ (top right), $g_P^*$ (bottom left) and $g_{\pi NN}$ (bottom right),
together with the experimental values and the other lattice QCD results.
The inner and outer error bars represent the statistical uncertainties and the total uncertainties given by adding both statistical and systematic errors in quadrature.
As for the systematic uncertainties,
we take into account the uncertainties stemming from the excited-state contamination and the renormalization.
In particular,
the systematic uncertainty of the excited-state contamination is examined by the combined analysis with the range of $1.0\  [\mathrm{fm}]\lesssim t_\mathrm{sep}\lesssim 1.2\  [\mathrm{fm}]$
for both the coarse lattice and fine lattice ensembles.

As for $g_A$, the accuracy of the lattice QCD results has improved
significantly, and all results obtained from each lattice study
including our results converge within a few percent of the experimental value. On the other hand, the lattice QCD results for
the axial radius $\sqrt{\langle r_A^2\rangle}$ are not sufficiently consistent with each other. 
Our two results of the axial radius, which are obtained from the coarse ($128^4$) and fine ($160^4$) lattice ensembles, barely agree with each other, albeit with their uncertainties.
However, the systematic uncertainty due to the discretization effect can be estimated as 11.1\% by a difference between the central values of the two results.
Recall that the similar estimate on $g_A$ is kept smaller than its statistical error of 2\%. Thus, the current discrepancy observed in the lattice QCD results of the axial radius can be attributed to the large discretization uncertainties on the axial radius.
The results of both $g_P^*$ and $g_{\pi NN}$ agree well with the other lattice QCD results, and also reproduce well the respective experimental values.

Needless to say that a comprehensive study of the lattice discretization effects, especially on the axial radius, and also the continuum limit extrapolation of all four target quantities require additional lattice simulations using 
the third PACS10 ensemble with a finer lattice spacing. 
Such a study is currently underway~\cite{Tsuji:2024scy}.

%
%
%

%
%
%

%
%
%

%
%
%

%
%
%

%
%

\subsection{Form factors: $G_E^v$ and $G_M^v$}
\label{ssec:the_ge_and_gm_form_factors}

In our previous studies~\cite{Shintani:2018ozy, Tsuji:2023llh, Tsuji:2024scy, Aoki:2025taf},
we also reported the \textit{isovector} nucleon form factors defined in the vector channel: electric ($G_E^v$) and magnetic ($G_M^v$) form factors.
(see Ref.~\cite{Tsuji:2023llh} for their definitions).
They are also contribute to Eq.~(\ref{eq:dcs}) and more accurately measured by the electron-nucleon scattering experiments compared to the three form factors that were previously discussed in Sec.~\ref{ssec:the_fa_fp_and_gp_form_factors}.

It should be noted that the $G_E^v$ form factor is not a complete counterpart to solving the proton radius puzzle.
For the evaluation of the ``proton" form factor,
it is necessary to calculate the disconnected contributions,
which are not treated in this study, but will be considered in our ongoing research.

Recall that we present results for the \textit{isovector} nucleon form factors that can be determined solely by the connected contributions in $2+1$ flavor QCD.
In this section,
let us highlight the results for the \textit{isovector} form factors and their root-mean-square (RMS) radii
$\sqrt{\langle r_{l}^2\rangle}$ determined from the slope of the corresponding form factors $G_{l}^{v}(q^2)$ ($l=\{E,M\}$) as
\begin{align}
\langle r_{l}^2\rangle=–\left.\frac{6}{G_{l}^{v}(0)}\frac{dG_l^v(q^2)}{dq^2}\right|_{q^2=0}
\end{align}
For the evaluation of $G_{E}^{v}(q^2)$ and $G_{M}^{v}(q^2)$,
the standard ratio method is employed. See Ref.~\cite{Tsuji:2023llh} for details.

\begin{figure*}
\centering
\includegraphics[width=0.45\textwidth,bb=0 0 792 692,clip]{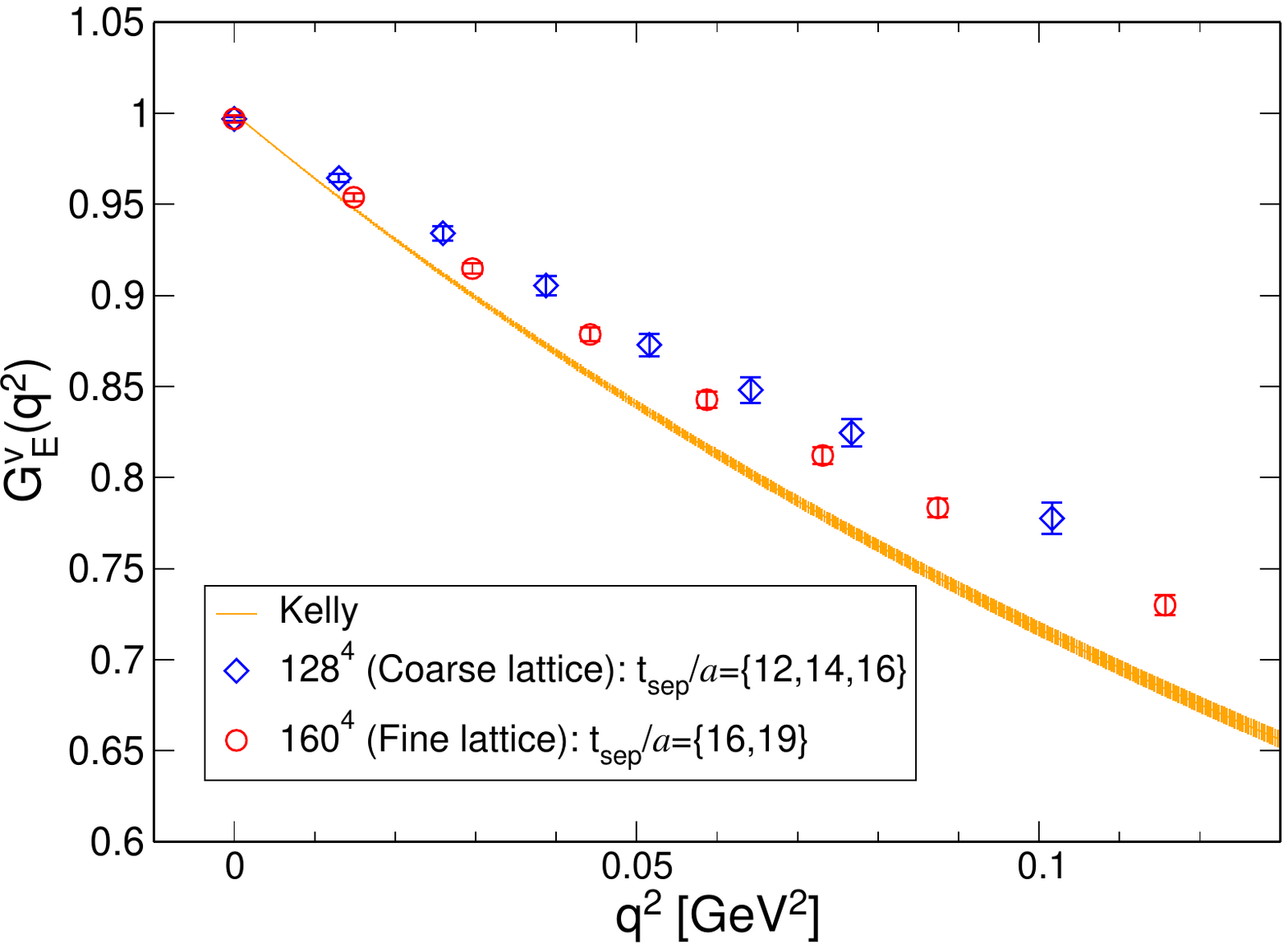}
\includegraphics[width=0.45\textwidth,bb=0 0 792 692,clip]{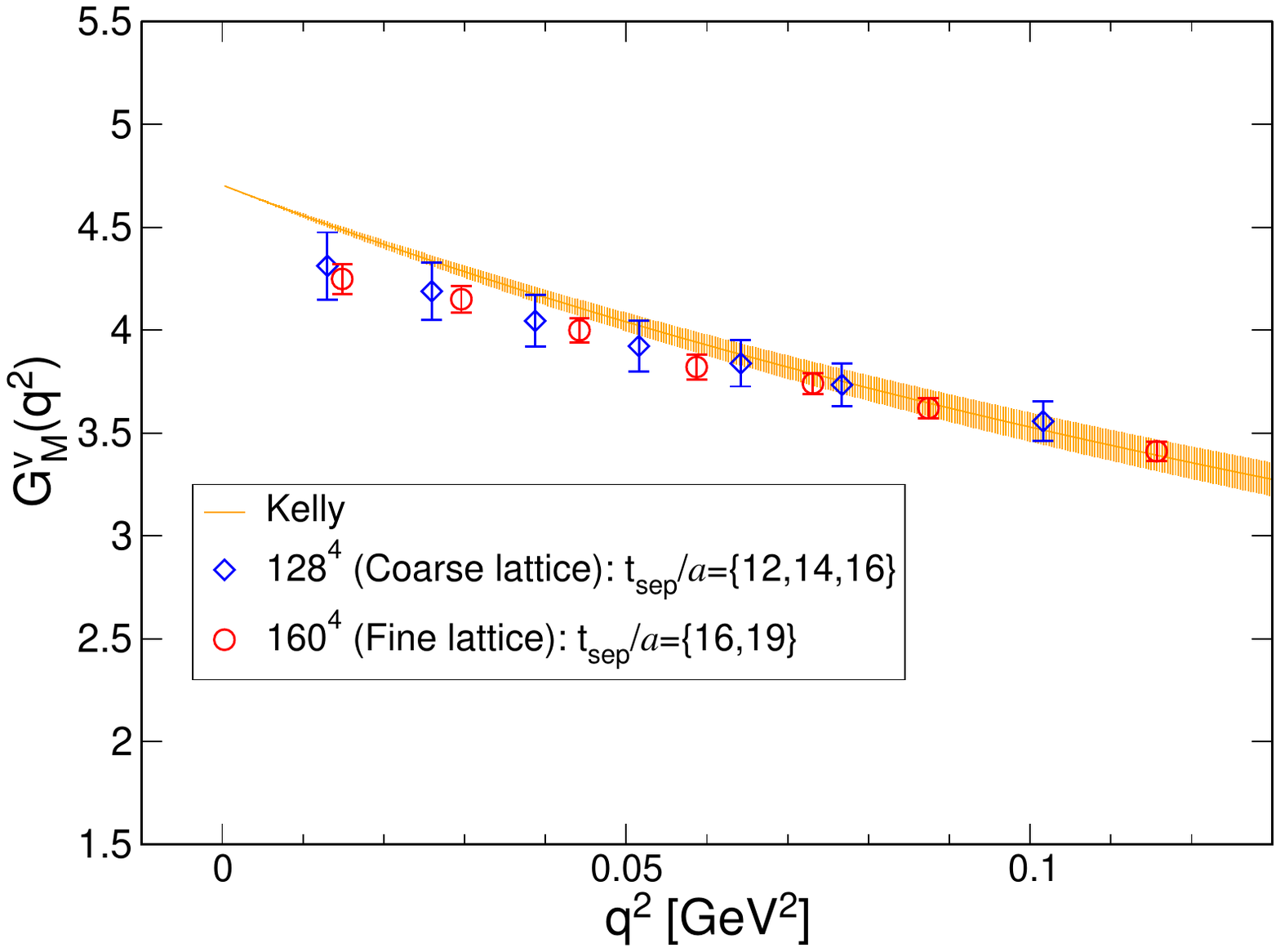}
\caption{
Results of the \textit{isovector} $G_E^v$ (left) and $G_M^v$ (right) form factors
as a functions of four-momentum squared $q^2$ for each combined data of $t_\mathrm{sep}/a=\{13,16,19\}$ for PACS10/L160 and $t_\mathrm{sep}/a=\{14,16\}$ for PACS10/L128.
The orange band represents Kelly's fit~\cite{Kelly:2004hm} as their ``experimental data''.
}
\label{fig:gegm_qsqr}
\end{figure*}

In Fig.~\ref{fig:gegm_qsqr},
the $q^2$ dependence of the renormalized $G_E^v(q^2)$ (left) and $G_M^v(q^2)$ (right)
is exhibited
together with the Kelly's fit~\cite{Kelly:2004hm} as their ``experimental data''.
For both quantities,
one can easily see that the results obtained from the fine lattice are located slightly above the Kelly's fit,
but appear systematically lower than the coarse lattice results.
However,
a quantitative comparison is not yet possible due to remaining systematic uncertainties associated with the lattice discretization effects as discussed for the $F_A$, $F_P$ and $G_P$ form factors in Sec.~\ref{ssec:the_fa_fp_and_gp_form_factors}.

\begin{figure*}
\centering
\includegraphics[width=0.32\textwidth,bb=0 0 528 612,clip]{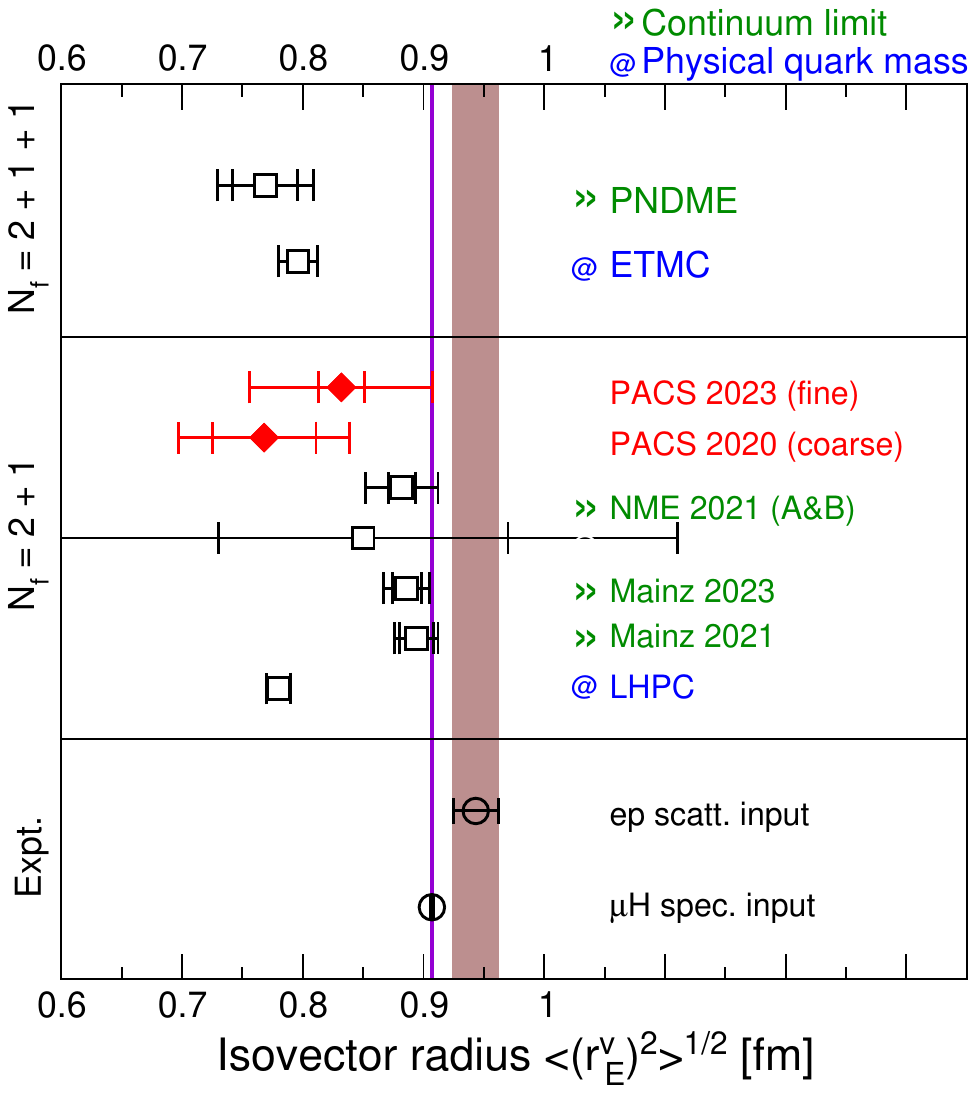}
\includegraphics[width=0.32\textwidth,bb=0 0 528 612,clip]{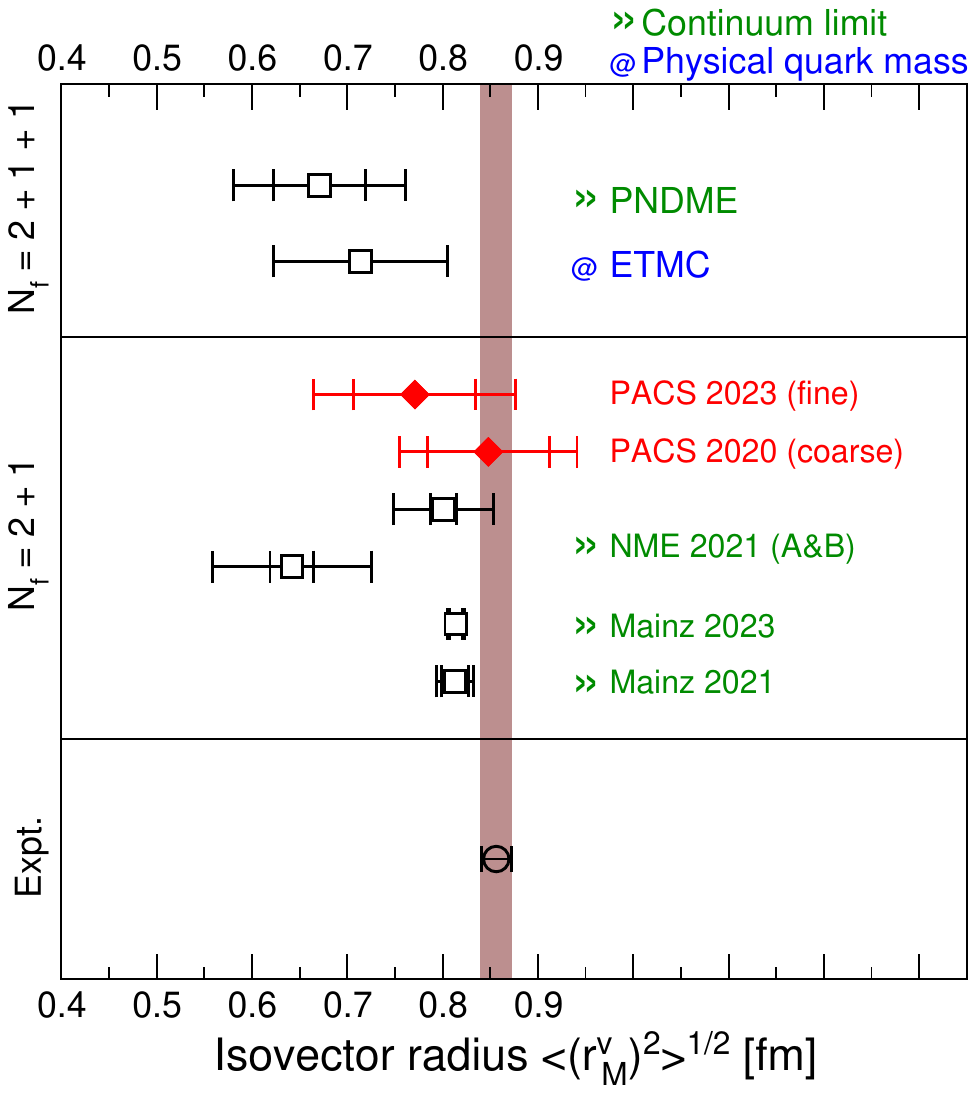}
\includegraphics[width=0.32\textwidth,bb=0 0 528 612,clip]{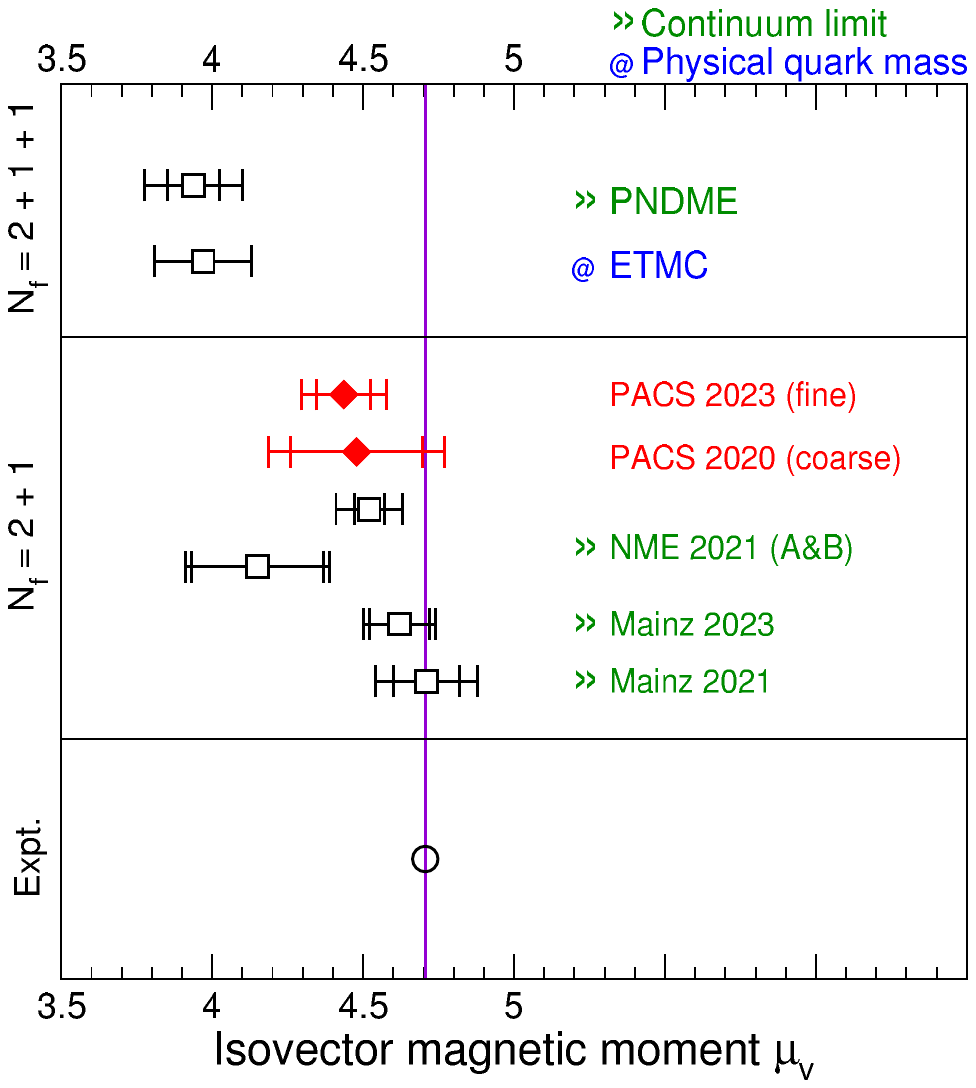}
\caption{Summary plot for our current status with the experimental values and other lattice QCD results: the \textit{isovector} electric radius $\sqrt{\langle(r_E^{v})^2\rangle}$ (left), the \textit{isovector} magnetic radius $\sqrt{\langle (r_M^v)^2\rangle}$ (middle) and \textit{isovector} magnetic moment $\mu_v$ (right).
The inner error bars represent the statistical errors,
while the outer error bars are evaluated by both the statistical and systematic errors added in quadrature.
The violet lines and brown bands appearing in each panel represent
the experimental values.
These figures are from Ref.~\cite{Tsuji:2023llh}.
}
\label{fig:rermmm_comparison}
\end{figure*}

Figure~\ref{fig:rermmm_comparison} presents summary for our results of $\sqrt{\langle (r_E^v)^2\rangle}$ (left), $\sqrt{\langle (r_m^v)^2\rangle}$ (middle), and the magnetic moment $\mu_v=G_M^v(q^2=0)$ (right),
together with the experimental values and the other lattice QCD results.
As in Fig.~\ref{fig:comparison},
the inner and outer error bars represent the statistical uncertainties and the total uncertainties given by adding both statistical and systematic errors in quadrature,
where the systematic uncertainties are evaluated in the same way as described in Sec~\ref{ssec:the_fa_fp_and_gp_form_factors}.
For the results of $\sqrt{\langle(r_E^v)^2\rangle}$ and $\sqrt{\langle(r_M^v)^2\rangle}$,
the systematic errors associated with the finite lattice spacing on the \textit{isovector} RMS radii are rather large, as much as 8-10\%.
Combining with the observation in $\sqrt{\langle r_A^2\rangle}$,
it was found that the discrepancy in the \textit{isovector} RMS radii for the different lattice spacings is observed as a possible discretization error of about 10\%, regardless of the channel.
On the other hand,
the small discretization error, which is less than 1\%,
is observed for the magnetic moment $\mu_v$, though the two results obtained from the coarse ($128^4$) and fine ($160^4$) lattices are both 5\%-6\% smaller than the experimental value.
However,
recall that the determination of $\mu_v$ potentially suffers from the large systematic uncertainty due to the $q^2$ extrapolation to the zero momentum point.
To avoid such uncertainty,
a direct calculation method without the extrapolation of $q^2\to 0$
was advocated in Ref.~\cite{Ishikawa:2021eut}. We will adopt this method in our future work.

%
%
%

%
%

\section{Numerical results II: The PCAC relation in the context of the nucleon correlation functions}
\label{sec:numerical_results_ii}

\begin{figure*}
\centering
\includegraphics[width=0.8\textwidth,bb=0 0 864 720,clip]{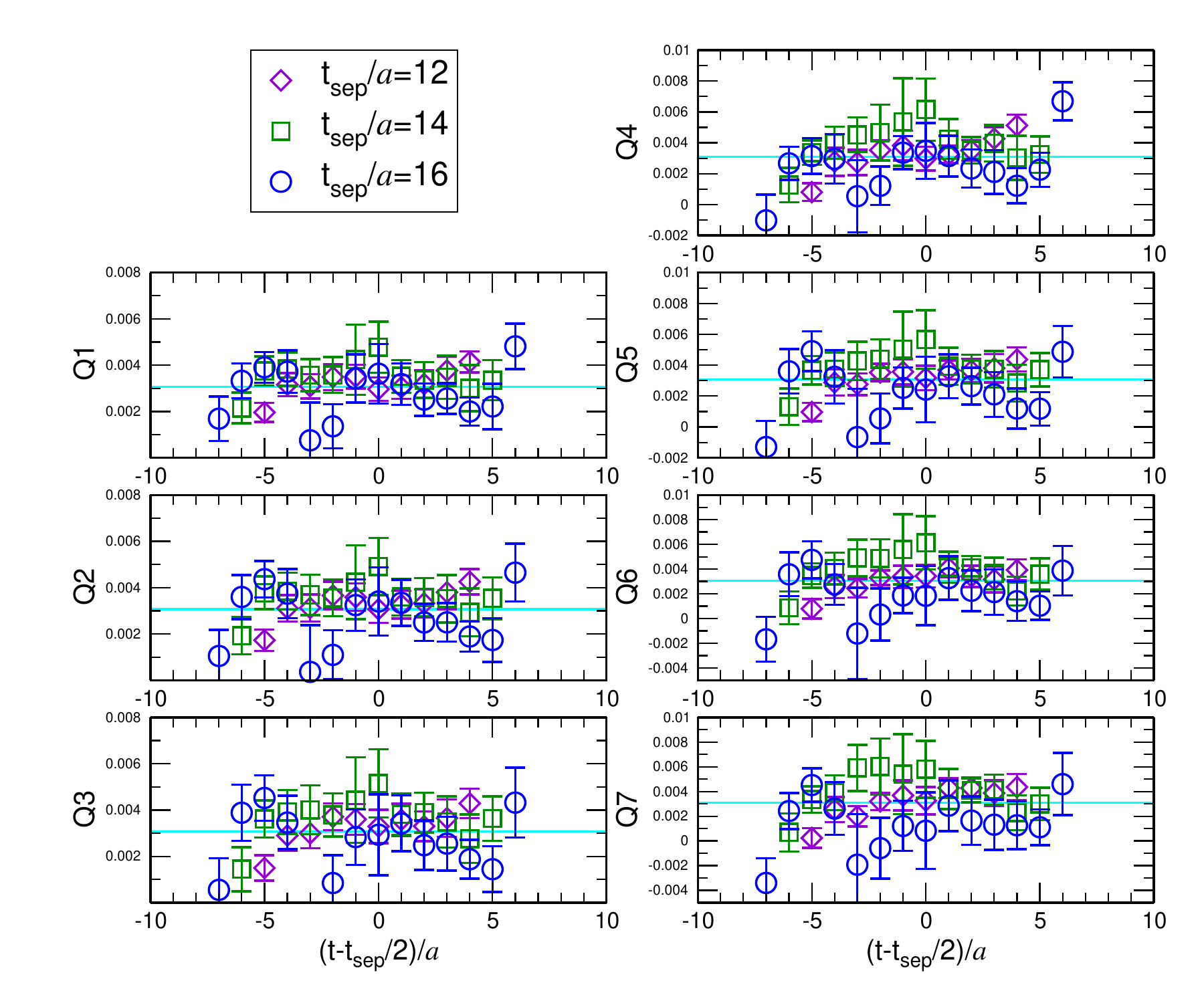}
\caption{
The values of $m_{\mathrm{PCAC}}^{\mathrm{nucl}}$
for all momentum transfers as functions of the current insertion time slice $t$ with PACS10/L128.
Results for $t_\mathrm{sep}/a=\{12,14,16\}$ are plotted as diamonds, squares and circles respectively.
In each panel, the value of $m_{\mathrm{PCAC}}^{\mathrm{pion}}$ is presented as a horizontal band.
}
\label{fig:mpcacnu_plateau_128}
\end{figure*}

\begin{figure*}
\centering
\includegraphics[width=0.8\textwidth,bb=0 0 864 720,clip]{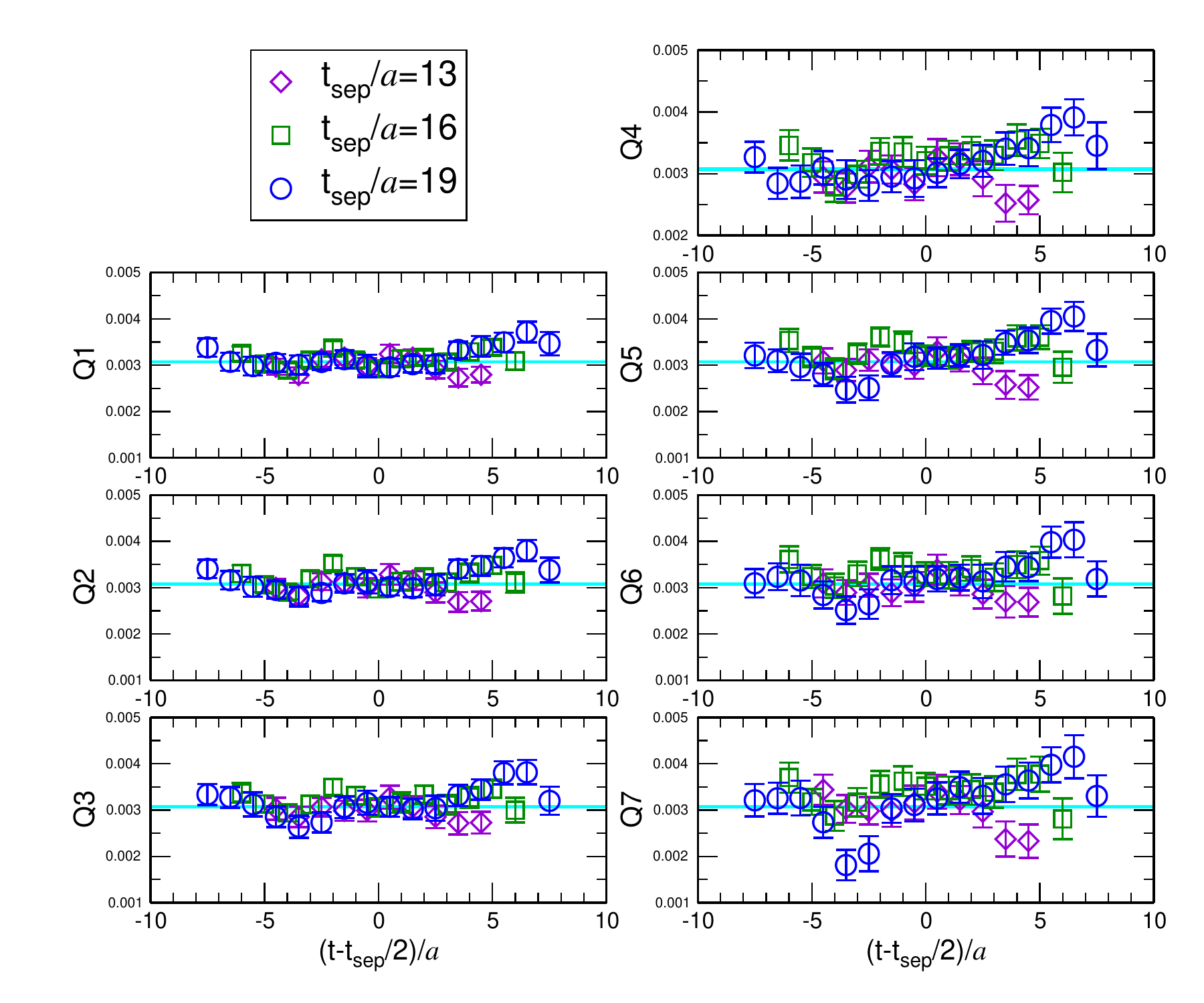}
\caption{
The values of $m_{\mathrm{PCAC}}^{\mathrm{nucl}}$
for all momentum transfers as functions of the current insertion time slice $t$ with PACS10/L160.
Results for $t_\mathrm{sep}/a=\{13,16,19\}$ are plotted as diamonds, squares and circles respectively.
In each panel, the value of $m_{\mathrm{PCAC}}^{\mathrm{pion}}$ is presented as a horizontal band.
This figure is reprinted from Ref.~\cite{Tsuji:2023llh}.
}
\label{fig:mpcacnu_plateau_160}
\end{figure*}

The PCAC relation derived from the AWT identity,
is strongly related to the low-energy relations and models that existed before QCD, as described in Sec.~\ref{ssec:axial_matrix_elements_of_the_weak_currents}.
Here let us consider the two types of the \textit{bare} quark mass:
$m_\mathrm{PCAC}^{\mathrm{pion}}$ and $m_\mathrm{PCAC}^\mathrm{nucl}$,
which are introduced in Sec.~\ref{ssec:the_pcac_relation_with_nucleon_correlation_functions},
to verify the PCAC relation in the context of the nucleon correlation functions.
Figures~\ref{fig:mpcacnu_plateau_128} and \ref{fig:mpcacnu_plateau_160} represent the ratios defined in Eq.~(\ref{eq:m_awti_pcac}) obtained from our $128^4$ and $160^4$ lattice ensembles, respectively.
All data show good plateaus regardless of all $q^2$ with all $t_\mathrm{sep}/a=\{12,14,16\}$ for PACS10/L128 and $t_\mathrm{sep}/a=\{13,16,19\}$ for PACS10/L160.
As long as the ratio defined in Eq.~(\ref{eq:m_awti_pcac}) exhibits $q^2$ independent
behavior as a function of $q^2$,
this ratio provides an alternative \textit{bare} quark mass, referred to as $m_\mathrm{PCAC}^{\mathrm{nucl}}$,
which should be identical to $m_\mathrm{PCAC}^\mathrm{pion}$ in the continuum limit.

\begin{figure*}
\centering
\includegraphics[width=0.8\textwidth,bb=0 0 792 692,clip]{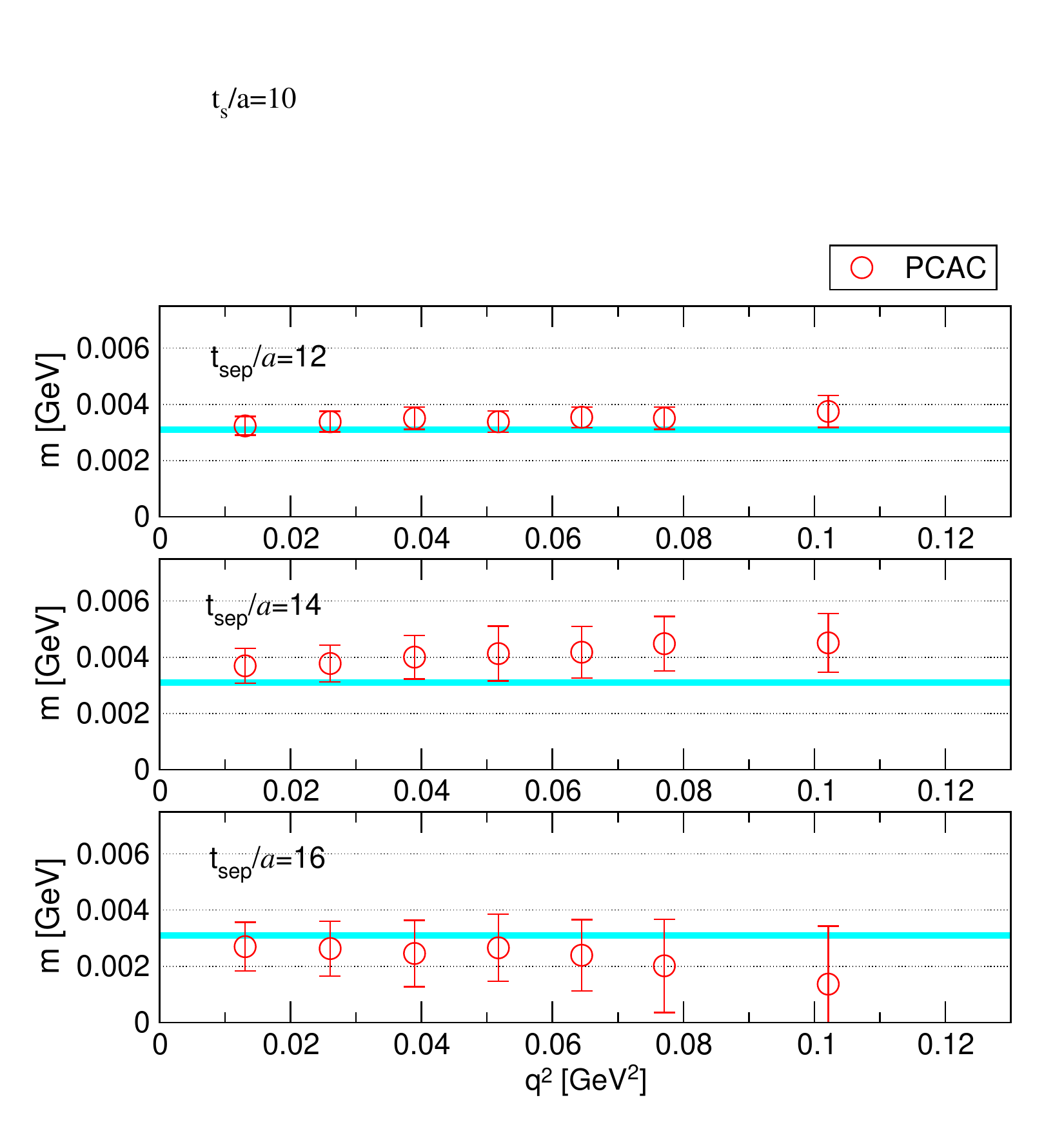}
\caption{
Comparison of the two types of {\textit{bare}} quark mass, $m_\mathrm{PCAC}^\mathrm{pion}$ (horizontal band) and $m_\mathrm{PCAC}^\mathrm{nucl}$ (open circles),
which are obtained from PACS10/L128.
Results for $t_\mathrm{sep}/a=\{12,14,16\}$ are plotted from top to bottom panels.
}
\label{fig:mpcacnu_qdep_128}
\end{figure*}

\begin{figure*}
\centering
\includegraphics[width=0.8\textwidth,bb=0 0 792 692,clip]{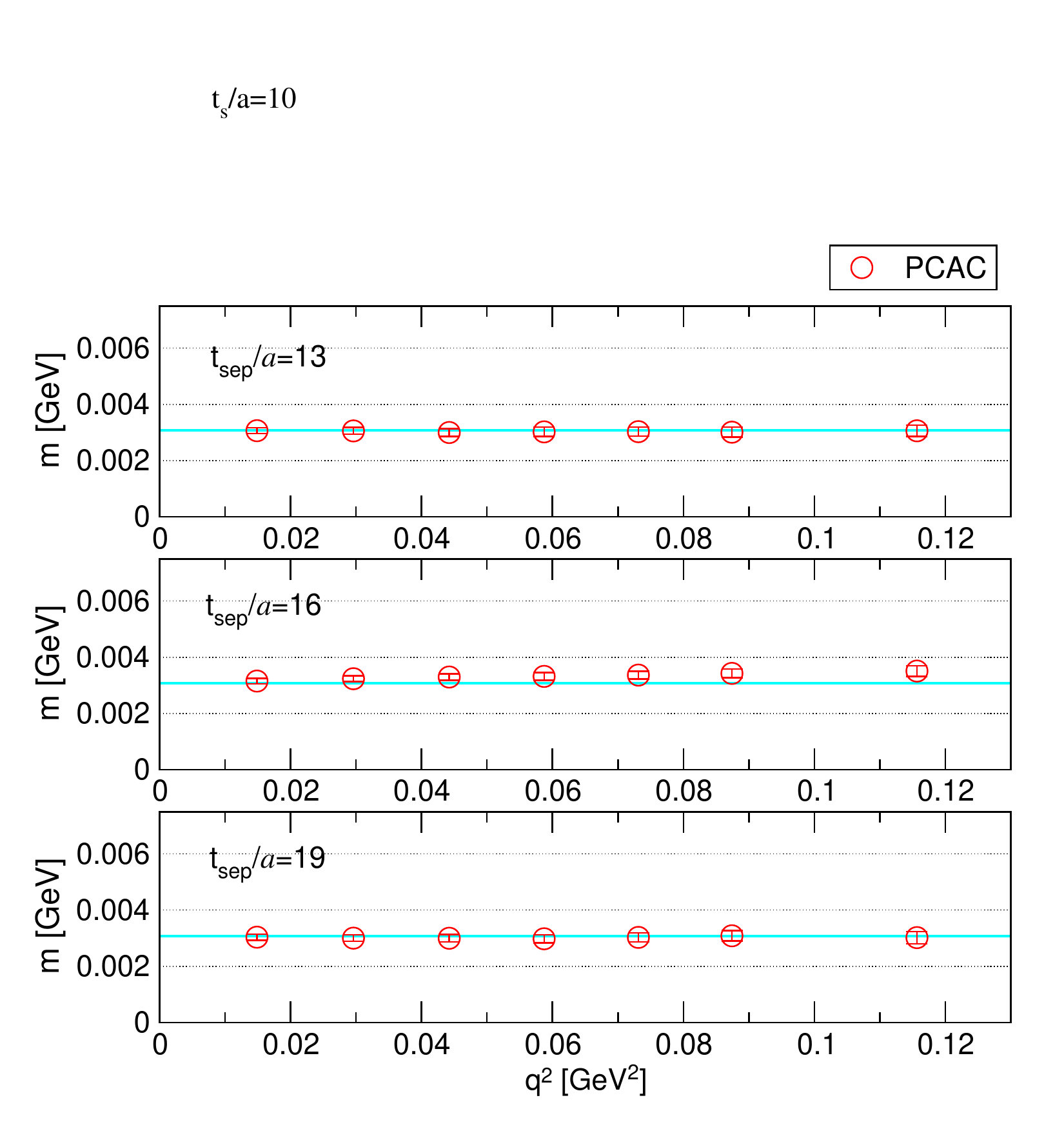}
\caption{
Comparison of the two types of {\textit{bare}} quark mass, $m_\mathrm{PCAC}^\mathrm{pion}$ (horizontal band) and $m_\mathrm{PCAC}^\mathrm{nucl}$ (open circles),
which are obtained from PACS10/L160.
Results for $t_\mathrm{sep}/a=\{13,16,19\}$ are plotted from top to bottom panels.
}
\label{fig:mpcacnu_qdep_160}
\end{figure*}

In Figs.~\ref{fig:mpcacnu_qdep_128} and \ref{fig:mpcacnu_qdep_160},
a direct comparison between 
$m_\mathrm{PCAC}^{\mathrm{pion}}$ (denoted by a horizontal line) and
$m_\mathrm{PCAC}^{\mathrm{nucl}}$ (denoted by circle symbols) 
is presented for each lattice ensemble of $128^4$ and $160^4$,
in which the values of $m_\mathrm{PCAC}^{\mathrm{nucl}}$ for each $q^2$ are evaluated by an error-weighted average of five data points for Figs.~\ref{fig:mpcacnu_plateau_128} and \ref{fig:mpcacnu_plateau_160} in the central range of $t/a$.
The resulting data of $m_\mathrm{PCAC}^{\mathrm{nucl}}$ show no significant $q^2$ dependence and
reproduce the value of $m_\mathrm{PCAC}^{\mathrm{pion}}$ regardless of the $q^2$ value.
This agreement between $m_\mathrm{PCAC}^{\mathrm{nucl}}$ and $m_\mathrm{PCAC}^{\mathrm{pion}}$ with any finite momentum transfer is highly non-trivial,
indicating the following two points:
1) the lattice QCD data correctly reproduce the physics in the continuum within statistical precision, and
2) the effect of the $O(a)$-improvement of the axial-vector current does not have a significant contribution.

Let us examine these two points numerically.
In this study,
the local axial-vector current is used to compute the matrix elements, 
which receive a larger discretization uncertainty, instead of using an $O(a)$-improved current of $\widetilde{A}^{\mathrm{imp}}_\alpha = \widetilde{A}_\alpha+ac_A\partial_\alpha \widetilde{P}$.
This implies that 
the definition of $m_{\mathrm{PCAC}}^{\mathrm{nucl}}$
defined in Eq.~(\ref{eq:m_awti_pcac}) does not take into account $O(a)$-improvement of the axial-vector current.
Indeed, the second term in the $O(a)$-improved axial-vector current yields the $O(a)$ correction, which is given by a positive constant shift for the ground-state contribution
on the value of $m_{\mathrm{PCAC}}^{\mathrm{pion}}$ as
\begin{align}
\label{eq:m_pion_pcac_Oa-imp}
(m_{\mathrm{PCAC}}^{\mathrm{pion}} )^{\mathrm{imp}}
=
m_{\mathrm{PCAC}}^{\mathrm{pion}} + \frac{aZ_Ac_A}{2} M_\pi^2,
\end{align}
where the correction term is proportional to $M_\pi^2$, 
since the pion two-point correlation function is projected onto zero three momentum.
On the other hand, $m_{\mathrm{PCAC}}^{\mathrm{nucl}}$,
which can be computed only with finite three momentum $\boldsymbol{q}$. 
Therefore, the $O(a)$ correction yields momentum dependent term for the ground-state contribution as
\begin{align}
\label{eq:m_awti_pcac_Oa-imp}
(m_{\mathrm{PCAC}}^{\mathrm{nucl}} )^{\mathrm{imp}}
&=
Z_A \times 
\frac{\partial_{\alpha}\left(
C^{5z}_{A_{\alpha}}(t;\boldsymbol{q}) +
a c_A \partial_{\alpha} C^{5z}_{P}(t;\boldsymbol{q})
\right)
}{
2 C^{5z}_{P}(t;\boldsymbol{q})
}
=
m_{\mathrm{PCAC}}^{\mathrm{nucl}} - \frac{aZ_Ac_A}{2} q^2,
\end{align}
where the correction term is proportional to
the square of momentum transfer $q^2$, which is given by
$q^2=2M_N(E_N(\boldsymbol{q})-M_N)$, with a negative sign.
Recall that
two types of the {\textit{bare}} quark mass of $m_{\mathrm{PCAC}}^{\mathrm{nucl}}$ and $m_{\mathrm{PCAC}}^{\mathrm{pion}}$ are supposed to be identical in the continuum limit
as described in Sec.~\ref{ssec:axial_matrix_elements_of_the_weak_currents}.
Therefore, the implications of our finding that $m_{\mathrm{PCAC}}^{\mathrm{nucl}}$ coincides $m_{\mathrm{PCAC}}^{\mathrm{pion}}$ within statistical precision without the $O(a)$ correction as shown in Figs.~\ref{fig:mpcacnu_qdep_128} and \ref{fig:mpcacnu_qdep_160}
are twofold: 1) the effect of the $O(a)$ correction is 
negligibly small in our calculations since the value of $c_A$ is likely to be nearly zero, 2) our calculations correctly inherit the low-energy physics associated with the AWT identity up to $q^2 \lesssim 0.1 [\mathrm{GeV}^2]$.

%
%
%

%
%
%

%
%
%

%
%

\section{Numerical results III: Generalized Goldberger-Trieman relation and pion-pole dominance model}
\label{sec:numerical_results_iii}

In Sec.~\ref{sec:numerical_results_ii},
it is observed that the PCAC relation is satisfied in the context of the nucleon correlation functions in a wide range of $q^2$ ($0.01\ [\mathrm{GeV}^2]\lesssim q^2\lesssim 0.12\ [\mathrm{GeV}^2])$ within statistical precision.
Next, we consider other low-energy relations,
such as the generalized Goldberger-Trieman (GGT) relation and the pion-pole dominance (PPD) model,
which are associated with the PCAC relation.

\subsection{The generalized Goldberger--Trieman relation and pion-pole dominance model}
\label{ssec:the_generalized_goldberger_trieman_relation_and_pion-pole_dominance_model}

According to Refs.~\cite{Jang:2023zts,Sasaki:2007gw,Yamazaki:2009zq},
three types of combinations of $\widetilde{F}_A$, $\widetilde{F}_P$ and $\widetilde{G}_P$ form factors are introduced to test whether the three form factors satisfy the GGT relation and the PPD model
described in Sec.~\ref{ssec:axial_matrix_elements_of_the_weak_currents}.
First the GGT relation of Eq.~(\ref{eq:ggt})
is rewritten as
\begin{align}
\label{eq:r1r2}
    R_1 + R_2 = 1
\end{align}
with
\begin{align}
    R_1 & \equiv
    \frac{q^2}{2M_N} \frac{\widetilde{F}_P(q^2)}{\widetilde{F}_A(q^2)},
    \\
    R_2 & \equiv
    \frac{1}{2M_N} \frac{2m_\mathrm{PCAC}\widetilde{G}_P(q^2)}{Z_A\widetilde{F}_A(q^2)}
\end{align}
with the \textit{bare} quark mass $m_\mathrm{PCAC}$,
where
$m_\mathrm{PCAC}^\mathrm{pion}$ is adopted for $m_\mathrm{PCAC}$ in this study based on the observation described in Sec.~\ref{sec:numerical_results_ii}.
To verify the PPD form of $F_P(q^2)$ given in Eq.~(\ref{eq:ppd}), the following
ratio of the $\widetilde{F}_P$ and $\widetilde{F}_A$ form factors
is introduced,
\begin{align}
\label{eq:r3}
    R_3 & \equiv
    \frac{q^2+M_\pi^2}{2M_N}\frac{\widetilde{F}_P(q^2)}{\widetilde{F}_A(q^2)}
\end{align}
with the simulated masses of $M_\pi$ and $M_N$~\footnote{
In Refs.~\cite{Sasaki:2007gw, Yamazaki:2009zq}, $R_3$ is introduced as $\alpha_{\mathrm{PPD}}$.}
In addition to verify the PPD form for $G_P(q^2)$ given in Eq.~(\ref{eq:ppd_gp}), the another ratio of the $\widetilde{G}_P$ and $\widetilde{F}_P$ form factors is also introduced,
\begin{align}
\label{eq:r4}
    R_4 & \equiv
    M_\pi^2 \frac{Z_A\widetilde{F}_P(q^2)}{2m_\mathrm{PCAC}\widetilde{G}_P(q^2)}.
\end{align}

Therefore, the following three relations
\begin{align}
\label{eq:tests}
    R_1 + R_2 = 1,\quad 
    R_3 = 1, \quad
    R_4 = 1
\end{align}
should be examined to verify the GGT relation and the PPD model, which are identified as a consequence of the low-energy physics associated with the PCAC relation.

\subsection{Numerical results}

First, let us examine the relation of $R_1+R_2=1$ with the two ratios of $R_1$ and $R_2$ defined in Eq.~(\ref{eq:r1r2}) to verify the GGT relation.
As explained in Sec.\ref{ssec:axial_matrix_elements_of_the_weak_currents}, the GGT relation is derived from the AWT identity in terms of the matrix elements of {\it the nucleon ground state}.
Therefore, verification of the GGT relation requires the three form factors extracted by isolating the ground-state contribution from the excited-state contributions in the nucleon three-point correlation functions. 
This point differs from the PCAC relation in terms of the nucleon correlation functions discussed in Sec.~\ref{sec:numerical_results_ii}, which can be satisfied 
without isolating the ground-state contribution.
In this context, testing the GGT relation is highly sensitive to the excited-state contamination.

\begin{figure*}
\centering
\includegraphics[width=0.8\textwidth,bb=0 0 792 692,clip]{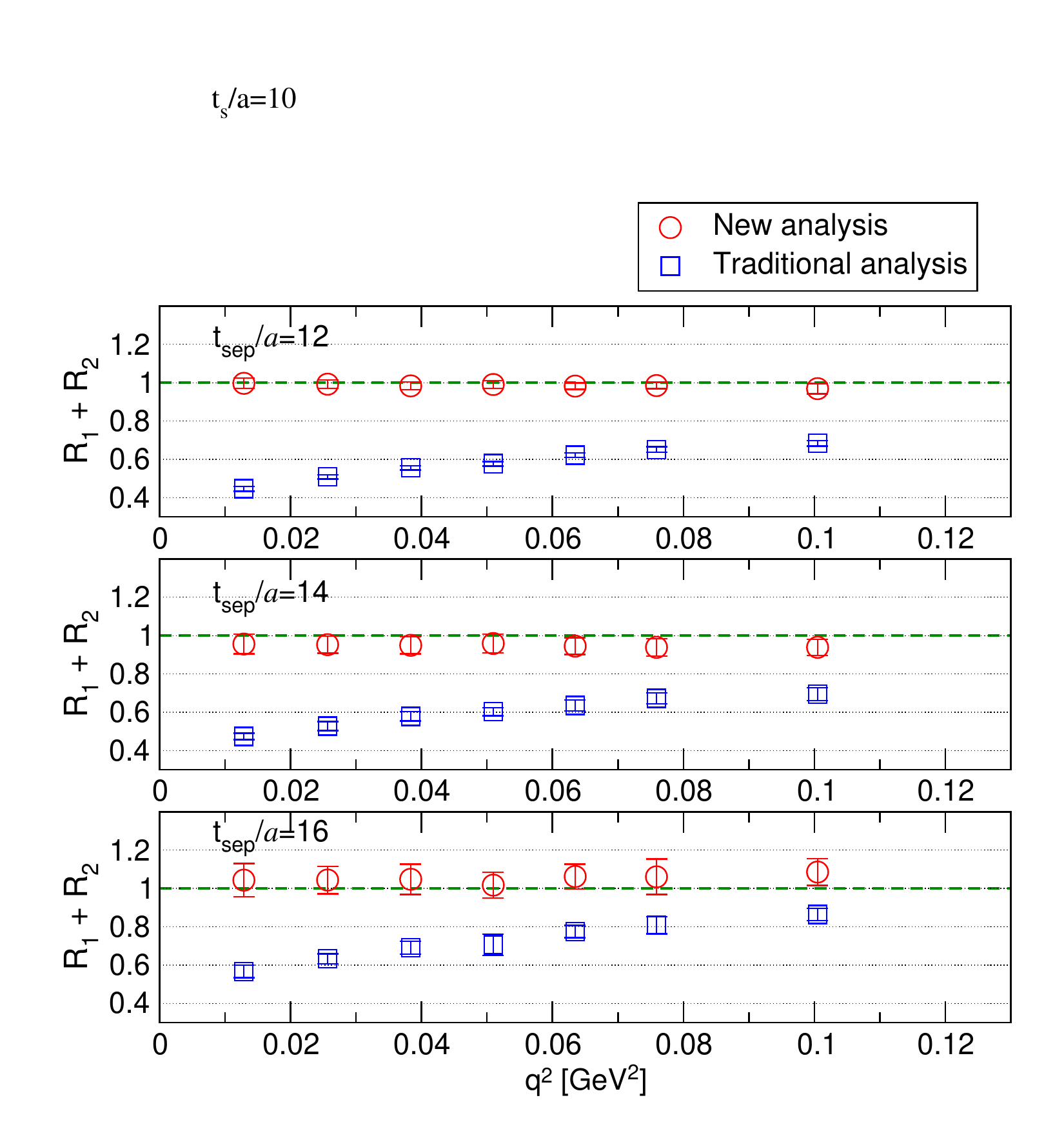}
\caption{
Comparisons of $R_1+R_2$ between our \textit{new analysis} (open circles) and the \textit{traditional analysis} (open squares) obtained from PACS10/L128,
while the dashed line represents the unity.
Results for $t_\mathrm{sep}/a=\{12,14,16\}$ are plotted from top to bottom panels.
}
\label{fig:R1R2_qdep_128}
\end{figure*}

\begin{figure*}
\centering
\includegraphics[width=0.8\textwidth,bb=0 0 792 692,clip]{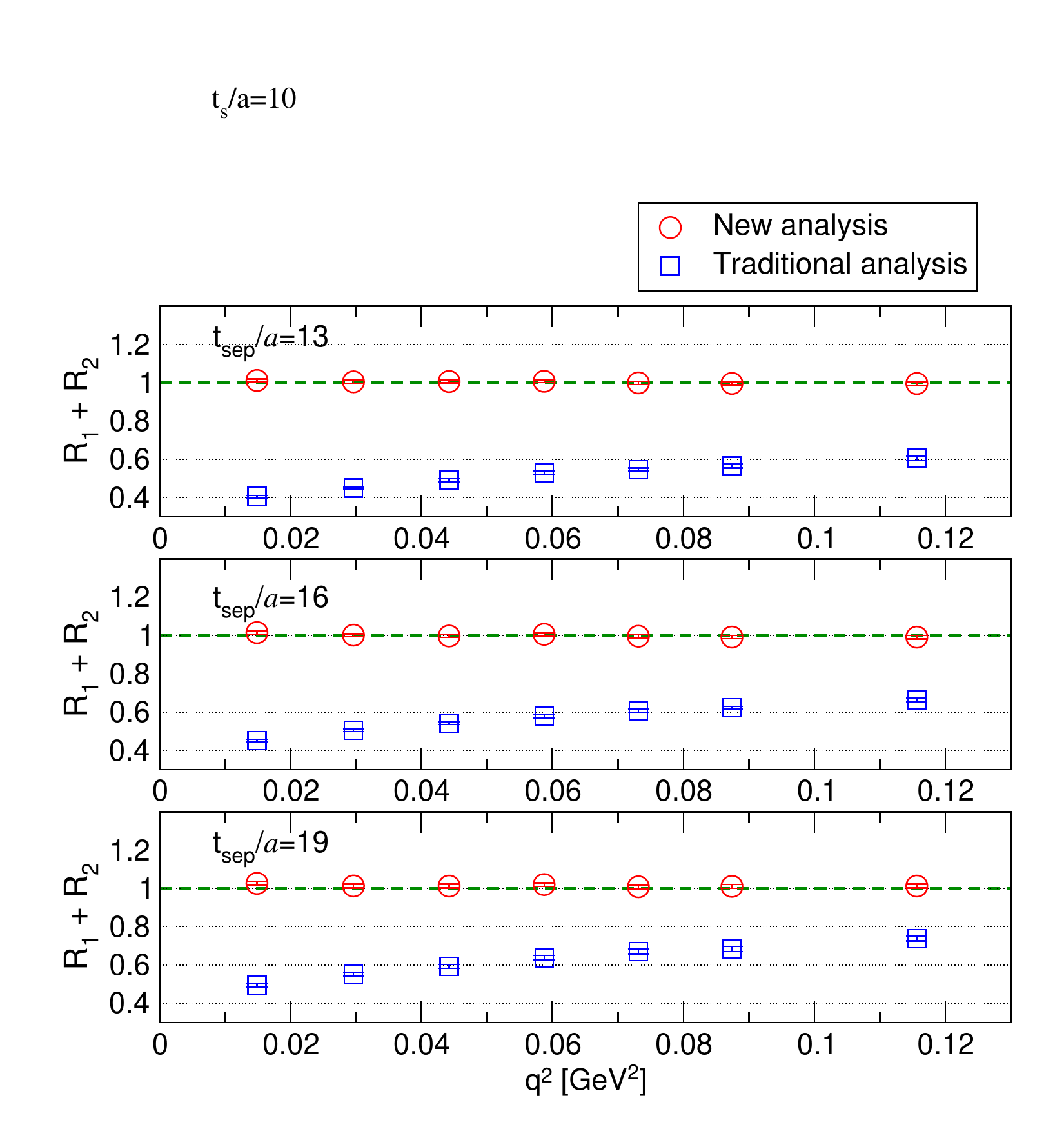}
\caption{
Comparisons of $R_1+R_2$ between our \textit{new analysis} (open circles) and the \textit{traditional analysis} (open squares) obtained from PACS10/L160,
while the dashed line represents the unity.
Results for $t_\mathrm{sep}/a=\{13,16,19\}$ are plotted from top to bottom panels.
Results for $t_\mathrm{sep}/a=\{12,14,16\}$ are plotted from top to bottom panels.
}
\label{fig:R1R2_qdep_160}
\end{figure*}

Figures~\ref{fig:R1R2_qdep_128} and \ref{fig:R1R2_qdep_160} show the values of $R_1+R_2$, calculated from the coarse ($128^4$) and fine ($160^4$) lattices, as a function of $q^2$.
In these figures, two results, which are obtained from the \textit{new analysis} (open circles) and the \textit{traditional analysis} (open squares), are included for comparison. Recall that the \textit{new analysis} uses the leading $\pi N$ subtraction method~\cite{Sasaki:2025qro,Aoki:2025taf}, which can remove the $\pi N$-state contribution from the $\widetilde{F}_P$ and $\widetilde{G}_P$ form factors given by the standard ratio method.

The results of the \textit{new analysis} demonstrate the successful reproduction of unity for all variations of $t_\mathrm{sep}$, regardless of the values of $q^2$ and the lattice spacing. In contrast, the data obtained from the \textit{traditional analysis} are substantially below unity, and the deviation from unity tends to be larger for smaller $q^2$. This is simply because the values of $\widetilde{F}_P(q^2)$ and $\widetilde{G}_P(q^2)$ obtained by the \textit{traditional analysis} are significantly underestimated in the low-$q^2$ region compared to the PPD model. A close look at the figures reveals that the deviation from unity found in the \textit{traditional analysis} tends to be smaller for larger $t_\mathrm{sep}$. 

This observation suggests that, although the GGT relation appears to be violated in the \textit{traditional analysis}, the GGT relation should strictly be valid 
if the residual contamination from the $\pi N$-state contribution is successfully eliminated in the determination of $\widetilde{F}_P(q^2)$ and $\widetilde{G}_P(q^2)$, as in the \textit{new analysis}.

\begin{figure*}
\centering
\includegraphics[width=0.8\textwidth,bb=0 0 792 692,clip]{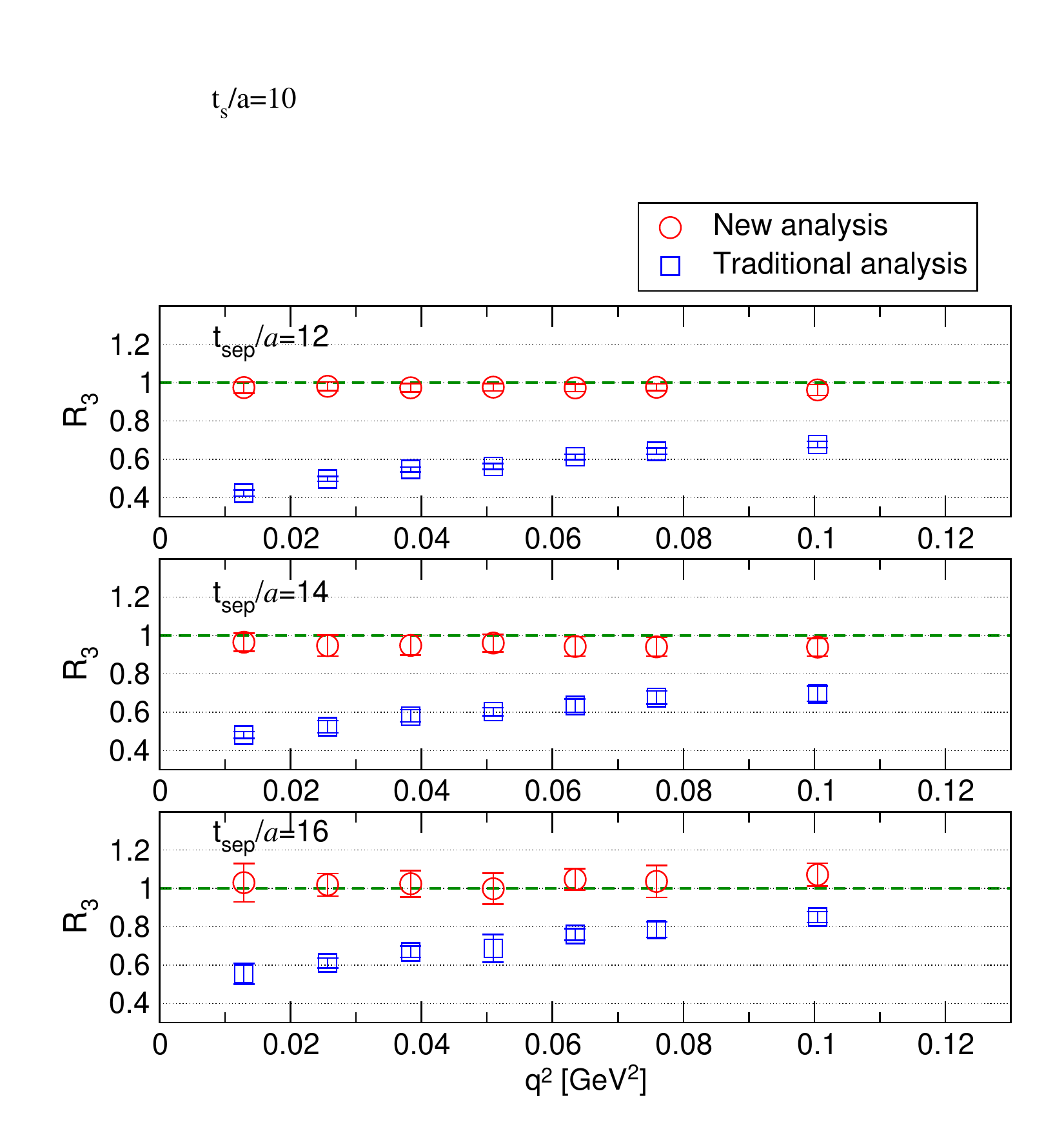}
\caption{
Comparisons of $R_3$ between our \textit{new analysis} (open circles) and the \textit{traditional analysis} (open squares) obtained from PACS10/L128,
while the dashed line represents the unity.
Results for $t_\mathrm{sep}/a=\{12,14,16\}$ are plotted from top to bottom panels.
}
\label{fig:R3_qdep_128}
\end{figure*}

\begin{figure*}
\centering
\includegraphics[width=0.8\textwidth,bb=0 0 792 692,clip]{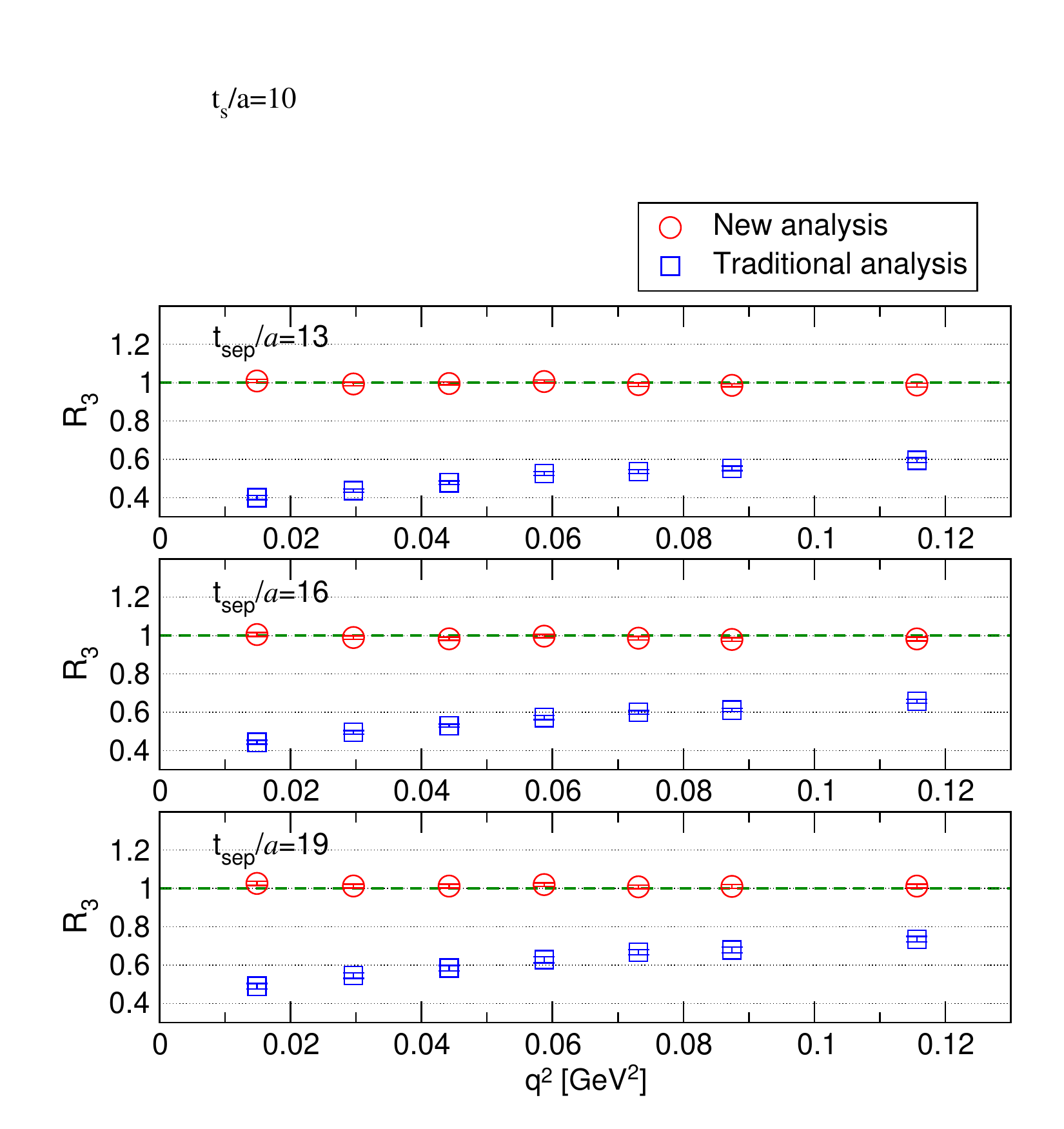}
\caption{
Comparisons of $R_3$ between our \textit{new analysis} (open circles) and the \textit{traditional analysis} (open squares) obtained from PACS10/L160,
while the dashed line represents the unity.
Results for $t_\mathrm{sep}/a=\{13,16,19\}$ are plotted from top to bottom panels.
}
\label{fig:R3_qdep_160}
\end{figure*}

Next, we examine the relation of $R_3=1$ with the ratio of $R_3$ defined in Eq.~(\ref{eq:r3}) to verify the PPD form for $F_P(q^2)$.
Figures~\ref{fig:R3_qdep_128} and \ref{fig:R3_qdep_160} show the values of $R_3$,
calculated from the coarse ($128^4$) and fine ($160^4$) lattices, as a function of $q^2$. Similar to the case of $R_1+R_2$, the results of the \textit{new analysis} 
show no significant $q^2$ dependence and are consistent with unity for 
all variations of $t_\mathrm{sep}$, regardless of the values of $q^2$ and the lattice spacing. On the other hand, the results obtained from the \textit{traditional analysis} are substantially below unity and the deviation
from unity tends to be smaller for larger $t_\mathrm{sep}$.

Unlike the GGT relation which is derived based on the AWT identity, the PPD model is justified only in the massless limit of $M_\pi$. Nevertheless, it is found that the PPD model reliably works at the physical point in the range of $q^2 \lesssim 0.1 \ [\mathrm{GeV}^2]$. This observation is consistent with what we observed in Fig.~\ref{fig:fp_2mgp_qsqr}. This finding provides the theoretical insight into the PPD model.

\begin{figure*}
\centering
\includegraphics[width=0.8\textwidth,bb=0 0 792 692,clip]{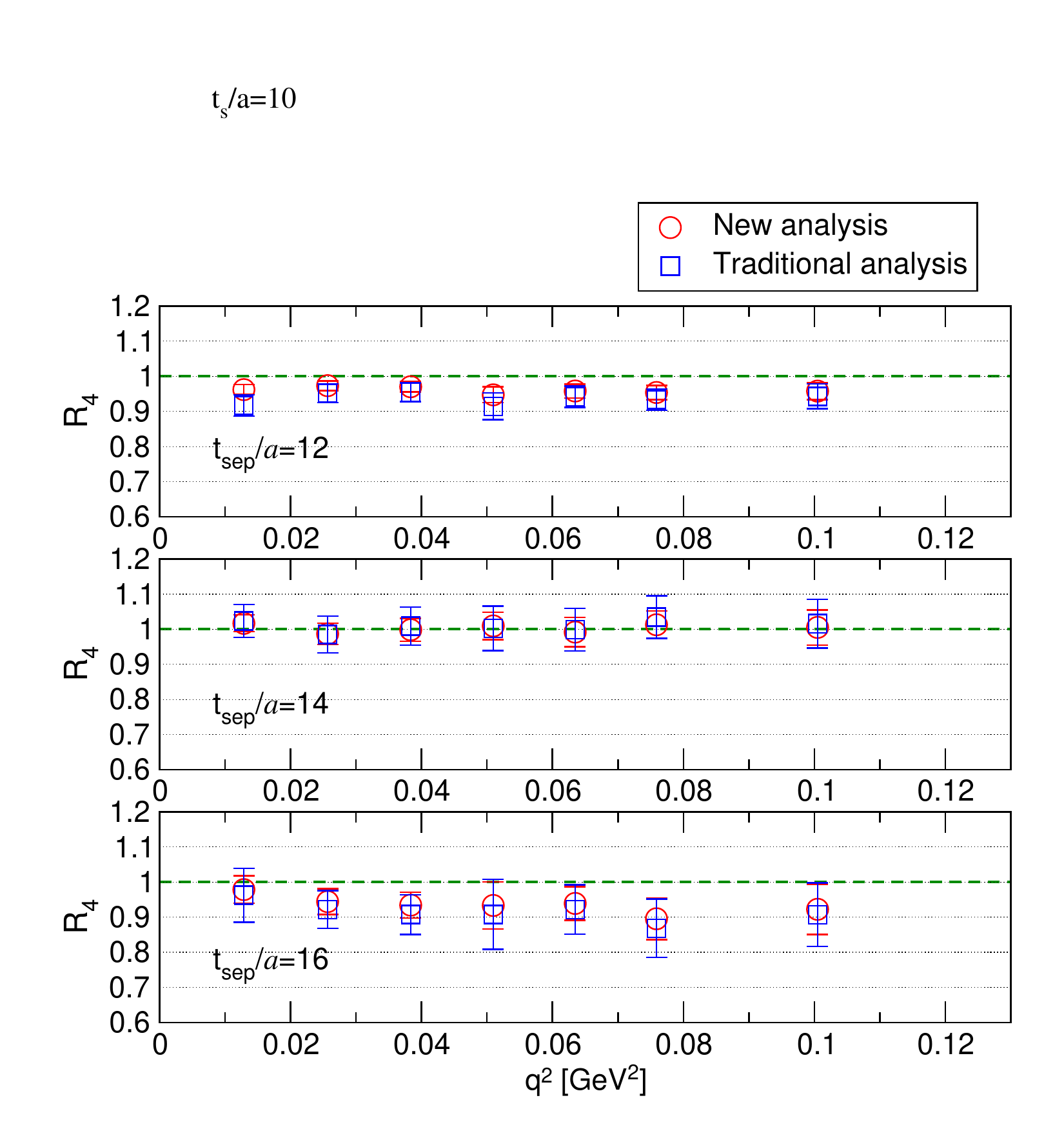}
\caption{
Comparisons of $R_4$ between our \textit{new analysis} (open circles) and the \textit{traditional analysis} (open squares) obtained from PACS10/L128,
while the dashed line represents the unity.
Results for $t_\mathrm{sep}/a=\{12,14,16\}$ are plotted from top to bottom panels.
}
\label{fig:R4_qdep_128}
\end{figure*}

\begin{figure*}
\centering
\includegraphics[width=0.8\textwidth,bb=0 0 792 692,clip]{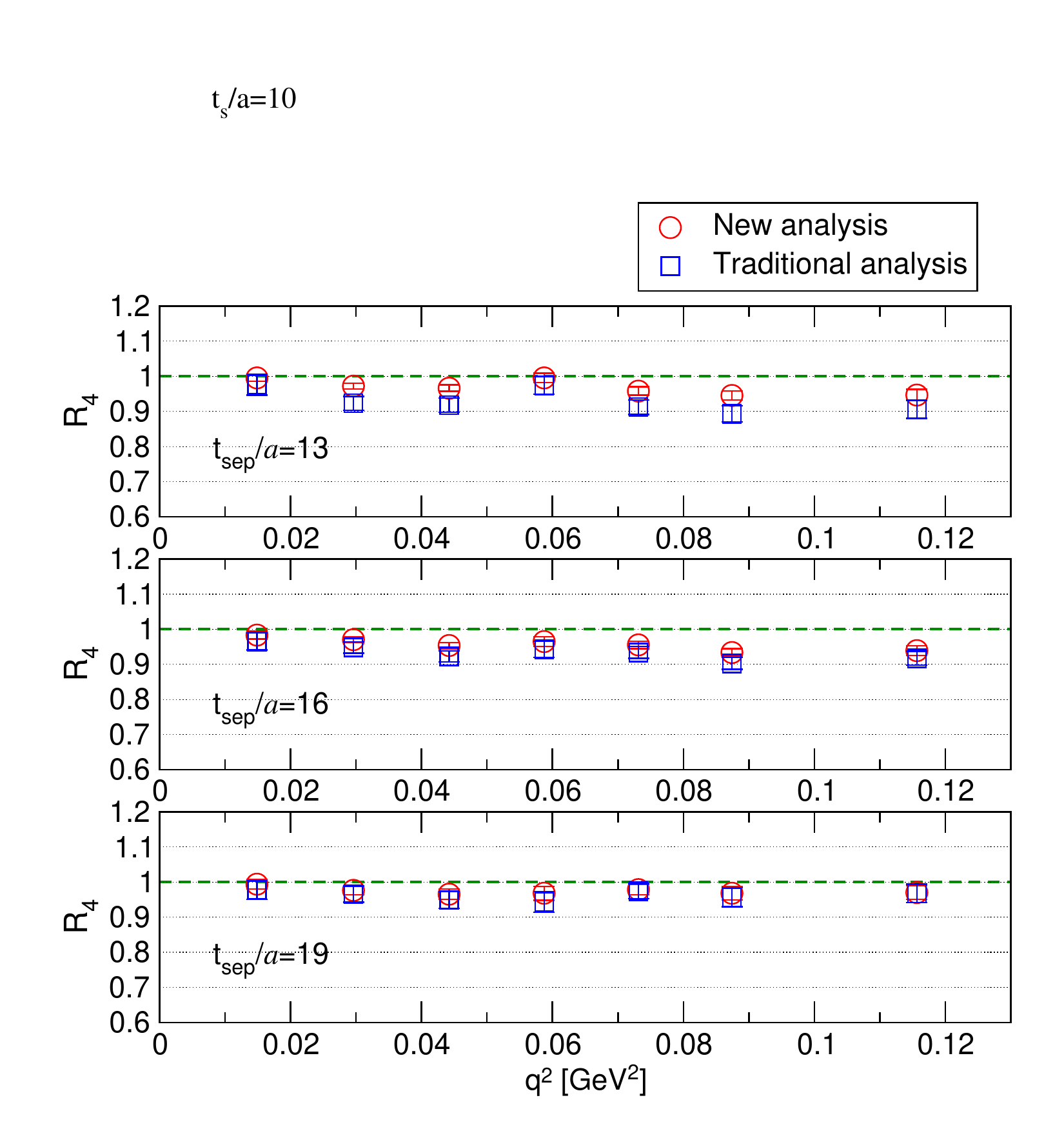}
\caption{
Comparisons of $R_4$ between our \textit{new analysis} (open circles) and the \textit{traditional analysis} (open squares) obtained from PACS10/L160,
while the dashed line represents the unity.
Results for $t_\mathrm{sep}/a=\{13,16,19\}$ are plotted from top to bottom panels.
}
\label{fig:R4_qdep_160}
\end{figure*}

Finally, we examine the relation of $R_4=1$ with the ratio of $R_4$ defined in Eq.~(\ref{eq:r4}) to verify the PPD form for $G_P(q^2)$, which is given by a combination of the GGT relation and the PPD model for $F_P(q^2)$.
Figures~\ref{fig:R4_qdep_128} and \ref{fig:R4_qdep_160} show the values of $R_4$ calculated from the coarse ($128^4$) and fine ($160^4$) lattices, as a function of $q^2$.
In contrast to the cases of $R_1+R_2$ and $R_3$, both of the \textit{new analysis} and the \textit{traditional analysis}
exhibit no appreciable dependence on $q^2$ for the coarse ($128^4$) and fine ($160^4$) lattice ensembles. 
The relation of $R_4=1$ is interpreted as 
\begin{align}
\label{eq:GP_FP_Ratio}
\frac{\widetilde{G}_P(q^2)}{F_P(q^2)} &= B_0
\end{align}
with the bare low-energy constant, $B_0=\frac{M_\pi^2}{2m_{\mathrm{PCAC}}}$, which
is associated with the GMOR relation~\cite{Sasaki:2007gw, Yamazaki:2009zq}.
As described in Eq.~(\ref{eq:gmor}), this relation can be satisfied when  
both the $\widetilde{F}_P(q^2)$ and $\widetilde{G}_P(q^2)$ simultaneously obey the PPD form. In fact, as shown in Fig.~\ref{fig:fp_2mgp_qsqr}, the $\widetilde{F}_P(q^2)$ and $\widetilde{G}_P(q^2)$ obtained from the \textit{new analysis}
are in good agreement with the PPD form. Thus, it is understandable that the relation of $R_4=1$ is obtained from  the \textit{new analysis}. What is the basis for $R_4\approx 1$
is obtained even from the \textit{traditional analysis}?
See Appendix E of Ref.~\cite{Aoki:2025taf} for a detailed discussion. 
The essential point is that the $\widetilde{F}_P$ and $\widetilde{G}_P$ form factors suffer from similar contaminations from the $\pi N$-state
contribution, which are connected by the AWT identity.

Numerical data for the ratios discussed in this section are summarized in Appendix.~\ref{app:numerical_data_for_the_ratios_versus_q^2}.

%
%
%

%
%
%

%
%
%

%
%
%

%
%
%

%
%

\section{Discussion: the applicable range of $q^2$ for the low-energy relations}
\label{sec:discussion}

As described in Secs.~\ref{sec:numerical_results_ii} and \ref{sec:numerical_results_iii},
our nucleon project employing the PACS10 gauge configurations for investigating the nucleon axial structure at the physical point
has successfully reproduced the low-energy relations, such as the PCAC and GGT relations, and the PPD model, within statistical precision in the range of $q^2\lesssim 0.1\ [\mathrm{GeV}^2]$.
This section is devoted to an investigation of the question of whether the low-energy relations are satisfied at the physical point in a much wider range of $q^2$ (beyond $q^2 = 0.1\ [\mathrm{GeV}^2]$).

For this purpose, we use existing data obtained from the $64^4$ lattice ensemble (PACS5/L64) at the coarse lattice spacing ($a=0.09\ \mathrm{fm}$)~\cite{Ishikawa:2021eut}, where large momentum calculations are more straightforward
due to the fact that the spatial extent is half that of the PACS10 ensemble as shown in Table~\ref{tab:simulation_details}.

Data of the nucleon three-point correlation functions
$C_{O_{\alpha}}^{5z}\left(t; \boldsymbol{q}\right)$
have been taken for the momenta $\boldsymbol{q}=\frac{2\pi}{aL}\boldsymbol{n}$ 
with nine types of $\boldsymbol{n}\neq\boldsymbol{0}$, which yield
a wide range of $q^2$ up to $q^2 \sim 0.42\ [\mathrm{GeV}^2]$, as listed in Table~\ref{tab:qsq_list_64}. 
For the computation of the nucleon three-point correlation functions,
$x$, $y$ and $z$ directions are chosen as the polarized direction for the $64^4$ calculation
(PACS5/L64), while only $z$ direction is chosen for the PACS10 ensembles (PACS10/L128 and PACS10/L128).
The average of all polarization directions is used to reduce statistical uncertainty in PACS5/L64.

First, let us examine the PCAC relation in the context of nucleon correlation functions, similar to what was done in Sec.~\ref{sec:numerical_results_ii}.
Recall that 
the PCAC relation of $m_\mathrm{PCAC}^\mathrm{nucl} = m_\mathrm{PCAC}^{\mathrm{pion}}$ is used in the determination of $\widetilde{G}_P(q^2)$ 
as a consequence of the most robust
low-energy relation in the continuum based on the AWT identity in the \textit{new analysis}.

\begin{figure*}
\centering
\includegraphics[width=0.8\textwidth,bb=0 0 864 720,clip]{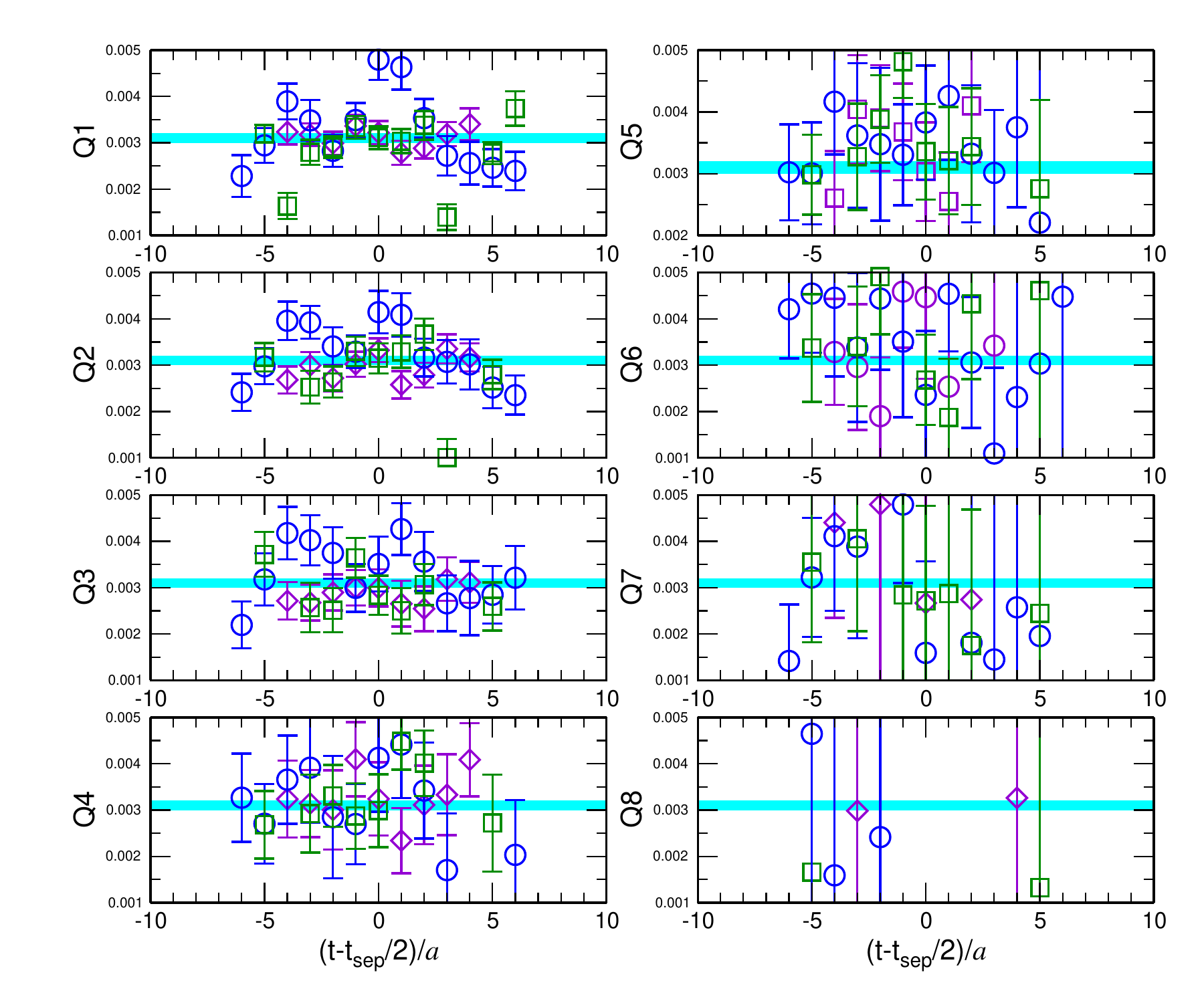}
\caption{
The values of $m_{\mathrm{PCAC}}^{\mathrm{nucl}}$
for all momentum transfers as functions of the current insertion time slice $t$ with PACS5/L64.
Results for $t_\mathrm{sep}/a=\{12,14,16\}$ are plotted as diamonds, squares and circles respectively.
In each panel, the value of $m_{\mathrm{PCAC}}^{\mathrm{pion}}$ is presented as a horizontal band.
}
\label{fig:mpcacnu_plateau_64}
\end{figure*}

Figure~\ref{fig:mpcacnu_plateau_64} shows the ratio defined in Eq.~(\ref{eq:m_awti_pcac}) with all $t_\mathrm{sep}/a=\{12,14,16\}$ for all eight non-zero momentum transfers (labeled from Q1 to Q8).
Only the three data (Q1, Q2 and Q3) in the range of  $q^2 \lesssim 0.15\ [\mathrm{GeV}^2]$ reveal a good plateau in the middle region of $t/a$,
while the other data in the range of $0.2\ [\mathrm{GeV}^2]\lesssim q^2 \lesssim 0.4\ [\mathrm{GeV}^2]$ only show plateau-like behavior due to large errors.
For $q^2 \gtrsim 0.4\ [\mathrm{GeV}^2]$, in particular, Q8 data show no valid signal except near the center of $t/a$ due to increased statistical noise.
The quark mass $m_\mathrm{PCAC}^{\mathrm{nucl}}$ can be evaluated using an error-weighted average of five data points in the middle range of $t/a$ for each $q^2$, except for Q8.
Therefore, Q8 data is not included in Fig.~\ref{fig:mpcacnu_plateau_64},
where a direct comparison of $m_\mathrm{PCAC}^\mathrm{nucl}$ (denoted by red circle symbols) and $m_\mathrm{PCAC}^\mathrm{pion}$ (denoted by cyan horizontal band) is presented.
Figure~\ref{fig:mpcacnu_plateau_64} shows that
the data points of $m_\mathrm{PCAC}^\mathrm{nucl}$ in the range of $q^2 \lesssim 0.4\ [\mathrm{GeV}^2]$ reproduce the value of $m_\mathrm{PCAC}^\mathrm{pion}$. 

It is noted here that since the Q8 data at $q^2 \sim 0.42\ [\mathrm{GeV}^2]$ are not precise enough to examine the PCAC relation, 
the discrepancy between $m_{\mathrm{PCAC}}^{\mathrm{nucl}}$ and $m_{\mathrm{PCAC}}^{\mathrm{pion}}$ cannot be determined within the current statistical precision
at $q^2 \sim 0.42\ [\mathrm{GeV}^2]$.
Therefore, the following \textit{new analysis}, where
the $\widetilde{F}_P$ and $\widetilde{G}_P$ form factors are calculated by the leading $\pi N$ subtraction method, uses the PCAC relation of $m_{\mathrm{PCAC}}^{\mathrm{nucl}}=m_\mathrm{PCAC}^\mathrm{pion}$ in the entire range of $q^2\lesssim 0.42\ [\mathrm{GeV}^2]$, including the Q8 data.

\begin{figure*}
\centering
\includegraphics[width=0.8\textwidth,bb=0 0 792 692,clip]{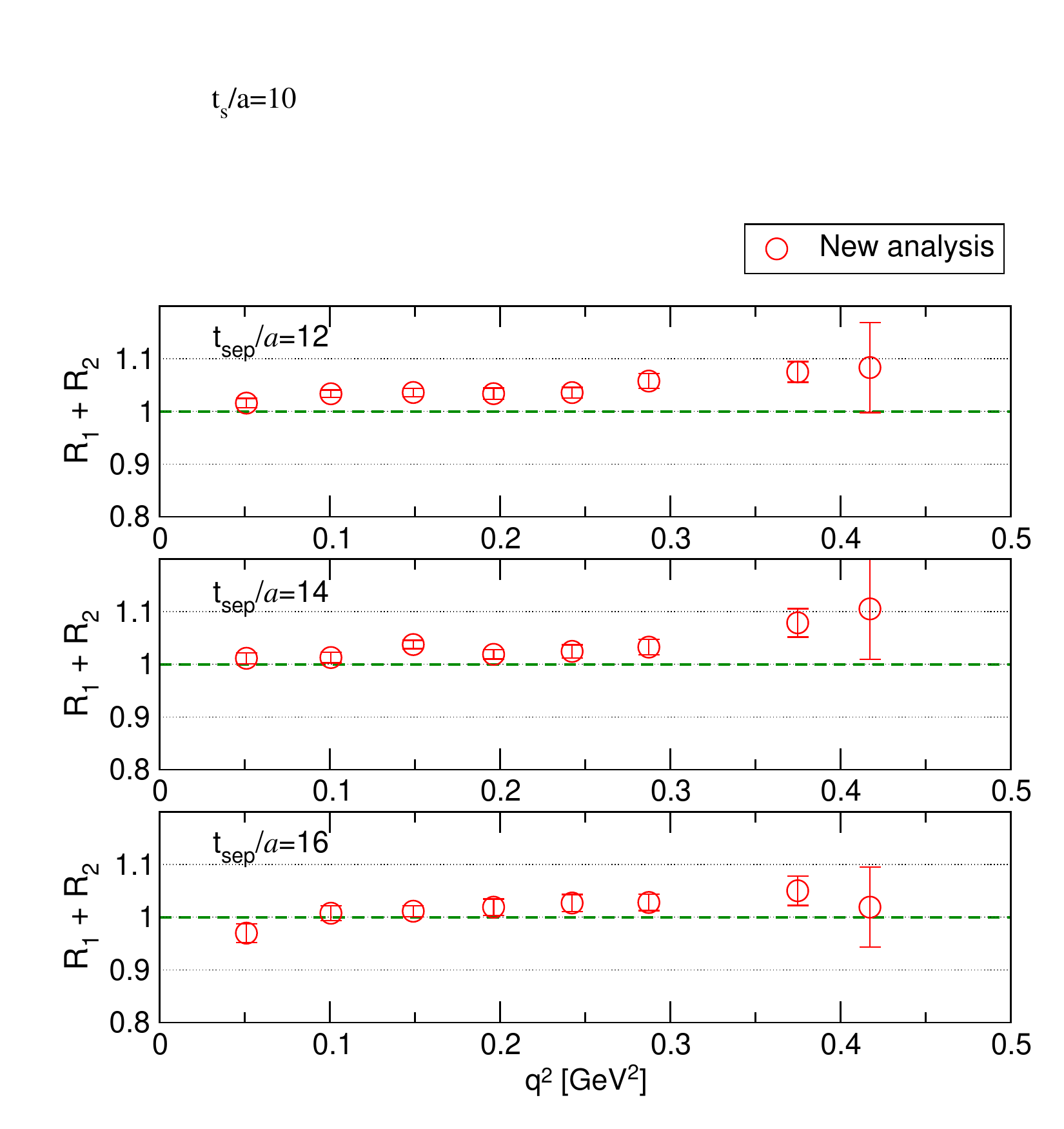}
\caption{
Results of $R_1+R_2$ (open circles) obtained from PACS5/L64 with our \textit{new analysis},
while the dashed line represents the unity.
Results for $t_\mathrm{sep}/a=\{12,14,16\}$ are plotted from top to bottom panels.
}
\label{fig:R1R2_qdep_64}
\end{figure*}

Next, we examine the applicable $q^2$ region of the GGT relation using the form factors obtained by the {\textit{new analysis}}.
Figure~\ref{fig:R1R2_qdep_64} shows 
the value of $R_1+R_2$ defined in Eq.~(\ref{eq:r1r2}) with all $t_\mathrm{sep}/a=\{12,14,16\}$ for all eight values of $q^2$ (including Q8).
The results of $R_1+R_2$ shown in the range of $q^2\lesssim 0.42\ [\mathrm{GeV}^2]$ nearly reproduce unity, although there remains at most 10\% deviation from unity at larger $q^2$, regardless of $t_\mathrm{sep}$. 

It is important to note that when the restriction is applied to the region of $q^2 \lesssim 0.1\ [\mathrm{GeV}^2]$, which corresponds to the region of the PACS10 ensemble results, the deviation from unity is reduced to the 5\% level.

\begin{figure*}
\centering
\includegraphics[width=0.8\textwidth,bb=0 0 792 692,clip]{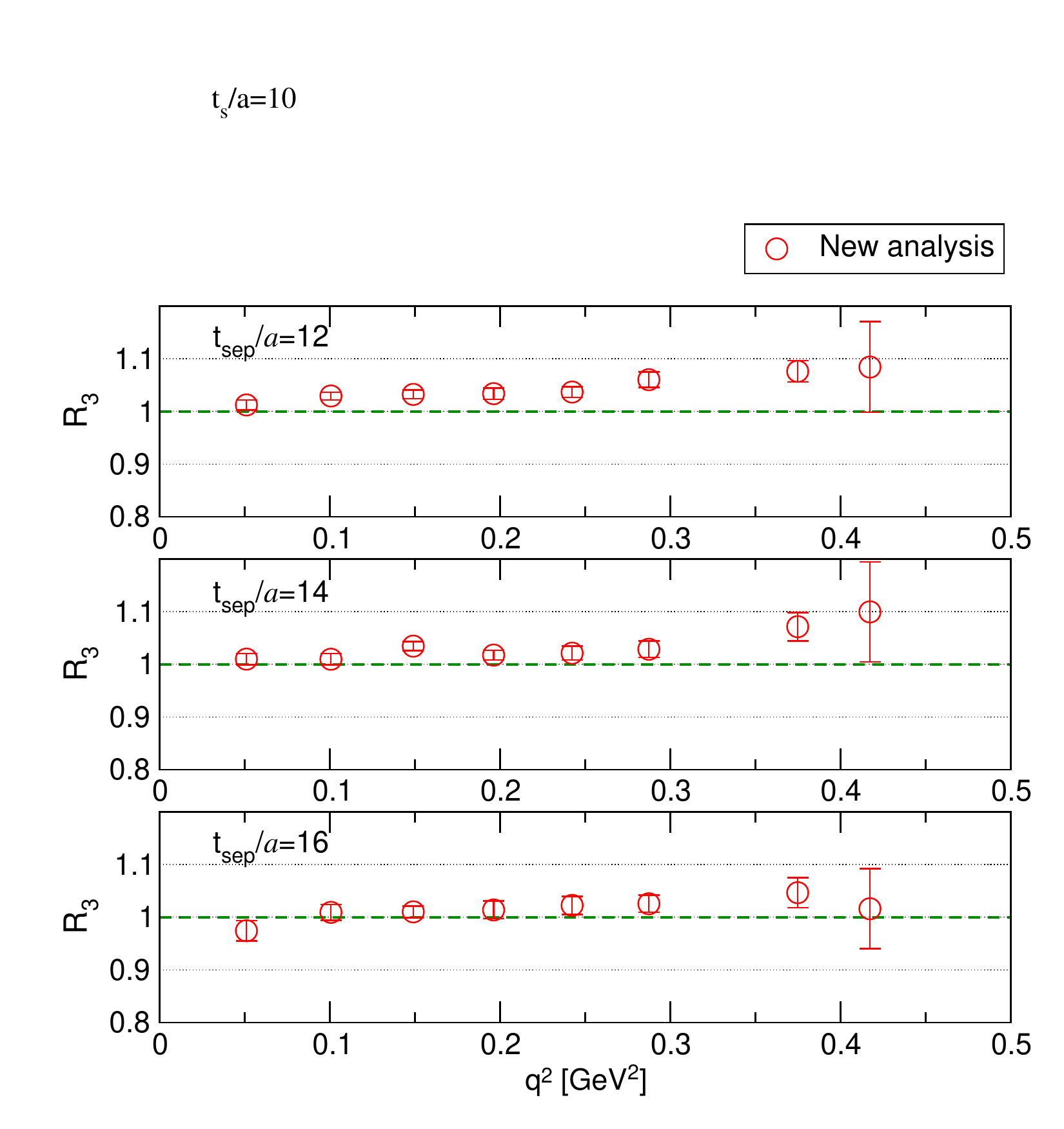}
\caption{
Results of $R_3$ (open circles) obtained from PACS5/L64 with our \textit{new analysis},
while the dashed line represents the unity.
Results for $t_\mathrm{sep}/a=\{12,14,16\}$ are plotted from top to bottom panels.
}
\label{fig:R3_qdep_64}
\end{figure*}

Next, let us consider $R_3$ defined in Eq.~(\ref{eq:r3}) to examine the applicable $q^2$ region of the PPD model for $F_P(q^2)$.
Figure~\ref{fig:R3_qdep_64} shows the values of $R_3$ with $t_\mathrm{sep}/a=\{12,14,16\}$ 
for all eight values of $q^2$ (including Q8).

As in the case of $R_1+R_2$ for testing the GGT relation, the results of
$R_3$ shown in the range of $q^2\lesssim 0.42\ [\mathrm{GeV}^2]$ nearly reproduce unity, although there remains at most 10\% deviation at larger $q^2$, regardless of $t_{\mathrm{sep}}$.

This observation indicates that the PPD model can reproduce the $q^2$ dependence of $F_P(q^2)$ determined by lattice QCD at the physical point up to $q^2\sim 0.42\ [\mathrm{GeV}^2]$ with an accuracy of less than 10\%, even though the PPD model is justified only in the massless limit of $M_\pi$.
In addition, for $q^2\lesssim 0.1\ [\mathrm{GeV}^2]$, the PPD model works with an accuracy of around 5\% level.

\begin{figure*}
\centering
\includegraphics[width=0.8\textwidth,bb=0 0 792 692,clip]{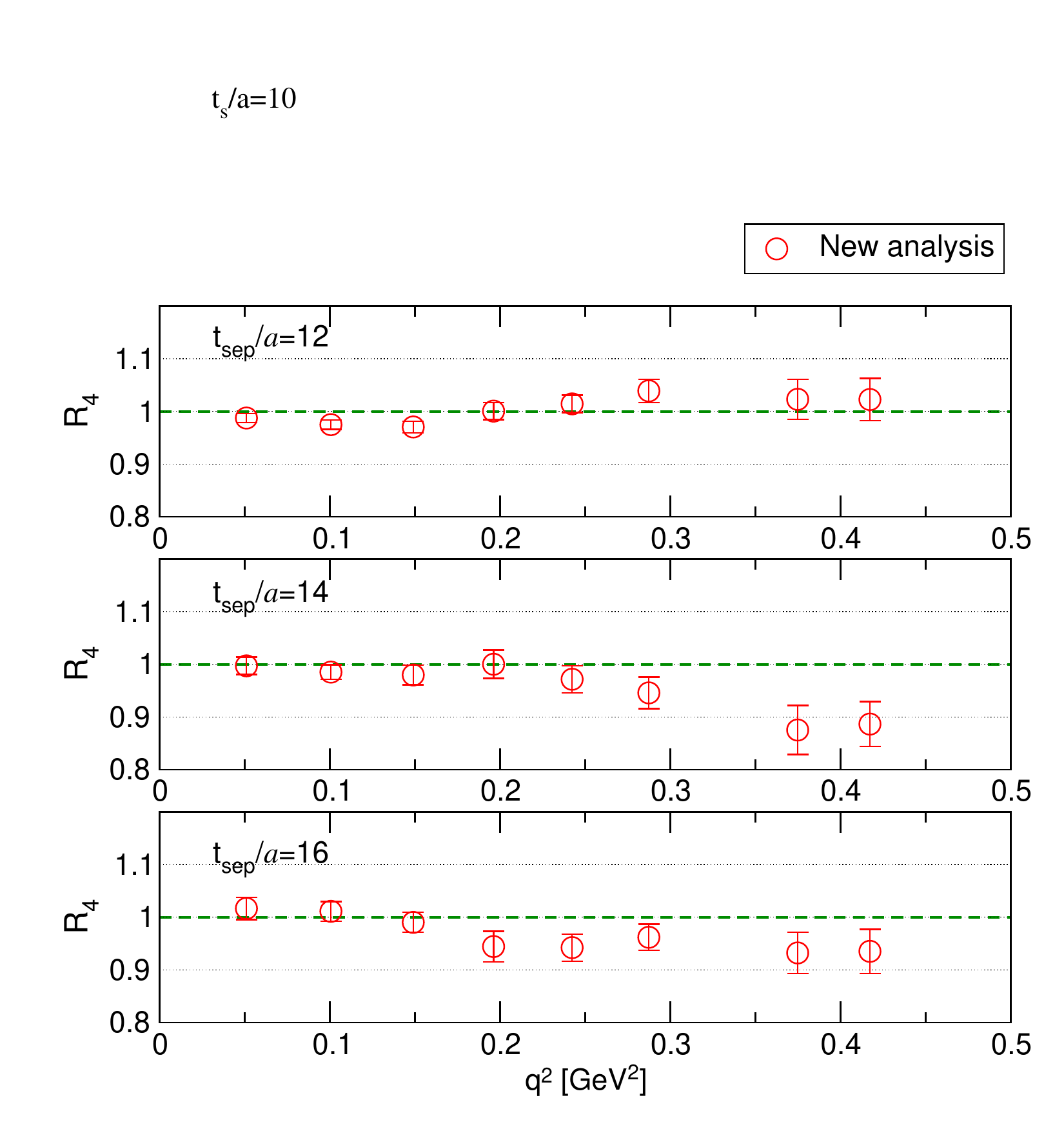}
\caption{
Results of $R_4$ (open circles) obtained from PACS5/L64 with our \textit{new analysis},
while the dashed line represents the unity.
Results for $t_\mathrm{sep}/a=\{12,14,16\}$ are plotted from top to bottom panels.
}
\label{fig:R4_qdep_64}
\end{figure*}

Finally, we examine the applicable $q^2$ region of the GMOR relation obtained as the ratio of $G_P(q^2)$ and $F_P(q^2)$ via the PPD model.
Figure~\ref{fig:R4_qdep_64} shows the values of
$R_4$ defined in Eq.~(\ref{eq:r4}) with all $t_\mathrm{sep}/a=\{12,14,16\}$ for all eight values of $q^2$ (including Q8).

The results of $R_4$ shown in the range of  $q^2\lesssim 0.3\ [\mathrm{GeV}^2]$ exhibit a relatively flat $q^2$ dependence,
and nearly reproduce unity with an accuracy of less than 10\%.
This observation suggests that
for $q^2\lesssim 0.3\ [\mathrm{GeV}^2]$, the GMOR relation remains holds with an accuracy of less than 10\%, while the same level of accuracy is maintained for the GGT relation and the PPD model up to $q^2\sim 0.42\ [\mathrm{GeV}^2]$.
These results are roughly consistent with those of previous studies reported in Ref.~\cite{Jang:2023zts}, where a sophisticated fitting strategy~\cite{Jang:2019vkm} was employed with large sets of $2+1+1$ flavor highly improved staggered quark (HISQ) ensembles generated by the MILC Collaboration.

In addition, it seems to be reasonable that the GMOR relation via the PPD model is applicable only in a narrower $q^2$ region than the other two low-energy relations, since a theoretical connection between
the $F_P$ and $G_P$ form factors requires taking the double limit of $M_\pi\to0$ and $q^2\to 0$ 
in the PPD model (see Appendix B of Ref.~\cite{Sasaki:2007gw}). On the other hand, there is no explicit limitation on either $q^2$ or $M_\pi$ in the GGT relation, while the PPD model for $F_P(q^2)$ may have the condition of $q^2\approx  M_\pi^2$ for small $M_\pi$.

%
%
\begin{table*}[tb]
    \caption{Choices for the nonzero spatial momenta for PACS5/L64: $\boldsymbol{q}=\frac{2\pi}{64a}\times \boldsymbol{n}$. The bottom row shows the degeneracy of $\boldsymbol{n}$ due to the permutation symmetry between $\pm x,\pm y,\pm z$ directions. 
    For $|\boldsymbol{n}|^2=9$, there are two types of integer vectors $\boldsymbol{n}=(2,2,1)$ and (3,0,0). We take the average of these two cases for Q8.
    \label{tab:qsq_list_64}}
\centering
{\scalebox{0.9}{
\begin{tabular}{ccccccccccc}
\hline \hline
    Label& Q0& Q1& Q2& Q3& Q4& Q5& Q6& Q7 & Q8 & Q8\\
          \hline
    $\boldsymbol{n}$& (0,0,0)& (1,0,0)& (1,1,0)& (1,1,1)& (2,0,0)& (2,1,0)& (2,1,1)& (2,2,0)& (2,2,1)& (3,0,0)\\
    $|\boldsymbol{n}|^2$ & 0& 1& 2& 3& 4& 5& 6& 8& 9& 9 \\
    Degeneracy& 1& 6& 12& 8& 6& 24& 24& 12& 24& 6 \\
    $q^2\ [\mathrm{GeV}^2]$ & 0 & 0.051 & 0.101& 0.149 & 0.196 & 0.242 & 0.288 & 0.375 & 0.418 & 0.418\\
\hline \hline
\end{tabular}
}}
\end{table*}
%

%
%
%

%
%
\begin{figure*}
\centering
\includegraphics[width=0.8\textwidth,bb=0 0 792 692,clip]{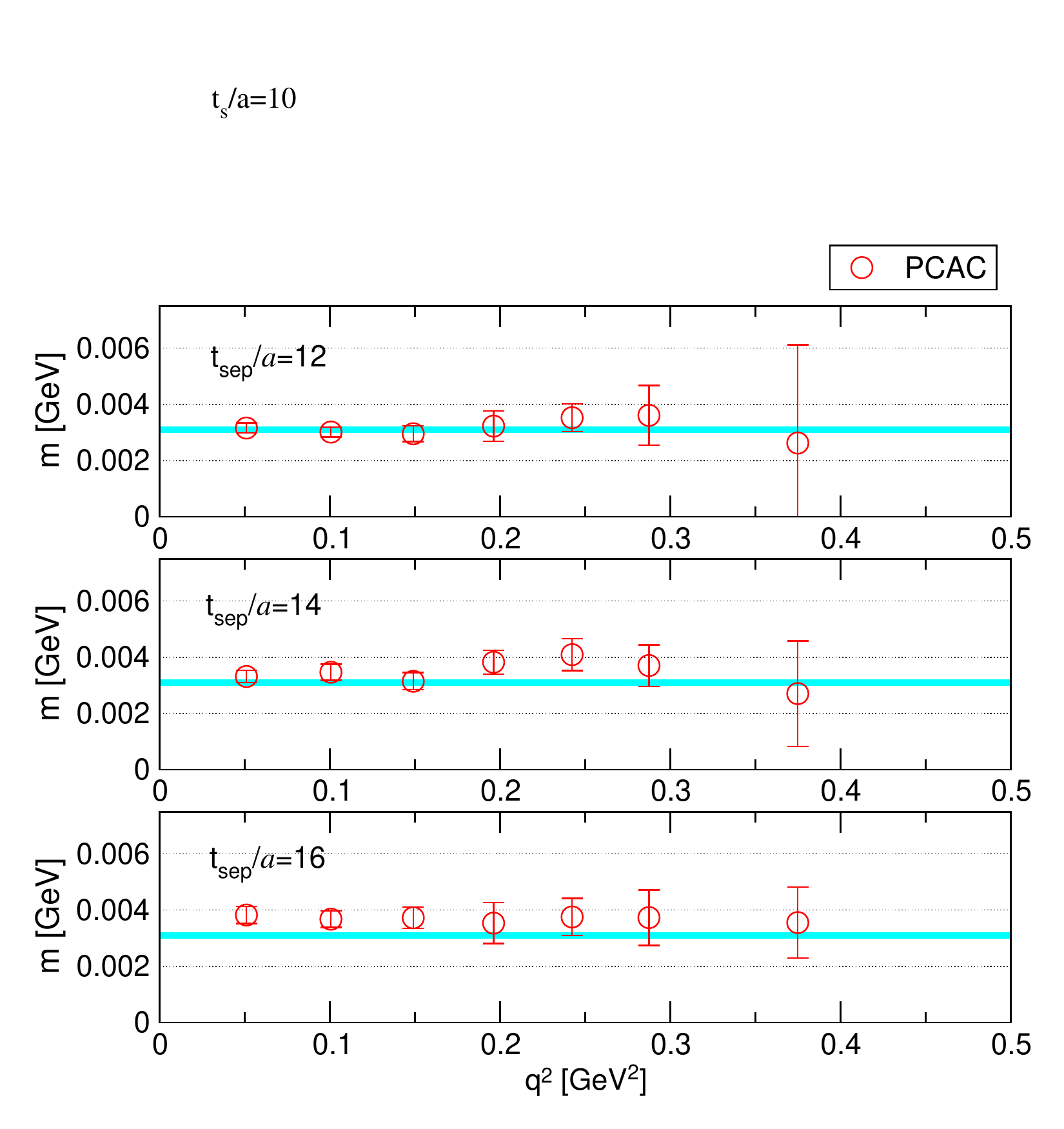}
\caption{
Comparison of the two types of {\textit{bare}} quark mass, $m_\mathrm{PCAC}^\mathrm{pion}$ (horizontal band) and $m_\mathrm{PCAC}^\mathrm{nucl}$ (open circles),
which are obtained from PACS5/L64.
Results for $t_\mathrm{sep}/a=\{12,14,16\}$ are plotted from top to bottom panels.
}
\label{fig:mpcacnu_qdep_64}
\end{figure*}
%

%
%
%

%
%
%

%
%

\section{Summary and outlook}
\label{sec:summary_and_outlook}

Lattice QCD simulations have recently enabled highly precise determinations of observables that can be compared to experimental measurements, including the \textit{isovector} electric and magnetic form factors, as well as the axial form factors of the weak currents. 
The precise calculations for the nucleon form factors are important to understand the dynamics of the strong interaction and provide a basis to explore more complicated quantities, such as the distribution of
mass, spin, and pressure structures inside the nucleon.

In this paper,
before proceeding to study of the mass, spin, and pressure distribution of the nucleon,
which is closely tied to the primary aim of the EIC physics,
we have studied axial matrix elements of the weak currents,
which are relevant to improving the accuracy of the current neutrino oscillation experiments.
We have reanalyzed the three form factors of $F_A(q^2)$, $F_P(q^2)$ and $G_P(q^2)$ by using two sets of the PACS10 ensembles (PACS10/L128 and PACS10/L160) and an additional ensemble with a smaller volume (PACS5/L64), which are reported in Refs.~\cite{Shintani:2018ozy, Ishikawa:2021eut, Tsuji:2023llh}.
The data are obtained from two of three sets of the $2+1$ flavor lattice QCD configurations
generated at the physical point with a spatial extent of over $10\ [\mathrm{fm}]$ (PACS10 project).
In addition, we also use the PACS5/L64 ensemble which has the same parameters as the PACS10/L128 ensemble with half the spatial extent.
The PACS gauge configurations are generated by the PACS Collaboration with the stout-smeared $O(a)$ improved Wilson quark action and Iwasaki gauge action~\cite{Iwasaki:1983iya}.
The AMA technique is used to significantly reduce the computational cost of multiple measurements and to achieve a much higher statistical accuracy.

In Sec.~\ref{sec:numerical_results_i},
we have presented a highlight of our results: the nucleon dispersion relation and five nucleon form factors (the axial ($F_A$), induced pseudoscalar ($F_P$), pseudoscalar ($G_P$), \textit{isovector} electric ($G_E^v$), and \textit{isovector} magnetic ($G_M^v$) form factors), 
which are involved in our previous studies~\cite{Shintani:2018ozy, Ishikawa:2021eut, Tsuji:2023llh, Aoki:2025taf}.
The results of the axial-vector coupling $g_A$ achieve control of all major systematic uncertainties from chiral extrapolation, finite-size effect, excited-state contamination, and discretization effect at the current statistical precision of less than 2\%, and reproduce the experimental value with an accuracy of a few percent.

The three RMS radii: the electric RMS radius, the magnetic RMS radius and the axial RMS radius, are evaluated with a slope of $G_E^v(q^2)$, $G_M^v(q^2)$ and $F_A(q^2)$ defined at $q^2=0$, respectively. 
It was found that the discrepancy in the RMS radii for the different lattice spacings is observed as a possible discretization error of approximately 10\%, regardless of the channel, while the effect on $g_A$ is kept smaller than the statistical error of 2\%. 

Concerning the axial structure of the nucleon,
we have succeeded in completely removing the leading $\pi N$ contribution from the analysis of the $F_P$ and $G_P$ form factors by our \textit{new analysis}. This success allows us to precisely determine 
the induced pseudoscalar coupling $g_P^*$ and the pion-nucleon coupling $g_{\pi NN}$ from $F_P(q^2)$ using the model independent z-expansion method. Our results, obtained at two lattice spacings, agree well with the experimental values. Furthermore these results also exhibit significantly smaller statistical and systematic errors for both $g_P^*$ and $g_{\pi NN}$ compared to the other lattice results.

The PCAC relation derived from the axial Ward--Takahashi identity, is strongly related to the low-energy relations and models such as the generalized Goldberger-Trieman (GGT) relation, the pion-pole dominance (PPD) model, and the Gell-Mann Oakes Renner (GMOR) relation.
We have introduced the two types of \textit{bare} quark mass: $m_\mathrm{PCAC}^{\mathrm{pion}}$ and $m_\mathrm{PCAC}^\mathrm{nucl}$ to verify the PCAC relation in the context of the nucleon correlation functions.
In Sec.\ref{sec:numerical_results_ii}, a direct comparison between $m_\mathrm{PCAC}^{\mathrm{pion}}$ and $m_\mathrm{PCAC}^\mathrm{nucl}$
is presented. Our numerical results show an excellent agreement between $m_\mathrm{PCAC}^{\mathrm{pion}}$ and $m_\mathrm{PCAC}^\mathrm{nucl}$ within statistical precision.
This observation
indicates that the PCAC relation,
which should hold in the low-$q^2$ range in the continuum limit,
is quantitatively satisfied in the range of $q^2 \lesssim 0.1 \ [\mathrm{GeV}^2]$ with finite lattice spacing.

In Sec.~\ref{sec:numerical_results_iii}, we then have 
investigated other low-energy relations, such as the GGT relation and the PPD model, which are associated with the PCAC relation.
For this purpose, we have examined the three tests, which are introduced in Ref.~\cite{Jang:2023zts}, for the $F_A$, $F_P$, and $G_P$ form factors. These form factors are evaluated with both of the \textit{new analysis} and the \textit{traditional analysis} for comparison. 
The three tests are designed to verify the GGT relation, the PPD model of $F_P(q^2)$ and the GMOR relation via the PPD model.
The results of the new analysis demonstrate that the three low-energy
relations are well satisfied in a wide range of $q^2$ ($0.01\ [\mathrm{GeV}^2]\lesssim q^2\lesssim 0.12\ [\mathrm{GeV}^2])$.
This observation suggests that, 
although the GGT relation appears to be violated in the \textit{traditional analysis}, the GGT relation should strictly be valid 
if the residual contamination from the $\pi N$-state contribution is successfully eliminated in the determination of $\widetilde{F}_P(q^2)$ and $\widetilde{G}_P(q^2)$, as in the \textit{new analysis}.
On the other hand, unlike the GGT relation, the PPD model is justified only in the massless limit of $M_\pi$.
Nevertheless, it is found that the PPD model reliably works at the physical point in the range of $q^2 \lesssim 0.1 \ [\mathrm{GeV}^2]$. 
This finding provides the theoretical insight into the PPD model.

Finally, 
Sec.~\ref{sec:discussion} was devoted to an investigation of the question of whether the low-energy relations are satisfied at the physical point in a much wider range of $q^2$ (beyond $q^2 = 0.1\ [\mathrm{GeV}^2]$). 
For this purpose, the $64^4$ lattice ensemble was used, since large momentum calculations are more straightforward due to the fact that the spatial extent is half that of the PACS10 ensemble.
Therefore, the low-energy relations can be verified up to $q^2 \sim 0.42\ [\mathrm{GeV}^2]$.

As for the PCAC relation in the context of the lattice nucleon correlation functions, we have confirmed that $m_\mathrm{PCAC}^\mathrm{nucl}$ sufficiently reproduces the value of $m_\mathrm{PCAC}^{\mathrm{pion}}$ in the range of $q^2 \lesssim 0.4 \ [\mathrm{GeV}^2]$, except the highest $q^2\sim 0.42\ [\mathrm{GeV}^2]$, where no valid signal of $m_\mathrm{PCAC}^\mathrm{nucl}$ is obtained due to increased statistical noise.
Although the PCAC relation of $m_\mathrm{PCAC}^\mathrm{nucl}=m_\mathrm{PCAC}^{\mathrm{pion}}$ was
not numerically confirmed at the highest $q^2\sim 0.42\ [\mathrm{GeV}^2]$, we simply used the leading $\pi N$ subtraction method, where
this relation is used to determine the $G_P$ form factor, in the entire range of $q^2 \lesssim 0.42\ [\mathrm{GeV}^2]$.

As a result, the \textit{new analysis} shows that 
the GGT relation and the PPD model for $F_P(q^2)$ remains holds 
up to $q^2\sim 0.42\ [\mathrm{GeV}^2]$ with an accuracy of less than 10\%, while the same level of accuracy is maintained for 
the GMOR relation via the PPD model only in the range of $q^2 \lesssim 0.3\ [\mathrm{GeV}^2]$.
These results are roughly consistent with those of previous studies reported in Ref.~\cite{Jang:2023zts}, where a sophisticated fitting strategy~\cite{Jang:2019vkm} was employed with large sets of $2+1+1$ flavor highly improved staggered quark (HISQ) ensembles generated by the MILC Collaboration.
Furthermore, the results of our study indicated that for the range $q^2 \lesssim 0.1\ [\mathrm{GeV}^2]$, the three low-energy relations are satisfied at approximately the 5\%-precision level.

Thus, the three form factors, the $F_A$, $F_P$ and $G_P$ form factors related to the nucleon axial structure can be calculated with a very high degree of accuracy using the PACS10 gauge configurations. Therefore, our lattice QCD
calculations can provide first-principles predictions relevant for current and future neutrino experiments.  
For this purpose, the axial radius is the most relevant quantity, though it should be noted that there is a possible discretization error of about 10\% in this quantity.
Therefore, it is necessary to continue our lattice QCD calculations for the axial structure of the nucleon with the third lattice spacing toward the continuum limit. Such a study is currently underway~\cite{Tsuji:2024scy}.

\section*{Acknowledgment}
We would like to thank members of the PACS Collaboration for useful discussions.
Numerical calculations in this work were performed on Oakforest-PACS in Joint Center for Advanced High Performance Computing (JCAHPC) and Cygnus  and Pegasus in Center for Computational Sciences at University of Tsukuba under Multidisciplinary Cooperative Research Program of Center for Computational Sciences, University of Tsukuba, and Wisteria/BDEC-01 in the Information Technology Center, the University of Tokyo. 
This research also used computational resources of the K computer (Project ID: hp180126) and the Supercomputer Fugaku (Project ID: hp20018, hp210088, hp230007, hp230007, hp230199, hp240028, hp240207, hp250037) provided by RIKEN Center for Computational Science (R-CCS), as well as Oakforest-PACS (Project ID: hp170022, hp180051, hp180072, hp190025, hp190081, hp200062),  Wisteria/BDEC-01 Odyssey (Project ID: hp220050) provided by the Information Technology Center of the University of Tokyo / JCAHPC.
This work is supported by the JLDG constructed over the SINET6 of NII.
This work was also supported in part by Grants-in-Aid for Scientific Research from the Ministry of Education, Culture, Sports, Science and Technology (Nos. 18K03605, 19H01892, 22K03612, 23H01195, 25KJ0404) and MEXT as ``Program for Promoting Researches on the Supercomputer Fugaku'' (Search for physics beyond the standard model using large-scale lattice QCD simulation and development of AI technology toward next-generation lattice QCD; Grant Number JPMXP1020230409).


%

\vspace{0.2cm}
\noindent


\appendix

\section{Numerical data for the ratios versus $q^2$}
\label{app:numerical_data_for_the_ratios_versus_q^2}

The three ratios defined in Eqs.~(\ref{eq:r1r2})-(\ref{eq:r4}) obtained from both lattice ensembles of $128^4$ and $160^4$
using our \textit{new analysis} to extract $F_A(q^2)$, $F_P(q^2)$ and $G_P(q^2)$
are summarized in Tables~\ref{tab:R1R2_qdep_128}-\ref{tab:R4_qdep_160}.
For an evaluation of the values of the ratio,
we construct the ratios with the values of the form factors evaluated by our \textit{new analysis} with the standard plateau method for each jackknife sample.

\begin{table}[!h]
\caption{The ratio $(R_1+R_2)$ obtained from PACS10/L128. The error represents the statistical error.}
\label{tab:R1R2_qdep_128}
\centering
\begin{tabular}{cccc}
\hline
\hline
{\bf PACS10/L128}&  $t_{\rm sep}/a=12$ & $t_{\rm sep}/a=14$  & $t_{\rm sep}/a=16$\\
\cline{2-4}
$q^2\ [\mathrm{GeV}^2]$ & $R_1+R_2$ & $R_1+R_2$ & $R_1+R_2$ \\
\hline
0.0129 &  0.997(27) & 0.956(51) & 1.043(86)  \\
0.0257 &  0.992(21) & 0.953(45) & 1.044(72) \\
0.0384 &  0.983(20) & 0.949(44) & 1.048(79) \\
0.0510 &  0.991(20) & 0.959(49) & 1.017(67) \\
0.0635 &  0.982(17) & 0.945(44) & 1.063(65) \\
0.0759 &  0.985(16) & 0.939(45) & 1.060(93) \\
0.1005 &  0.969(27) & 0.939(42) & 1.086(70) \\
\hline
\hline
\end{tabular}
\end{table}
\begin{table}[!h]
\caption{The ratio $(R_1+R_2)$ obtained from PACS10/L160. The error represents the statistical error.}
\label{tab:R1R2_qdep_160}
\centering
\begin{tabular}{cccc}
\hline
\hline
{\bf PACS10/L160} & $t_{\rm sep}/a=13$ & $t_{\rm sep}/a=16$  & $t_{\rm sep}/a=19$\\
\cline{2-4}
$q^2\ [\mathrm{GeV}^2]$ & $R_1+R_2$ & $R_1+R_2$ & $R_1+R_2$ \\
\hline
0.0149 & 1.012(8) & 1.015(8) & 1.026(11) \\
0.0296 & 1.004(8) & 1.002(8) & 1.013(10) \\
0.0442 & 1.006(7) & 0.998(7) & 1.013(10) \\
0.0587 & 1.007(7) & 1.006(7) & 1.020(9) \\
0.0731 & 0.999(8) & 0.996(8) & 1.009(9) \\
0.0874 & 0.997(7) & 0.992(8) & 1.011(10) \\
0.1156 & 0.996(9) & 0.992(9) & 1.012(11) \\
\hline
\hline
\end{tabular}
\end{table}
\begin{table}[!h]
\caption{The ratio $R_3$ obtained from PACS10/L128. The error represents the statistical error.}
\label{tab:R3_qdep_128}
\centering
\begin{tabular}{cccc}
\hline
\hline
{\bf PACS10/L128} & $t_{\rm sep}/a=12$ & $t_{\rm sep}/a=14$  & $t_{\rm sep}/a=16$\\
\cline{2-4}
$q^2\ [\mathrm{GeV}^2]$ & $R_3$ & $R_3$ & $R_3$ \\
\hline
0.0129 & 0.975(29) & 0.966(48) & 1.030(100) \\
0.0257 & 0.981(23) & 0.948(54) & 1.019(60) \\
0.0384 & 0.974(21) & 0.949(51) & 1.024(69) \\
0.0510 & 0.977(19) & 0.961(46) & 0.999(81) \\
0.0635 & 0.973(19) & 0.943(50) & 1.047(56) \\
0.0759 & 0.976(18) & 0.941(49) & 1.037(83) \\
0.1005 & 0.962(29) & 0.940(46) & 1.072(60) \\
\hline
\hline
\end{tabular}
\end{table}
\begin{table}[!h]
\caption{The ratio $R_3$ obtained from PACS10/L160. The error represents the statistical error.}
\label{tab:R3_qdep_160}
\centering
\begin{tabular}{cccc}
\hline
\hline
{\bf PACS10/L160} & $t_{\rm sep}/a=13$ & $t_{\rm sep}/a=16$  & $t_{\rm sep}/a=19$\\
\cline{2-4}
$q^2\ [\mathrm{GeV}^2]$ & $R_3$ & $R_3$ & $R_3$ \\
\hline
0.0149 & 1.009(9)  & 1.006(11) & 1.023(14) \\
0.0296 & 0.993(9)  & 0.990(9)  & 1.003(12) \\
0.0442 & 0.996(8)  & 0.983(8)  & 1.002(11) \\
0.0587 & 1.006(8)  & 0.997(8) & 1.012(11) \\
0.0731 & 0.990(9)  & 0.987(9) & 1.004(10) \\
0.0874 & 0.986(8)  & 0.980(9)  & 1.005(11) \\
0.1156 & 0.988(10) & 0.982(10) & 1.008(12) \\
\hline
\hline
\end{tabular}
\end{table}
\begin{table}[!h]
\caption{The ratio $R_4$ obtained from PACS10/L128. The error represents the statistical error.}
\label{tab:R4_qdep_128}
\centering
\begin{tabular}{cccc}
\hline
\hline
{\bf PACS10/L128} & $t_{\rm sep}/a=12$ & $t_{\rm sep}/a=14$  & $t_{\rm sep}/a=16$\\
\cline{2-4}
$q^2\ [\mathrm{GeV}^2]$ & $R_4$ & $R_4$ & $R_4$ \\
\hline
0.0129 & 0.962(14) & 1.017(24) & 0.978(39) \\
0.0257 & 0.973(14) & 0.987(30) & 0.944(37) \\
0.0384 & 0.970(15) & 0.998(33) & 0.934(36) \\
0.0510 & 0.947(22) & 1.009(39) & 0.933(67) \\
0.0635 & 0.958(20) & 0.992(41) & 0.939(48) \\
0.0759 & 0.954(20) & 1.013(40) & 0.896(59) \\
0.1005 & 0.958(24) & 1.005(50) & 0.923(72) \\
\hline
\hline
\end{tabular}
\end{table}
\begin{table}[!h]
\caption{The ratio $R_4$ obtained from PACS10/L160. The error represents the statistical error.}
\label{tab:R4_qdep_160}
\centering
\begin{tabular}{cccc}
\hline
\hline
{\bf PACS10/L160} & $t_{\rm sep}/a=13$ & $t_{\rm sep}/a=16$  & $t_{\rm sep}/a=19$\\
\cline{2-4}
$q^2\ [\mathrm{GeV}^2]$ & $R_4$ & $R_4$ & $R_4$ \\
\hline
0.0149 & 0.995(10) & 0.983(11) & 0.994(13) \\
0.0296 & 0.972(8)  & 0.970(8)  & 0.976(13) \\
0.0442 & 0.966(9)  & 0.954(9)  & 0.966(14) \\
0.0587 & 0.996(13) & 0.965(14) & 0.968(20)\\
0.0731 & 0.958(12) & 0.955(11) & 0.978(14)\\
0.0874 & 0.945(13) & 0.934(11) & 0.968(19) \\
0.1156 & 0.947(16) & 0.939(14) & 0.970(20) \\
\hline
\hline
\end{tabular}
\end{table}

\let\doi\relax


\clearpage
\bibliographystyle{ptephy}
\bibliography{ptephy}

\end{document}